\documentclass[11pt,a4paper]{article}
\usepackage{authblk}
\usepackage{lmodern}
\usepackage{jheppub}
\usepackage{amsmath}
\usepackage{mathtools}
\usepackage{tikz}
\usepackage{empheq}
\usepackage{enumitem}
\usepackage{graphics}
\usepackage{wrapfig}
\usepackage{caption}

\usepackage{natbib}
\usepackage{xcolor}
\usepackage{framed}
\usepackage{array}
\definecolor{shadecolor}{gray}{0.925}

\usepackage[cbgreek]{textgreek}

\def\sideremark#1{\ifvmode\leavevmode\fi\vadjust{\vbox to0pt{\vss
 \hbox to 0pt{\hskip\hsize\hskip1em
 \vbox{\hsize3cm\tiny\raggedright\pretolerance10000
 \noindent #1\hfill}\hss}\vbox to8pt{\vfil}\vss}}}%

\newcommand{\bi}{\begin{itemize}}
\newcommand{\ei}{\end{itemize}}
\newcommand{\bea}{\begin{align}}
\newcommand{\eea}{\end{align}}
\newcommand{\be}{\begin{equation}}
\newcommand{\ee}{\end{equation}}

\newcommand{\pl}{{\partial}}

\newcommand{\tcb}{\textcolor{blue}}

\makeatletter
\renewcommand*\env@matrix[1][\arraystretch]{%
  \edef\arraystretch{#1}%
  \hskip -\arraycolsep
  \let\@ifnextchar\new@ifnextchar
  \array{*\c@MaxMatrixCols c}}
\makeatother

\author[\ensuremath{a},\ensuremath{b}]{Charlotte SLEIGHT}
\author[\ensuremath{c},\ensuremath{d}]{\quad Massimo TARONNA}

\affiliation[\ensuremath{a}]{Centre for Particle Theory and Department of Mathematical Sciences, \\ Durham University, Durham, DH1 3LE, U.K.}

\affiliation[\ensuremath{b}]{School of Natural Sciences, Institute for Advanced Study,\\
1 Einstein Drive, Princeton, NJ 08540}

\affiliation[\ensuremath{c}]{Dipartimento di Fisica ``Ettore Pancini'', Universit\`a degli Studi di Napoli Federico II, \\Monte S. Angelo, Via Cintia, 80126 Napoli, Italy}

\affiliation[\ensuremath{d}]{INFN, Sezione di Napoli, Monte S. Angelo, Via Cintia, 80126 Napoli, Italy}

\emailAdd{charlotte.sleight@durham.ac.uk, massimo.taronna@unina.it,}

\title{\centering \huge On the consistency of (partially-)massless\\ matter couplings in de Sitter space}

\abstract{We study the consistency of the cubic couplings of a (partially-)massless spinning field to two scalars in $\left(d+1\right)$-dimensional de Sitter space. Gauge invariance of observables with external (partially)-massless spinning fields translates into Ward-Takahashi identities on the boundary. Using the Mellin-Barnes representation for boundary correlators in momentum space, we give a systematic study of Ward-Takahashi identities for tree-level 3- and 4-point processes involving a single external (partially-)massless field of arbitrary integer spin-$J$. 3-point Ward-Takahashi identities constrain the mass of the scalar fields to which a (partially-)massless spin-$J$ field can couple. 4-point Ward-Takahashi identities then constrain the corresponding cubic couplings. For massless spinning fields, we show that Weinberg's flat space results carry over to $\left(d+1\right)$-dimensional de Sitter space: For spins $J=1,2$ gauge-invariance implies charge-conservation and the equivalence principle while, assuming locality, higher-spins $J>2$ cannot couple consistently to scalar matter. This result also applies to anti-de Sitter space. For partially-massless fields, restricting for simplicity to those of depth-2, we show that there is no consistent coupling to scalar matter in local theories. Along the way we also give a detailed account of how contact amplitudes with and without derivatives are represented in the Mellin-Barnes representation. Various new explicit expressions for 3- and 4-point functions involving (partially-)massless fields and conformally coupled scalars in dS$_4$ are given.}

\begin{document}

\begin{flushright}    
\texttt{}
\end{flushright}

\maketitle

\newpage

\section{Introduction}

In the past decades the Holographic principle has seen a number of key developments in the study of observables in Quantum Gravity, especially in the context of the AdS/CFT correspondence \cite{Maldacena:1997re}. Scattering processes in $\left(d+1\right)$-dimensional asymptotically anti-de Sitter (AdS) space can be re-cast as correlation functions of local operators in a $d$-dimensional Conformal Field Theory (CFT), which are defined non-perturbatively by a combination of conformal symmetry, unitarity and a consistent operator product expansion. These are the three main pillars of the Conformal Bootstrap programme, which aims to carve out the space of consistent CFTs using symmetry and mathematical consistency \cite{Simmons-Duffin:2016gjk,Poland:2018epd} and is a spectacular fully-functioning revival of the Bootstrap philosophy put forward in the early days of S-matrix theory \cite{jacob1964strong,LR1967707}. 

Recent years have seen a renewed interested in the challenge to extend the successes of the Bootstrap and the AdS/CFT paradigm to more general backgrounds, in particular those that are closer to real world. A natural testing ground is provided by de Sitter (dS) space, which shares the isometry group with Euclidean AdS. Indeed, it has long been known that correlators on the boundary of dS are constrained by conformal symmetry \cite{Antoniadis:2011ib,Creminelli:2011mw,Creminelli:2012ed,Bzowski:2011ab,Mata:2012bx,Kundu:2014gxa,Kundu:2015xta,Pajer:2016ieg,Shukla:2016bnu,Farrow:2018yni}. This has since evolved into the Cosmological Bootstrap \cite{Arkani-Hamed:2015bza,Arkani-Hamed:2017fdk,Arkani-Hamed:2018kmz,Kim:2019wjo,Sleight:2019mgd,Sleight:2019hfp,Baumann:2019oyu,Green:2020ebl,Baumann:2020dch,Sleight:2020obc,Goodhew:2020hob,Cespedes:2020xqq,Pajer:2020wxk,Kim:2021pbr,Jazayeri:2021fvk,Melville:2021lst,Goodhew:2021oqg}, which aims to identify the symmetries and consistency criteria that should be satisfied by boundary correlators in dS (which, in contrast, for the AdS case above are known) and apply them to constrain (or even completely determine) the form that such boundary correlators can take.

One of the first instances in which symmetry and consistency were used to successfully deduce model independent properties of quantum theories is Weinberg's seminal 1964 result \cite{Weinberg:1964ew,Weinberg:1965rz} on the couplings of massless particles of arbitrary integer spin to scalar matter in flat space.\footnote{This was extended to Fermions and supersymmetric theories in \cite{Grisaru:1976vm,Grisaru:1977kk}.} Weinberg showed that locality and unitarity constrain the S-matrix for the emission of a single massless spin-$J$ particle with momentum $q$ to take the following form in the soft limit $q\to 0$:
\begin{equation}\label{softwein}
    S\left(p_1,\ldots, p_N, q, \epsilon \right) \approx \left[\sum\limits^N_{a=1} \frac{g_a\left(\epsilon \cdot p_a\right)^J}{-2 q \cdot p_a}\right] \times S\left(p_1,\ldots, p_N \right),
\end{equation}
where $S\left(p_1,\ldots, p_N \right)$ is the S-matrix for the process before the emission of the massless spin-$J$ particle, $g_a$ is its coupling to the $a$-th external particle and $\epsilon$ its polarization vector which is null $\left(\epsilon^2=0\right)$ and transverse $\left(q \cdot \epsilon =0\right)$. This however gives a redundant description of the massless spin-$J$ particle and (gauge invariance) requires that the spurious longitudinal components decouple, viz.
\begin{equation}\label{ginvwein}
  \left(q \cdot D_{\epsilon}\right)  S\left(p_1,\ldots, p_N, q, \epsilon \right) = 0.
\end{equation}
From the form \eqref{softwein} for the S-matrix in the limit $q\to 0$, this gives the constraint
\begin{equation}
    \sum\limits^N_{a=1} g_a\left(\epsilon \cdot p_a\right)^{J-1} = 0, \qquad \forall \, p_a.
\end{equation}
For spin $J=1$ this constraint implies conservation of charge:
\begin{equation}\label{ccintro}
    \sum\limits^N_{a=1} g_a=0,
\end{equation}
for $J=2$ it implies energy-momentum conservation and the principle of equivalence
\begin{equation}\label{eqintro}
   \sum\limits^N_{a=1} p_a=0 \qquad \text{and} \qquad g_1=\ldots = g_N,
\end{equation}
and for $J>2$ that there is no coupling of a massless higher-spin particle  to scalar matter,
\begin{equation}\label{hsintro}
    g_1=\ldots = g_N=0.
\end{equation}
The beauty of this argument lies in its universality: Fundamental features of theories of massless spinning particles, such as charge conservation and the equivalence principle, follow in a model independent way as a simple consequence of locality and unitarity. It also has the strength to rule out (in local theories) interactions of certain collections of particles altogether.

Extending Weinberg's analysis to a curved background has long faced various difficulties, mostly related to the problem of defining an S-matrix on spaces with non-vanishing curvature. Given the recent developments in adapting S-matrix techniques to boundary correlators in (A)dS space however, we are increasingly in a position to start tackling this problem concretely and, in the specific case of couplings to conformally coupled scalars in dS$_4$,\footnote{Note that, results for scalars of certain masses -- which includes massless scalars relevant for inflation -- can be obtained from those for conformally coupled scalars by acting with a differential "weight-shifting" operator \cite{Arkani-Hamed:2015bza,Arkani-Hamed:2018kmz,Baumann:2020dch}.} we have already seen significant progress \cite{Baumann:2020dch} for massless particles of spins $J=1,2$. The story in dS space is moreover particularly rich from a phenomenological perspective, where unitary irreducible representations of the de Sitter isometry group admit not only massless but also \emph{partially-}massless spinning particles \cite{Deser:1983mm,Higuchi:1986py,Deser:2001pe,Deser:2001us,Deser:2001us,Dolan:2001ih,Deser:2003gw}, whose consistent scattering observables also require the decoupling of (a subset of) the longitudinal components. A natural question is then if the couplings of such partially-massless particles can be similarly constrained in a model-independent way.

In this work we demonstrate how the Mellin-Barnes formalism for boundary correlators in momentum space introduced in \cite{Sleight:2019mgd,Sleight:2019hfp} can be used to extend Weinberg's analysis to (A)dS$_{d+1}$, including the matter couplings of partially-massless fields peculiar to the dS case. For boundary correlators in (A)dS, the gauge-invariance constraint \eqref{ginvwein} on S-matrix elements is replaced by a Ward-Takahashi identity which relates the longitudinal components to lower-point correlators with the massless external field removed. As we shall see, Ward-Takahashi identities at the level of the Mellin-Barnes representation are encoded in a particular form of polynomial \eqref{fansatz} in the Mellin variables. Upon computing three- and four-point functions of a (partially)-massless field with scalars (see figures \ref{fig::3pt} and \ref{fig::exch}) this feature allows us to systematically study the constraints from the Ward-Takahashi identities. 

By studying constraints from Ward-Takahashi identities at the three-point level, we recover the results \cite{Joung:2012rv,Joung:2012hz} that gauge-invariance constrains \eqref{depthcond} the scaling dimensions of the scalar fields to which a (partially-)massless field can couple. At four-point, we find that the Ward-Takahashi identity is generally violated by terms that are singular in the total energy $E_T$. This observation was also made in \cite{Baumann:2020dch} where, for couplings to conformally coupled scalars in dS$_4$, it was shown that charge conservation \eqref{ccintro} and the equivalence principle \eqref{eqintro} ensure that the Ward-Takahashi identity is satisfied. One should also rule out the possibility that the total energy singularities violating the Ward-Takahashi identity cannot be compensated by adding local quartic contact terms -- i.e. those generated by quartic vertices with a finite number of derivatives involving one massless spin-$J$ field and the three external scalars. In flat space it is clear, since local quartic contact amplitudes do not contribute singularities in $q$ and are therefore subleading in the soft limit \eqref{softwein}. In (A)dS the separation appears less sharp. Contact amplitudes associated to local quartic vertices in (A)dS tend to dominate in the limit $E_T \to 0$, with the singularity in $E_T$ increasing with the number of derivatives.\footnote{These have been classified in \cite{Arkani-Hamed:2018kmz} for quartic contact diagrams involving only scalar fields.} 
By translating the problem into the Mellin-Barnes representation, where the local contact terms are associated to polynomials in the Mellin variables of a minimum degree, we are able to establish that there are total energy singularities violating the Ward-Takahashi identity which cannot be compensated by adding local quartic vertices to the theory. In particular, there are singularities in $E_T$ that violate the Ward-Takahashi identity that are of a too low degree to be generated by a local quartic vertex. From this, combined with our results for four-point exchanges with a single external (partially)-massless spin-$J$ field, one can establish the following:

\begin{itemize}
    \item {\bf Massless spin-$J$ fields in (A)dS$_{d+1}$:} Weinberg's conclusions on the couplings of massless spinning fields to scalar matter carry over to (A)dS. In particular, the four-point Ward-Takahashi identities require: Charge conservation \eqref{ccintro} for $J=1$, the equivalence principle \eqref{eqintro} for $J=2$ and that, in local theories, there is no coupling of massless higher-spin fields $J>2$ to scalar matter \eqref{hsintro}.
    \item {\bf Partially-massless spin-$J$ field of depth-2 in dS$_{d+1}$:}\footnote{For simplicity we focused on the scalar matter couplings to depth-2 partially-massless fields, though from considering various examples for other depths we expect that this result holds for all non-zero depths.} The four-point Ward-Takahashi identities imply that partially-massless fields of depth-2, which can have spin-3 and higher, cannot couple consistently to scalar matter in local theories. 
\end{itemize}

The paper is organised as follows. In section \ref{sec::3pt} we introduce the Mellin-Barnes representation of three-point boundary correlators in (A)dS$_{d+1}$,\footnote{In the case of dS$_{d+1}$, by boundary correlators we mean late time in-in correlators computed within the in-in/Schwinger-Keldysh formalism \cite{Maldacena:2002vr,Weinberg:2005vy} (for a review see \cite{Chen:2017ryl}). These should not be confused with wavefunction coefficients which are sometimes (with an abuse of terminology) are referred to in the literature as correlators.} focusing on the case of correlators involving two scalars and a spin-$J$ field. For (partially-)massless spin-$J$ field we study how gauge-invariance manifests itself in the Mellin-Barnes representation and how it constrains the masses of the scalar fields to which it can couple. We derive three-point Ward-Takahashi identities for massless spinning fields and partially-massless spinning fields of depths 1 and 2. We also provide a double check of these results by deriving the Ward-Takahashi identities in $d=3$ for the case that the scalars are conformally coupled, where the Mellin-Barnes integrals in the three-point correlators can be lifted completely. We also give various new explicit expressions for three-point correlators of (partially-)massless fields with conformally coupled scalars, including all lower helicity components.

In section \ref{sec::4ptfunc} we introduce the Mellin-Barnes representation of four-point functions, focusing on tree-level processes -- namely, four-point exchanges and quartic contact diagrams (including those of derivative interactions). We show how quartic contact diagrams can be packaged as improvement terms to cubic vertices in a four-point exchange and how this is naturally described within the Mellin-Barnes formalism.

In section \ref{sec::4ptWT}, focusing on four-point functions involving three scalars and a single (partially)-massless spin-$J$ field in (A)dS$_{d+1}$, we explore the constraints coming from gauge-invariance. We show that the Ward-Takahashi identity, for generic cubic couplings, is violated by quartic contact terms and argue that this cannot be restored by the addition of local quartic vertices -- thus leading to a constraint on the cubic couplings of (partially)-massless fields to scalars. We verify this explicitly for the case of massless spin-$J$ fields and partially-massless spin-$J$ fields of depth 2, deriving the corresponding constraints on the cubic couplings. As for the three-point functions in section \ref{sec::3pt}, we provide a check of these results in $d=3$ in the case that the scalars are conformally coupled, where the Mellin-Barnes integrals can be lifted completely. We give various explicit expressions for four-point exchanges involving conformally coupled scalars and a single external massless spinning field, including all lower helicity components.

Various technical details are relegated to the appendices.

\paragraph{Notation and conventions.} Throughout we denote scalar fields by the symbol $\phi$ and spinning fields by $\varphi$. A scalar operator on the boundary with scaling dimension $\Delta=\frac{d}{2}+i\nu$ is denoted by $O_\nu$. If the operator instead has spin-$J$ it is denoted by $O_{\nu,J}$. The $d$-dimensional spatial vector ${\bf x}$ parameterises the boundary directions and ${\bf k}$ denotes the boundary momentum. These have magnitudes denoted by $x=|{\bf x}|$ and $k=|{\bf k}|$. In dS$_{d+1}$ we work with metric signature $\left(-\,+ \ldots +\,+\,\right)$.

\section{Three-point functions}
\label{sec::3pt}

We begin in section \ref{subsec::MBrep} by reviewing and extending the relevant aspects of the Mellin-Barnes representation for three-point functions in momentum space introduced in \cite{Sleight:2019mgd,Sleight:2019hfp}. In section \ref{subsec::WSop} we introduce some useful differential operators which can be used to derive relations between correlators with operator scaling dimensions and spins that differ by integer shifts, as well as correlators generated by derivative interactions. In section \ref{sec:3ptWard} we consider three-point functions of a (partially)-massless field and two scalars in (A)dS$_{d+1}$. We describe how the constraints from gauge-invariance manifest themselves in the Mellin-Barnes representation and derive explicit expressions for the corresponding three-point Ward-Takahashi identities. In section \ref{subsec::3ptimp} we detail how the freedom to add improvement (i.e. on-shell vanishing) terms to cubic vertices can be used to simplify the Mellin-Barnes representation of three-point functions. In section \ref{subsec::cc3pt} we consider the special case in which the two scalar fields are conformally coupled in $d=3$. In this case the Mellin-Barnes representation is not required to describe the correlator completely and the integrals can be lifted to give explicit closed form expressions for the three-point function of a (partially-)massless field and two conformally coupled scalars. These results provide a cross-check of the three-point Ward-Takahashi identities we derived in section \ref{sec:3ptWard}.

Momentum space three-point functions of (partially)-massless fields in (A)dS have been studied in various works. For a (most-likely) incomplete list see \cite{Maldacena:2002vr,Maldacena:2011nz,Bzowski:2011ab,Mata:2012bx,Bzowski:2013sza,Arkani-Hamed:2015bza,Bzowski:2017poo,Anninos:2017eib,Coriano:2018bbe,Goon:2018fyu,Coriano:2018bsy,Bzowski:2018fql,Farrow:2018yni,Isono:2019ihz,Sleight:2019hfp,Lipstein:2019mpu,Baumann:2020dch}.

\subsection{Mellin-Barnes representation}
\label{subsec::MBrep}

The \emph{Mellin-Barnes representation} of a generic three-point conformal correlation function in $d$-dimensional momentum space is defined as 
\begin{multline}\label{3ptcormom}
    \langle O_{\nu_1,J_1}\left({\bf k}_1\right)O_{\nu_2,J_2}\left({\bf k}_2\right)O_{\nu_3,J_3}\left({\bf k}_3\right)  \rangle^\prime \\ = \int^{i\infty}_{-i\infty} \left[ds\right]_3\, \langle O_{\nu_1,J_1}\left({\bf k}_1\right)O_{\nu_2,J_2}\left({\bf k}_2\right)O_{\nu_3,J_3}\left({\bf k}_3\right)  \rangle^\prime_{s_1,s_2,s_3},
\end{multline}
where in the usual way the prime denotes the correlator with the momentum conserving delta function stripped off,
\begin{equation}\nonumber
  \hspace*{-0.5cm}  \langle O_{\nu_1,J_1}\left({\bf k}_1\right)O_{\nu_2,J_2}\left({\bf k}_2\right)O_{\nu_3,J_3}\left({\bf k}_3\right)  \rangle = \left(2\pi\right)^d \delta^{\left(d\right)}\left({\bf k}_1+{\bf k}_2+{\bf k}_3\right)  \langle O_{\nu_1,J_1}\left({\bf k}_1\right)O_{\nu_2,J_2}\left({\bf k}_2\right)O_{\nu_3,J_3}\left({\bf k}_3\right)  \rangle^\prime.
\end{equation}
The operator $O_{\nu_j,J_j}$ has spin $J_j$ and its scaling dimension $\Delta^+_j$ parameterised as $\Delta^+_j=\frac{d}{2}+i\nu_j$, so that the shadow scaling dimension is given by sending $\nu_j \to -\nu_j$ i.e. $\Delta^-_j=\frac{d}{2}-i\nu_j$.\footnote{Note that, throughout, the parameters $\nu_j$ are not necessarily real. The constraint $\nu_j \in \mathbb{R}$ defines Principle series representations and at the level of the Mellin-Barnes representation ensures that the integration contours do not get pinched. Other representations can be obtained from the Principle Series by analytic continuation and careful treatment of any divergences, for which we refer the reader to \cite{Sleight:2019hfp}. See \cite{Basile:2016aen} for a nice overview of unitary irreducible representations in (anti-)de Sitter space.} We refer to the variables $s_j$ as \emph{Mellin variables}, which are assigned to each momentum ${\bf k}_j$.\footnote{Later on Mellin variables will be divided into \emph{external} and \emph{internal} Mellin variables, associated to external and internal momenta respectively. The $s_j$ above are therefore external Mellin variables.} The Mellin-Barnes representation can be expressed in the form: 
\begin{multline}\label{MBrep3pt}
  \langle O_{\nu_1,J_1}\left({\bf k}_1\right)O_{\nu_2,J_2}\left({\bf k}_2\right)O_{\nu_3,J_3}\left({\bf k}_3\right)  \rangle^\prime_{s_1,s_2,s_3}  = {\cal A}_{\nu_1,J_1;\nu_2,J_2;\nu_3,J_3}\left(s_1,{\bf k}_1,\boldsymbol{\epsilon}_1;s_2,{\bf k}_2,\boldsymbol{\epsilon}_2;s_3,{\bf k}_3,\boldsymbol{\epsilon}_3\right)  \\  \times \rho_{\nu_1,\nu_2,\nu_3}\left(s_1,s_2,s_3\right)\prod^3_{j=1}\left(\frac{k_j}{2}\right)^{-2s_j+i\nu_j},
\end{multline}
where the $\boldsymbol{\epsilon}_j$ are null auxiliary vectors $\boldsymbol{\epsilon}_j \cdot \boldsymbol{\epsilon}_j = 0$ encoding the tensor structure (as in e.g. \cite{Costa:2011mg}):
\begin{equation}
    O_{\nu,J} = \left(O_{\nu}\right)_{i_1 ...i_J} \epsilon^{i_1} \ldots \epsilon^{i_J}, \qquad i_a = 1, \ldots, d, \qquad a = 1, \ldots, J.
\end{equation}
The function $\rho_{\nu_1,\nu_2,\nu_3}\left(s_1,s_2,s_3\right)$ carries two infinite families of poles for each Mellin variable,
\begin{equation}\label{rho3}
    \rho_{\nu_1,\nu_2,\nu_3}\left(s_1,s_2,s_3\right)=\prod^3_{j=1}\frac{1}{2\sqrt{\pi}}\Gamma\left(s_j+\tfrac{i\nu_j}{2}\right)\Gamma\left(s_j-\tfrac{i\nu_j}{2}\right).
\end{equation}
These poles are associated to the Mellin-Barnes representation of the corresponding bulk-boundary propagators, which are given by a type of Bessel functions \cite{Raju:2010by}. The function ${\cal A}_{\nu_i,J_i}\left(s_i,{\bf k}_i,\boldsymbol{\epsilon}_i\right)$ is what we refer to throughout as the \emph{Mellin-Barnes amplitude}, 
\begin{multline}\label{MBamplitude}
    {\cal A}^{\left(x\right)}_{\nu_1,J_1;\nu_2,J_2;\nu_3,J_3}\left(s_j,{\bf k}_j,\boldsymbol{\epsilon}_j\right) = i \pi \delta\left(\tfrac{x}{4}-s_1-s_2-s_3\right)\\ \times  \mathfrak{C}_{\nu_1,J_1;\nu_2,J_2;\nu_3,J_3}\left(s_1,s_2,s_3|\boldsymbol{\epsilon}_k \cdot {\bf k}_j,\boldsymbol{\epsilon}_k \cdot \boldsymbol{\epsilon}_j\right),
\end{multline}
where the function $\mathfrak{C}_{\nu_1,J_1;\nu_2,J_2;\nu_3,J_3}\left(s_1,s_2,s_3|\boldsymbol{\epsilon}_k \cdot {\bf k}_j,\boldsymbol{\epsilon}_k \cdot \boldsymbol{\epsilon}_j\right)$ encodes the tensorial structure, which is a polynomial in the contractions $\left(\boldsymbol{\epsilon}_k \cdot {\bf k}_j\right)$ and $\left(\boldsymbol{\epsilon}_k \cdot \boldsymbol{\epsilon}_j\right)$ and a rational function of the Mellin variables $s_j$. We shall give some explicit examples below. In the next section \ref{subsec::WSop} it will be shown that this function can always be transformed into a polynomial in the $s_j$ through the appropriate change of Mellin integration variables, so that all poles in the $s_j$ are encoded in functions \eqref{rho3}. 

The Dirac delta function in \eqref{MBamplitude} enforces a constraint among the Mellin variables that is analogous to momentum conservation ${\bf k}_1+{\bf k}_2+{\bf k}_3=0$. In particular, analogous to how translation invariance implies momentum conservation, the Dilatation Ward identity (see e.g. \cite{Bzowski:2013sza} for its form in momentum space) requires
\begin{equation}
    s_1+s_2+s_3=\frac{x}{4}, \qquad x=d+2N,\label{scalecond}
\end{equation}
where $N$ is the degree of the polynomial $\mathfrak{C}_{\nu_1,J_1;\nu_2,J_2;\nu_3,J_3}\left(s_1,s_2,s_3|\boldsymbol{\epsilon}_k \cdot {\bf k}_j,\boldsymbol{\epsilon}_k \cdot \boldsymbol{\epsilon}_j\right)$ in the contractions $\left(\boldsymbol{\epsilon}_k \cdot {\bf k}_j\right)$. From a holographic perspective, the Dirac delta function can be expressed as an integral over the bulk radial co-ordinate, which in Poincar\'e co-ordinates
\begin{equation}
    ds^2_{\text{EAdS}} = \frac{dz^2+d{\bf x}^2}{z^2},
\end{equation}
with radial co-ordinate $z$, reads:\footnote{For scalar operators, this integral is precisely the integral in the triple-$K$ integral representation \cite{Bzowski:2013sza} for conformal correlation functions of scalar operators in momentum space. In that case the three Mellin variables $s_j$ arise from the Mellin-Barnes representation for each $K$, which is modified Bessel function of the second kind.}
\begin{equation}\label{ddeta}
  i \pi  \delta\left(\tfrac{x}{4}-s_1-s_2-s_3\right)= \lim_{z_0 \to 0}\int^{\infty}_{z_0}\frac{dz}{z}\, z^{\frac{x}{2}-2\left(s_1+s_2+s_3\right)}.
\end{equation}
Boundary terms are therefore encoded in the Mellin-Barnes amplitude \eqref{MBamplitude} by terms that vanish on the constraint \eqref{scalecond}, since:
\begin{equation}\label{MBbdryterm}
     \left(\tfrac{x}{4}-s_1-s_2-s_3\right)i \pi \delta\left(\tfrac{x}{4}-s_1-s_2-s_3\right) = \lim_{z_0 \to 0}\int_{z_0}^{\infty}dz\,\partial_{z}\left[ z^{\frac{x}{2}-2\left(s_1+s_2+s_3\right)}\right].
\end{equation}
Dirac delta functions \eqref{ddeta} which are not accompanied by a factor $\left(\tfrac{x}{4}-s_1-s_2-s_3\right)$ as in \eqref{MBbdryterm} above are therefore the fingerprint of genuine bulk contact interactions, and hence of the presence of a singularity in the total energy variable $E_T = k_1+k_2+k_3$ as $E_T \to 0$.\footnote{Bulk contact terms only have singularities in $E_T$ and are characterised by the order of the pole in $E_T$ \cite{Maldacena:2011nz,Raju:2012zr}.} Boundary terms do not have a singularity in $E_T$. The most general boundary term is given by \eqref{MBbdryterm} dressed with a polynomial in $s_1$, $s_2$ and $s_3$. This is analogous to the representation of the momentum conserving delta function as an integral over the boundary co-ordinates ${\bf x}$,
\begin{equation}
    \left(2\pi\right)^d \delta^{(d)}\left({\bf k}_1+{\bf k}_2+{\bf k}_3\right) = \int d^d{\bf x}\, e^{i {\bf x} \cdot \left({\bf k}_1+{\bf k}_2+{\bf k}_3\right)},
\end{equation}
where boundary terms are encoded in terms that vanish by momentum conservation:
\begin{equation}
   \left({\bf k}_1+{\bf k}_2+{\bf k}_3\right) \left(2\pi\right)^d \delta^{(d)}\left({\bf k}_1+{\bf k}_2+{\bf k}_3\right) = \int d^d{\bf x}\, i \partial_{{\bf x}}\left[e^{i {\bf x} \cdot \left({\bf k}_1+{\bf k}_2+{\bf k}_3\right)}\right].
\end{equation}
Given the above parallels between the Mellin-Barnes and momentum space representation of conformal correlators it is tempting to regard the Mellin-Barnes representation as an analogue of momentum space for the bulk radial direction.

 We will often find it useful to work with the Mellin-Barnes amplitude at the level of the integrand in the bulk radial co-ordinate $z$, which can be immediately read off from \eqref{ddeta}
 \begin{align}\label{MBAz1}
     {\cal A}_{\nu_1,J_1;\nu_2,J_2;\nu_3,J_3}\left(s_i,{\bf k}_i,\boldsymbol{\epsilon}_i\right) &= \lim_{z_0 \to 0}\left[\int_{z_0}^{\infty}\frac{dz}{z}\, {\cal A}_{\nu_1,J_1;\nu_2,J_2;\nu_3,J_3}\left(s_i,{\bf k}_i,\boldsymbol{\epsilon}_i| z \right)\right],
\end{align}
where we defined
\begin{multline}\label{MBAz2}
  {\cal A}_{\nu_1,J_1;\nu_2,J_2;\nu_3,J_3}\left(s_i,{\bf k}_i,\boldsymbol{\epsilon}_i| z \right) =z^{\frac{x}{2}-2\left(s_1+s_2+s_3\right)}\mathfrak{C}_{\nu_1,J_1;\nu_2,J_2;\nu_3,J_3}\left(s_i|\boldsymbol{\epsilon}_k \cdot {\bf k}_j,\boldsymbol{\epsilon}_k \cdot \boldsymbol{\epsilon}_j\right).  
\end{multline}

The Mellin-Barnes amplitude for the corresponding in-in 3pt function in dS$_{d+1}$ is obtained from its EAdS$_{d+1}$ counterpart by multiplying with the following constant sinusoidal factor:\footnote{The factor ${\cal N}_3$ accounts for the change in two-point function normalization as we move from EAdS to dS. The details of this procedure can be found in \cite{Sleight:2019hfp}, see e.g. equation (2.93) of the latter reference.}
\begin{equation}\label{sinefactor3pt}
   {\cal N}_3 \sin \left(\pi \left(\tfrac{x}{4}+\tfrac{i\left(\nu_1+\nu_2+\nu_3\right)}{2}\right)\right).
\end{equation}
This factor combines the contributions from the $+$ and $-$ in-in contour branches, which have equal and opposite phases generated by analytic continuation from EAdS$_{d+1}$ -- for details see \cite{Sleight:2019hfp}.

Having outlined the general framework for the Mellin-Barnes representation of conformal 3pt functions above, below we will give some examples.

\paragraph{Example 1: Three scalars.} The simplest example is given by boundary three-point correlation functions generated by the following simple non-derivative bulk cubic vertex of scalar fields $\phi_i$ 
\begin{equation}\label{000V}
    {\cal V}_{0,0,0}=g\,\phi_1\phi_2\phi_3,
\end{equation}
with coupling $g$. For bulk fields $\phi_i$ in EAdS$_{d+1}$ the Mellin-Barnes amplitude \eqref{MBamplitude} of the dual operators $O_{\nu_i,0}$ simply reads \cite{Sleight:2019hfp}:
\begin{equation}\label{scalar3pt}
     {\cal A}^{(d)}_{\nu_1,0;\nu_2,0;\nu_3,0}\left(s_j,{\bf k}_j\right) = g\, i \pi \delta\left(\tfrac{d}{4}-s_1-s_2-s_3\right),
\end{equation}
where for scalar 3pt contact diagrams $x=d$. Correlators of the shadow operators $O_{-\nu_j}$ with scaling dimensions $\Delta^-_j=\frac{d}{2}-i\nu_j$ are obtained from the above by sending $\nu_j \to -\nu_j$. To obtain the corresponding result in dS$_{d+1}$ one simply multiplies by the factor \eqref{sinefactor3pt} with $x=d$.

The corresponding correlator \eqref{3ptcormom} is, up to normalisation, the unique solution to the Conformal Ward identities for scalar operators, which for the generic scaling dimensions considered above is given by Appell's $F_4$ function \cite{Coriano:2013jba,Bzowski:2013sza}. The bulk counterpart of the uniqueness of this conformal structure is that the vertex $\phi_1\phi_2\phi_3$ generating it is unique on-shell. Other cubic vertices involving $\phi_1$, $\phi_2$ and $\phi_3$ differ from the latter by terms that vanish on-shell, so-called improvement terms, which generate boundary terms \eqref{MBbdryterm} that give a vanishing contribution to the three-point function.

\paragraph{Example 2: two scalars and a spin $J$.} The cubic vertex involving a single spin-$J$ field $\varphi_J$ and the two scalars $\phi_{1,2}$ is also unique on-shell, taking the following form up to integration by parts and the free equations of motion:
\begin{equation}\label{00Jcanonicalcoupling}
    {\cal V}_{0,0,J} = g\,\left(\phi_1 \nabla^{\mu_1} \ldots \nabla^{\mu_J} \phi_2\right)\varphi_{\mu_1 \ldots \mu_J}\,.
\end{equation}
For bulk fields in EAdS$_{d+1}$ the Mellin-Barnes amplitude it generates for the three-point correlation function of the dual spin-$J$ operator $O_{\nu_3,J}$ with auxiliary vector $\boldsymbol{\epsilon}_3$ and two scalar operators $O_{\nu_{1,2},0}$ is (see section 3.2 of \cite{Sleight:2019hfp}):
\begin{multline}\label{00J3pt}
 \hspace*{-0.5cm}    {\cal A}^{\left(d+2J\right)}_{\nu_1,0;\nu_2,0;\nu_3,J}\left(s_1,{\bf k}_1;s_2,{\bf k}_2;s_3,{\bf k}_3,\boldsymbol{\epsilon}_3\right) = g^{\left(J\right)}_{12}\, i \pi \delta\left(\tfrac{d+2J}{4}-s_1-s_2-s_3\right)\, \\ \times  \mathfrak{C}_{\nu_1,0;\nu_2,0;\nu_3,J}\left(s_1,s_2,s_3|\boldsymbol{\epsilon}_3 \cdot {\bf k}_1,\boldsymbol{\epsilon}_3 \cdot {\bf k}_2,\boldsymbol{\epsilon}_3 \cdot {\bf k}_3\right).
\end{multline}
To obtain the corresponding result in dS$_{d+1}$ one simply multiplies by the factor \eqref{sinefactor3pt}. The tensorial structure $\mathfrak{C}_{\nu_1,0;\nu_2,0;\nu_3,J}\left(s_j|\boldsymbol{\epsilon}_3 \cdot {\bf k}_j\right)$ is a degree $J$ polynomial in the contractions $\left(\boldsymbol{\epsilon}_3 \cdot {\bf k}_j\right)$:
\begin{multline}\label{tensor00J}
 \mathfrak{C}_{\nu_1,0;\nu_2,0;\nu_3,J}\left(s_j|\boldsymbol{\epsilon}_3 \cdot {\bf k}_j\right)=\sum^J_{\alpha=0}\binom{J}{\alpha} \left(- \boldsymbol{\epsilon}_3 \cdot {\bf k}_3\right)^\alpha \sum^\alpha_{\beta=0}\binom{\alpha}{\beta} {\cal Y}^{\left(J\right)}_{\nu_1,\nu_2,\nu_3|\alpha,\beta}\left(\boldsymbol{\epsilon}_3 \cdot {\bf k}_1,\boldsymbol{\epsilon}_3 \cdot {\bf k}_2\right)\\ \times  H_{\nu_1,\nu_2,\nu_3|\alpha,\beta}\left(s_1,s_2,s_3\right),
\end{multline}
where
\begin{equation}\label{Hab}
    H_{\nu_1,\nu_2,\nu_3|\alpha,\beta}\left(s_1,s_2,s_3\right) = \frac{\left(s_1+\frac{i\nu_1}{2}\right)_{\alpha-\beta}\left(s_2+\frac{i\nu_2}{2}\right)_{\beta}}{\left(s_3+\frac{i\nu_3}{2}-\alpha\right)_\alpha},
\end{equation}
and ${\cal Y}^{\left(J\right)}_{\nu_1,\nu_2,\nu_3|\alpha,\beta}\left(\boldsymbol{\epsilon}_3 \cdot {\bf k}_1,\boldsymbol{\epsilon}_3 \cdot {\bf k}_2\right)$ is a degree $J-\alpha$ polynomial in $\left(\boldsymbol{\epsilon}_3 \cdot {\bf k}_{1,2}\right)$ which is independent of the Mellin variables $s_j$ and whose explicit form is reviewed in appendix \ref{APP::00J}. Vertices that differ from the canonical choice \eqref{00Jcanonicalcoupling} by on-shell vanishing terms generate the same three-point correlation function \eqref{00J3pt} modulo boundary terms \eqref{MBbdryterm} that give a vanishing contribution to the three-point function.
In the next section we will see how rational functions of the Mellin variables such as \eqref{Hab} can be translated into a polynomial via an appropriate change of integration variables.

Note that above we have taken the fields participating in the vertex \eqref{00Jcanonicalcoupling} to be generic. When the spin-$J$ field is (partially-)massless, gauge invariance constrains the masses of the scalar fields $\phi_1$ and $\phi_2$ to which it can couple (see \cite{Joung:2012rv} section 3.2). In section \ref{sec:3ptWard} we will see how such constraints manifest themselves in the Mellin-Barnes formalism.

\subsection{Weight shifting operators}
\label{subsec::WSop}

The Mellin-Barnes representation has the virtue of making manifest certain useful recursion relations that hold between correlators with operator scaling dimensions and spins that differ by integer shifts,\footnote{Note that such positive integer shifts of the operator dimensions parameterised by $\Delta=\frac{d}{2}\pm i\nu$ can be naturally interpreted as shifts in the dimension $d$ of the space-like de Sitter boundary.} as well as correlators with and without derivative interactions. See section \tcb{4.4} of \cite{Sleight:2019hfp}, which we review and expand upon in the following. See e.g. \cite{Isono:2018rrb,Arkani-Hamed:2018kmz,Isono:2019ihz,Baumann:2019oyu,Isono:2019wex,Baumann:2020dch} for other works on weight-shifting operators in momentum space CFTs. 

Let us first consider three-point functions of scalar operators. Given a three-point function of scalar operators in general boundary dimensions $d$, the scaling dimension of the operator $O_{\nu_j}$ can be \emph{lowered} by one unit by shifting $d \to d-2$ and acting with a simple differential operator ${\sf M}_{k_j}$ in the momentum $k_j$ on the three-point function:\footnote{In this section \ref{subsec::WSop}, all expressions are for correlators with the momentum conserving delta function stripped off. To avoid notational clutter, we shall often leave implicit the ${}^\prime$ which denotes this. The superscript $(\bullet)$ in $\langle O_{\nu_1}\left({\bf k}_1\right)O_{\nu_2}\left({\bf k}_2\right)O_{\nu_3}\left({\bf k}_3\right)  \rangle{}^{(\bullet)}$ denotes the boundary dimension.}
\begin{equation}\label{Mlowcor}
   \langle O_{\nu_1}\left({\bf k}_1\right)O_{\nu_2}\left({\bf k}_2\right)O_{\nu_3}\left({\bf k}_3\right)  \rangle{}^{(d)}\Big|_{\nu_j \to \nu_j+i} =  {\sf M}_{k_j}\left[\langle O_{\nu_1}\left({\bf k}_1\right)O_{\nu_2}\left({\bf k}_2\right)O_{\nu_3}\left({\bf k}_3\right)  \rangle{}^{(d-2)} \right],
\end{equation}
with 
\begin{equation}\label{Mlow}
    {\sf M}_{k_j} = -4 \partial_{k^2_j},
\end{equation}
which \emph{lowers} by one unit the scaling dimension of $O_{\nu_j}$ while \emph{raising} by two units the boundary dimension $d$. This relation is straightforward to establish from the Mellin-Barnes representation \eqref{scalar3pt}, where shifts in the dimension $d$ induce shifts in the parameters $\nu_j$ through a re-definition of the Mellin variable $s_j$. This generates the Mellin-Barnes representation of the three-point function with the new, shifted, $\nu_j$ dressed with a polynomial in $s_j$. The latter is then naturally recast as a differential operator \eqref{Mlow} in the momentum $k_j$. 

Likewise, the scaling dimension of the operator $O_{\nu_j}$ can be \emph{raised} by one unit upon shifting $d \to d-2$ in \eqref{scalar3pt} and acting with a simple differential operator ${\sf P}_{k_j}$:
\begin{equation}\label{Praisecor}
   \langle O_{\nu_1}\left({\bf k}_1\right)O_{\nu_2}\left({\bf k}_2\right)O_{\nu_3}\left({\bf k}_3\right)  \rangle{}^{(d)}\Big|_{\nu_j \to \nu_j-i} =  {\sf P}_{k_j}\left[\langle O_{\nu_1}\left({\bf k}_1\right)O_{\nu_2}\left({\bf k}_2\right)O_{\nu_3}\left({\bf k}_3\right)  \rangle{}^{(d-2)} \right],
\end{equation}
where
\begin{equation}\label{Praise}
    {\sf P}_{k_j}\left[\bullet\right] = - k^{2i\left(\nu_j-i\right)}_j\partial_{k^2_j}\left(k^{-2i\nu_j}_j \bullet \right),
\end{equation}
which instead \emph{raises} by one unit the scaling dimension of $O_{\nu_j}$ while also \emph{raising} by two units the boundary dimension $d$. The operators \eqref{Mlow} and \eqref{Praise} can then be used recursively to obtain any integer shift $\Delta_j \to \Delta_j \mp n$ in the scaling dimensions, which compose simply as
\begin{equation}
   {\sf M}^n_{k_j} = \left(-4\right)^n \partial^n_{k^2_j}, \qquad {\sf P}^{n}_{k_j}\left[\bullet\right] = \left(-1\right)^n k^{2i\left(\nu_j-i n\right)}_j\partial^n_{k^2_j}\left(k^{-2i\nu_j}_j \bullet \right).
\end{equation}

More generally, any polynomial in the Mellin variables $s_j$ that dresses the Mellin-Barnes representation of the scalar three-point function \eqref{scalar3pt} can be translated into the action of a differential operator. This can be achieved by expressing the polynomial as a sum of Pochammer factors $\left(s_j\pm\tfrac{i\nu_j}{2}\right)_n$, which in turn can be absorbed into the action of the following differential operators:
\begin{subequations}\label{PochhOp}
\begin{align}
    \mathcal{O}_{k,\nu}^{(n)}[\bullet]&=(-1)^n k^{2n+2i\nu}\pl_{x}^n\left(x^{-i\nu}\bullet\Big|_{k=\sqrt{x}}\right)\Big|_{x=k^2}\,,\\
    \widetilde{\mathcal{O}}_{k,\nu}^{(n)}[\bullet]&=(-1)^n k^{2n}\pl_{x}^n\left(\bullet\Big|_{k=\sqrt{x}}\right)\Big|_{x=k^2},
\end{align}
\end{subequations}
which have the property
\begin{subequations}\label{actionop}
\begin{align}
  \hspace*{-0.25cm}  \mathcal{O}_{k_j,\nu_j}^{(n)}\left[\langle O_{\nu_1}\left({\bf k}_1\right)O_{\nu_2}\left({\bf k}_2\right)O_{\nu_3}\left({\bf k}_3\right)  \rangle^\prime_{s_1,s_2,s_3}\right]&=\left(s_j+\tfrac{i\nu_j}{2}\right)_n\langle O_{\nu_1}\left({\bf k}_1\right)O_{\nu_2}\left({\bf k}_2\right)O_{\nu_3}\left({\bf k}_3\right)  \rangle^\prime_{s_1,s_2,s_3},\\
   \hspace*{-0.25cm}  \widetilde{\mathcal{O}}_{k_j,\nu_j}^{(n)}\left[\langle O_{\nu_1}\left({\bf k}_1\right)O_{\nu_2}\left({\bf k}_2\right)O_{\nu_3}\left({\bf k}_3\right)  \rangle^\prime_{s_1,s_2,s_3}\right]&=\left(s_j-\tfrac{i\nu_j}{2}\right)_n
   \langle O_{\nu_1}\left({\bf k}_1\right)O_{\nu_2}\left({\bf k}_2\right)O_{\nu_3}\left({\bf k}_3\right)  \rangle^\prime_{s_1,s_2,s_3}.
\end{align}
\end{subequations}
As will become clear, the relations \eqref{actionop} are particularly useful when dealing with three-point functions generated by derivative interactions and also operators with spin. In particular, in the previous section we saw that the Mellin-Barnes representation for spinning correlators \eqref{00J3pt} differs from that for the corresponding scalar correlator \eqref{scalar3pt} by a rational function \eqref{tensor00J} of the Mellin variables $s_j$ that encodes the tensorial structure and a shift in the boundary dimension $d \to d + 2J$. The key point is that the function encoding the tensorial structure can always be transformed into a polynomial in both the Mellin variables $s_j$ and the contractions $\boldsymbol{\epsilon}_3 \cdot {\bf k}_j$ through a change of variables. For example, for the three-point function involving a single spin-$J$ operator \eqref{00J3pt} this is achieved for each term in the finite sum over $\alpha$ by redefining $s_3 \to s_3 + \alpha$, which gives: 
\begin{multline}\label{00Jpoly}
  \hspace*{-1cm} \langle O_{\nu_1}\left({\bf k}_1\right)O_{\nu_2}\left({\bf k}_2\right)O_{\nu_3,J}\left({\bf k}_3;\boldsymbol{\epsilon}_3\right)  \rangle^{(x)}_{s_1,s_2,s_3}  = \sum^J_{\alpha=0}\left(\frac{-2 \boldsymbol{\epsilon}_3 \cdot {\bf k}_3}{k^2_3}\right)^\alpha\binom{J}{\alpha}\sum^{\alpha}_{\beta=0}\binom{\alpha}{\beta}  {\cal Y}^{\left(J\right)}_{\nu_1,\nu_2,\nu_3|\alpha,\beta}\left(\boldsymbol{\epsilon}_3 \cdot {\bf k}_1,\boldsymbol{\epsilon}_3 \cdot {\bf k}_2\right) \\ \times  \left(s_1+\tfrac{i\nu_1}{2}\right)_{\alpha-\beta}\left(s_2+\tfrac{i\nu_2}{2}\right)_{\beta}\left(s_3-\tfrac{i\nu_3}{2}\right)_{\alpha}\langle O_{\nu_1}\left({\bf k}_1\right)O_{\nu_2}\left({\bf k}_2\right)O_{\nu_3}\left({\bf k}_3\right)  \rangle^{\left(x-\alpha\right)}_{s_1,s_2,s_3},
\end{multline}
where we recall that $x=d+2J$. Using the relations \eqref{actionop}, the Pochhammer factors on the second line dressing the scalar 3pt conformal structure with $x-\alpha$ boundary dimensions can be absorbed into the action of the differential operators \eqref{PochhOp}:
\begin{multline}\label{diffop123}
    \left(s_1+\tfrac{i\nu_1}{2}\right)_{\alpha-\beta}\left(s_2+\tfrac{i\nu_2}{2}\right)_{\beta}\left(s_3-\tfrac{i\nu_3}{2}\right)_{\alpha}\langle O_{\nu_1}\left({\bf k}_1\right)O_{\nu_2}\left({\bf k}_2\right)O_{\nu_3}\left({\bf k}_3\right)  \rangle^{\left(x-\alpha\right)}_{s_1,s_2,s_3}\\
    = \mathcal{O}_{k_1,\nu_1}^{(\alpha-\beta)} \circ \mathcal{O}_{k_2,\nu_2}^{(\beta)} \circ \widetilde{\mathcal{O}}_{k_3,\nu_3}^{(\alpha)}\left[\langle O_{\nu_1}\left({\bf k}_1\right)O_{\nu_2}\left({\bf k}_2\right)O_{\nu_3}\left({\bf k}_3\right)  \rangle^{\left(x-\alpha\right)}_{s_1,s_2,s_3}\right].
\end{multline}
This establishes that the Mellin-Barnes representation of three-point functions for spinning operators can be reduced to that \eqref{scalar3pt} of the scalar operators with the same scaling dimensions as their spinning counterparts up to a shift in the boundary dimension.

By re-instating the Mellin-Barnes integrals via the definition \eqref{3ptcormom}, the identity \eqref{00Jpoly} above combined with \eqref{diffop123} furthermore gives a decomposition of the three-point function involving a single spin-$J$ operator into a sum of three-point functions involving only scalar operators which are acted upon by the operators \eqref{actionop}. Using the Mellin-Barnes representation we can also express the three-point function involving a single spin-$J$ operator as a differential operator acting on a \emph{single} three-point function of scalar operators in which one of the scalar operators has scaling dimension shifted by the spin-$J$:
\begin{multline}\label{spinJ3fromsc}
   \hspace*{-0.65cm} \langle O_{\nu_1}\left({\bf k}_1\right)O_{\nu_2}\left({\bf k}_2\right)O_{\nu_3,J}\left({\bf k}_3;\boldsymbol{\epsilon}_3\right)  \rangle^\prime  =\sum^J_{\alpha=0}\binom{J}{\alpha} \left(\frac{-2 \boldsymbol{\epsilon}_3 \cdot {\bf k}_3}{k^2_3}\right)^\alpha \sum^\alpha_{\beta=0}\binom{\alpha}{\beta} {\cal Y}^{\left(J\right)}_{\nu_1,\nu_2,\nu_3|\alpha,\beta}\left(\boldsymbol{\epsilon}_3 \cdot {\bf k}_1,\boldsymbol{\epsilon}_3 \cdot {\bf k}_2\right)\\ \times \mathcal{O}_{k_1,\nu_1}^{(\alpha-\beta)}\circ\mathcal{O}_{k_2,\nu_2}^{(\beta)}\circ\mathcal{O}_{k_3,\bar{\nu}_3}^{(J-\alpha)}\left[ \langle O_{\nu_1}\left({\bf k}_1\right)O_{\nu_2}\left({\bf k}_2\right)O_{{\bar \nu}_3}\left({\bf k}_3\right)  \rangle^\prime\right],
\end{multline}
where $\bar{\nu}_3=\nu_3+iJ$, so that the correlator with spin-$J$ operator with scaling dimension $\Delta_3 = \frac{d}{2}+i\nu_3$ is obtained by acting with the above differential operator on the correlator where it is replaced by a scalar operator with scaling dimension $\Delta_3 -J$. Note that $\tau_3 = \Delta_3 -J$ is the twist of the spin-$J$ operator, meaning that if two correlators where the operators in one correlator have the same twist as their counterparts in the other\footnote{E.g. conserved operators (which are dual to massless spinning fields) all have the same twist $\tau = d-2$, as do partially-conserved operators of the same depth $r$ which have twist $\tau=d-2-r$.}, both correlators are obtained from the same correlation function of scalar operators in this way. The identity \eqref{spinJ3fromsc} is straightforward to establish from the Mellin-Barnes representation \eqref{00J3pt} by making the change of variables $s_3 \to s_3 + \frac{J}{2}$ and using \eqref{PochhOp}.

\subsection{Ward-Takahashi identities}\label{sec:3ptWard}

\begin{figure}[t]
    \centering
    \captionsetup{width=0.95\textwidth}
    \includegraphics[width=0.4\textwidth]{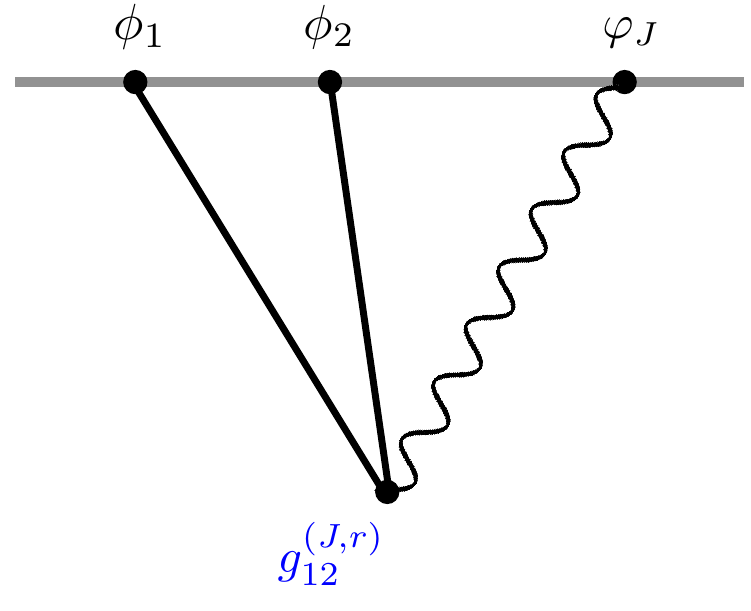}
    \caption{Boundary three-point function of two scalars $\phi_{1,2}$ and a spin-$J$ partially massless field $\varphi_J$ of depth-$r$ in de Sitter space, with coupling $g^{\left(J,r\right)}_{12}$.}
    \label{fig::3pt}
\end{figure}

Correlation functions involving conserved currents are further constrained by Ward-Takahashi identities. These restrict scaling dimension of the operators that can appear in correlators involving conserved currents at the three-point level \cite{Joung:2012rv,Joung:2012hz}. In this section we detail how these features manifest themselves in the Mellin-Barnes representation, focusing on three-point functions of a (partially)-massless field and two scalars. See figure \ref{fig::3pt}.

The spin-$J$ primary operator $O_{\nu_3,J}$ is a conserved current at the following special values of $\nu_3$:
\begin{equation}\label{pmnu3}
    \nu_3 = -i\left(\frac{x}{2}-2-r\right), \qquad r = 0, 1, 2, \ldots, J-1,
\end{equation}
where we refer to the parameter $r$ as the \emph{depth}.\footnote{Sometimes in the literature another definition of depth, $t$, is given and is related to $r$ above via: $t=J-1-r$.} For these values of $\nu_3$ the operator satisfies the conservation condition \cite{Dolan:2001ih,Deser:2003gw}:
\begin{equation}\label{conscondt}
    \left({\bf k}_3 \cdot D_{{\boldsymbol{\epsilon}}_3}\right)^{r+1}O_{\nu_3,J}\left({\bf k}_3,\boldsymbol{\epsilon}_3\right) = 0.
\end{equation}
Operators satisfying \eqref{conscondt} with depth $r>0$ are often referred to in the literature as \emph{partially}-conserved, with the terminology ``conserved current'' reserved for those with depth $r=0$. For $J=2$ the latter is familiar as the stress tensor. When inserted into a correlator, the above conservation condition relates the longitudinal components to lower point correlators of the other operators. For instance, for the three-point function of $O_{\nu_3,J}$ with two scalar operators $O_{\nu_{1},0}$ and $O_{\nu_{2},0}$, whose Mellin-Barnes amplitude we gave in \eqref{00J3pt}, we have
\begin{multline}\label{WT00J}
   \left({\bf k}_3 \cdot D_{{\boldsymbol{\epsilon}}_3}\right)^{r+1} \langle O_{\nu_{1},0}\left({\bf k}_1\right) O_{\nu_{2},0}\left({\bf k}_2\right) O_{\nu_3,J}\left({\bf k}_3,\boldsymbol{\epsilon}_3\right) \rangle^\prime \\= -\left\langle O_{\nu_1,0}(-\mathbf{k}_1) \delta_{{\boldsymbol{\epsilon}}_3}O_{\nu_1,0}(\mathbf{k}_1)\right\rangle-\left\langle\delta_{{\boldsymbol{\epsilon}}_3}O_{\nu_2,0}(\mathbf{k}_2)\,O_{\nu_2,0}(-\mathbf{k}_2)\right\rangle,
\end{multline}
where $\delta_{{\boldsymbol{\epsilon}}_3}$ denotes the action of the charge associated to the current \eqref{conscondt}. 

The Ward-Takahashi identities are intimately related to invariance under gauge transformations of the corresponding field $\varphi_J$ in the bulk. In particular, the Ward-Takahashi identity \eqref{WT00J} is equivalent to the following gauge invariance condition (see e.g. \cite{Berends:1984rq}):\footnote{The notation $(n)$ signifies that the corresponding term is power $n$ in the fields.} 
\begin{equation}\label{cubiccon}
    \delta^{(0)}_{\xi}S^{\left(3\right)}\left[\phi_1,\phi_2,\varphi_J\right]+\delta^{(1)}_{\xi}S^{\left(2\right)}\left[\phi_1\right]+\delta^{(1)}_{\xi}S^{\left(2\right)}\left[\phi_2\right]=0,
\end{equation}
which relates the cubic coupling in the action $S^{\left(3\right)}$ generating the three-point function in \eqref{WT00J} to the kinetic term $S^{\left(2\right)}$ of the scalar fields $\phi_{1,2}$ that are dual to the operators $O_{\nu_{1},0}$ and $O_{\nu_{2},0}$. The $\delta^{(0)}_{\xi}$ is the linearised gauge transformation of the spin-$J$ field with gauge parameter $\xi$, which for depth $r$ is: 
\begin{equation}
    \delta^{(0)}_{\xi}\varphi_{\mu_1 \ldots \mu_J} = \nabla_{\left(\mu_1\right.} \ldots \nabla_{\mu_{r+1}} \xi_{\left. \mu_{r+1} \ldots \mu_{J}\right)}.
\end{equation}
The field $\varphi_J$ therefore has helicities ranging over $\left\{J-r, J-r+1, \ldots, J\right\}$. Since massless spin-$J$ fields have helicity $J$, fields with depth $r>0$ are known as \emph{partially}-massless. The $\delta^{(1)}_{\xi}$ is the transformation of the scalar fields $\phi_{1,2}$ induced by the cubic vertex $S^{\left(3\right)}$ and is linear in $\phi_{1,2}$.

Since the Ward-Takahashi identities are a constraint on the longitudinal components, it is useful consider the helicity decomposition:
\begin{equation}\nonumber
    \langle O_{\nu_1,0}\left({\bf k}_1\right)O_{\nu_2,0}\left({\bf k}_2\right)O_{\nu_3,J}\left({\bf k}_3\right)  \rangle^\prime =\sum^J_{m=0}\Upsilon_{J-m}(\boldsymbol{\epsilon}_3,{\bf k}_3){}^{\left(m\right)}\langle O_{\nu_1,0}\left({\bf k}_1\right)O_{\nu_2,0}\left({\bf k}_2\right)O_{\nu_3,J}\left({\bf k}_3\right)  \rangle^\prime,
\end{equation}
where the helicity-$m$ component is obtained by acting on the 3pt function with the differential operator \eqref{helicity_proj}:
\begin{multline}
  {}^{\left(m\right)}\langle O_{\nu_1,0}\left({\bf k}_1\right)O_{\nu_2,0}\left({\bf k}_2\right)O_{\nu_3,J}\left({\bf k}_3\right)  \rangle^\prime \\ = \widehat{\mathcal{E}}_{J,m}^{\left({\bf k}_3\right)}\left[\langle O_{\nu_1,0}\left({\bf k}_1\right)O_{\nu_2,0}\left({\bf k}_2\right)O_{\nu_3,J}\left({\bf k}_3\right)  \rangle^\prime\right]\Big|_{\boldsymbol{\epsilon}_3 \to \zeta_3}.
\end{multline}
The function $\Upsilon_{J-m}(\boldsymbol{\epsilon}_3,{\bf k}_3)$ encodes the $J-m$ longitudinal indices and is given by the following Gegenbauer polynomial
\begin{equation}
    {\Upsilon}_{n}(\boldsymbol{\epsilon},{\bf k})=\frac{n!}{2^{n}\left(\frac{d-2}{2}+J-n\right)_n}\,\boldsymbol{\epsilon}^n\,C_{n}^{(\frac{d}{2}+J-n-1)}\left(\hat{\boldsymbol{\epsilon}}\cdot\hat{{}\bf k}\right),
\end{equation}
whose derivation is given in appendix \ref{appendix::helciityproj}. The helicity-$m$ component ${}^{\left(m\right)}{\cal A}_{\nu_1,0;\nu_2,0;\nu_3,J}$ of the corresponding Mellin-Barnes amplitude \eqref{MBamplitude} is defined as
\begin{multline}\label{helicitymcorr}
    {}^{\left(m\right)}\langle O_{\nu_1,0}\left({\bf k}_1\right)O_{\nu_2,0}\left({\bf k}_2\right)O_{\nu_3,J}\left({\bf k}_3\right)  \rangle^\prime_{s_1,s_2,s_3} =   {}^{\left(m\right)}{\cal A}_{\nu_1,0;\nu_2,0;\nu_3,J}\left(s_1,{\bf k}_1;s_2,{\bf k}_2;s_3,{\bf k}_3,\zeta_3\right)\\ \times k^{m-J}_3\,\rho_{\nu_1,\nu_2,\nu_3}\left(s_1,s_2,s_3\right)\prod^3_{j=1}\left(\frac{k_j}{2}\right)^{-2s_j+i\nu_j}.
\end{multline}
By definition this is a monomial of degree-$m$ in the contraction $\zeta_3 \cdot {\bf k}_{12}$, where $\zeta_3$ replaces $\boldsymbol{\epsilon}_3$ as an auxiliary vector which is also transverse, i.e. $\zeta_3 \cdot {\bf k}_3=0$, in addition to being null. In particular,
\begin{multline}\label{helicitym}
    {}^{\left(m\right)}{\cal A}^{\left(x\right)}_{\nu_1,0;\nu_2,0;\nu_3,J}\left(s_1,{\bf k}_1;s_2,{\bf k}_2;s_3,{\bf k}_3,\zeta_3\right) = g^{\left(J,r\right)}_{12}\,i\pi\,  \delta\left(\tfrac{x-4\left(J-m\right)}{4}-s_1-s_2-s_3\right) \\ \times \left(-\frac{i}{2}\right)^m \left(\zeta_3 \cdot {\bf k}_{12}\right)^m\,f^{\left(\nu_1,\nu_2,\nu_3\right)}_{J-m}\left(s_1,s_2,s_3\right),
\end{multline}
where $f^{\left(\nu_1,\nu_2,\nu_3\right)}_{J-m}\left(s_1,s_2,s_3\right)$ is a polynomial in the Mellin variables $s_j$ and $g^{\left(J,r\right)}_{12}$ is the coupling of the spin-$J$ (partially-)massless field of depth $r$ to scalars $\phi_1$ and $\phi_2$. The polynomial $f^{\left(\nu_1,\nu_2,\nu_3\right)}_{J-m}\left(s_1,s_2,s_3\right)$ becomes more and more involved as the helicity $m$ decreases. In particular, for the helicity $m=J$ and $m=J-1$ components we have:
\begin{subequations}\label{polyf}
\begin{align}
    f^{\left(\nu_1,\nu_2,\nu_3\right)}_{0}\left(s_1,s_2,s_3\right)&=1,\\
    f^{\left(\nu_1,\nu_2,\nu_3\right)}_{1}\left(s_1,s_2,s_3\right)&=\left[\left(s_1-s_2\right)\left(2\left(s_1+s_2+s_3\right)-i \nu_3\right) +\frac{1}{2} \left(\nu_1-\nu_2\right)\left(\nu_1+\nu_2\right)\right. \label{f1} \\ & \hspace*{-1cm}\left.-i\frac{2\left(\nu_1-\nu_2\right)\left(s_1+s_2\right)(2 s_3-i \nu_3)}{2 i \nu_3+x-4}-i\frac{(\nu_1-\nu_2) (2 i \left(\nu_1+\nu_2\right)+4 s_3+4-x)}{2\left(2 i \nu_3+x-4\right)} \right],\nonumber \\
    & \vdots \nonumber
\end{align}
\end{subequations}
In section \ref{subsec::3ptimp} we will see how the explicit form of these polynomials can be simplified using the freedom to add improvement (on-shell vanishing) terms to the cubic vertices. A useful feature of the functions $f^{\left(\nu_1,\nu_2,\nu_3\right)}_{J-m}\left(s_1,s_2,s_3\right)$ is that they depend on the spin-$J$ and the boundary dimension $d$ only through the combination $x = d+2J$. This implies that to extract the helicity-$m$ component it is sufficient to extract it for spin-$\left(J-m\right)$.\footnote{This in particular means that the helicity-$\left(J-1\right)$ component can be extracted from that of the three-point function with $J=1$; the helicity-$\left(J-2\right)$ from that with $J=2$, ... and so on. For the higher helicity components this property of $f^{\left(\nu_1,\nu_2,\nu_3\right)}_{J-m}\left(s_1,s_2,s_3\right)$ simplifies the task enormously.}

For generic values of $\nu_3$, the Dirac delta distribution in each helicity component \eqref{helicitymcorr} indicates the presence of a bulk contact term. Or equivalently, a singularity in the total energy variable $E_T = k_1+k_2+k_3$ as $E_T \to 0$. See section \ref{subsec::MBrep}. For the values \eqref{pmnu3} of $\nu_3$ corresponding to (partially-)massless fields, for them to couple consistently to scalar matter the bulk contact terms must be absent starting from the helicity-$\left(J-1-r\right)$ component of the correlator down to helicity-$0$. This requires that for $m=0,\ldots,J-1-r$ the helicity-$m$ component \eqref{helicitymcorr} of the Mellin-Barnes amplitude takes the following form at the (partially-)massless points \eqref{pmnu3}:
\begin{equation}\label{fansatz}
  f^{\left(\nu_1,\nu_2,\nu_3\right)}_{J-m}\left(s_1,s_2,s_3\right)=  \left(\frac{x-4\left(J-m\right)}{4}-s_1-s_2-s_3\right)p^{\text{W-T}}_{J-m}\left(s_1,s_2,s_3\right),
\end{equation}
 where $p^{\text{W-T}}_{J-m}\left(s_1,s_2,s_3\right)$ is a polynomial in $s_1$, $s_2$ and $s_3$, and the factor that multiplies it coincides with the argument of the Dirac delta function in \eqref{helicitymcorr}. This requirement places constraints on the values of $\nu_1$ and $\nu_2$ for the scalar fields that admit consistent cubic couplings with (partially)-massless fields. As we saw in section \ref{subsec::MBrep}, the above form corresponds to a local boundary term \eqref{MBbdryterm} in the Mellin-Barnes representation and therefore does not encode a singularity in the $E_T \to 0$ limit. In particular, inserting the expression \eqref{ddeta} for the Dirac delta function as an integral over the bulk radial coordinate gives the following representation of the helicity-$m$ component \eqref{helicitym}:
 \begin{multline}\label{helmeta}
     {}^{\left(m\right)}{\cal A}^{\left(x\right)}_{\nu_1,0;\nu_2,0;\nu_3,J}\left(s_1,{\bf k}_1;s_2,{\bf k}_2;s_3,{\bf k}_3,\zeta_3\right) \\ = g^{\left(J, r\right)}_{12}\,\lim_{z_0 \to 0}\left[ \left(-\frac{i}{2}\right)^m \left(\zeta_3 \cdot {\bf k}_{12}\right)^m\,f^{\left(\nu_1,\nu_2,\nu_3\right)}_{J-m}\left(s_1,s_2,s_3\right) \frac{z_0^{\tfrac{x-4\left(J-m\right)}{2}-2\left(s_1+s_2+s_3\right)}}{\tfrac{x-4\left(J-m\right)}{4}-s_1-s_2-s_3}\right].
 \end{multline}
In this form the presence of a bulk contact singularity is indicated by a simple pole at
\begin{equation}\label{helmconstr}
    \frac{x-4\left(J-m\right)}{4}-s_1-s_2-s_3 = 0.
\end{equation}
For consistent couplings of the (partially-)massless points \eqref{pmnu3} to scalar matter, this pole is cancelled by the corresponding factor \eqref{fansatz} in the polynomial $f^{\left(\nu_1,\nu_2,\nu_3\right)}_{J-m}\left(s_1,s_2,s_3\right)$, generating a boundary term. For the case of two scalars and a (partially-)massless field this boundary term is actually non-zero and generates a non-trivial Ward-Takahashi identity, hence the ``W-T" in \eqref{fansatz}, which we discuss in the following.

Consistent couplings involving (partially-)massless fields come in two types \cite{Berends:1984rq}: Those which are exactly gauge invariant under the corresponding gauge transformations,
\begin{equation}\label{exactgi}
    \delta^{\left(0\right)}_{\xi}S^{(3)} = 0,
\end{equation}
and those which are gauge-invariant up to terms proportional to the free equations of motion,
\begin{equation}\label{nonexactgi}
    \delta^{\left(0\right)}_{\xi}S^{(3)} \approx 0.
\end{equation}
In order to satisfy the cubic order gauge invariance condition \eqref{cubiccon}, the latter induce a non-trivial deformation $\delta^{\left(1\right)}_\xi$ in the linearised gauge transformation via \eqref{cubiccon}. Exactly gauge invariant cubic couplings \eqref{exactgi} are instead non-deforming, $\delta^{\left(1\right)}_\xi = 0$. Correspondingly, three-point functions generated by exactly gauge invariant cubic couplings are exactly conserved, while those generated by couplings that induce deformations in the gauge transformations give non-trivial Ward-Takahashi identities. Cubic vertices of two scalars and a (partially-)massless field in (A)dS$_{d+1}$ are of the latter type \cite{Berends:1985xx,Bekaert:2010hk,Joung:2012hz}. At the level of the Mellin-Barnes representation, the latter can only be generated by the residues of the poles \eqref{rho3} satisfying the constraint \eqref{helmconstr} with non-zero residue. These give a finite contribution in the limit $z_0 \to 0$, since the constraint \eqref{helmconstr} sets to zero the exponent of $z_0$ in \eqref{helmeta}. They are, 
\begin{equation}\label{wtpoles}
   s_1 = \pm \frac{i\nu_1}{2}-n_1, \qquad s_2 = \mp \frac{i\nu_2}{2}-n_2, \qquad s_3 =  \left(\frac{x-4-2r}{4}\right)-n_3, 
\end{equation}
where $n_j=0,1,2,3, \ldots$ and:
\begin{equation}\label{relbdryterm}
    \pm i\nu_1 \mp i\nu_2+r=2\left(n_1+n_2+n_3\right)-2\left(J-r-1-m\right).
\end{equation}
This makes clear that, for a given spin $J$ and depth $r$, only a finite number of poles \eqref{rho3} contribute to a non-trivial Ward-Takahashi identity, which furthermore only emerges for scaling dimensions $\nu_{1,2}$ satisfying \eqref{relbdryterm}. As we shall see below, this is the case for all values \cite{Joung:2012rv,Joung:2012hz} of $\nu_{1,2}$ for which a consistent cubic coupling to a partially-massless field exists. This is to be expected since such couplings are only gauge invariant on-shell \eqref{nonexactgi} and therefore induce a non-trivial $\delta^{(1)}$. In the following we derive the corresponding three-point Ward-Takahashi identities, considering couplings to massless fields in section \ref{subsec::massless3pt} and partially-massless fields in section \ref{subsec::pmmassless3pt}. This simply consists of extracting the function \eqref{fansatz} for each helicity component and evaluating the residues \eqref{wtpoles}, which can be implemented for a given helicity in Mathematica. Note that at the 3pt level the cubic coupling $g^{\left(J,r\right)}_{ij}$ is not constrained by the Ward-Takahashi identity, only the masses of the scalar fields $\phi_i$ and $\phi_j$ that can couple to a (partially-)massless field of spin-$J$ are.\footnote{This can be understood from the constraints \eqref{exactgi} and \eqref{nonexactgi} imposed by gauge invariance, which are homogeneous equations for $S^{\left(3\right)}$.} These are instead constrained by the four-point Ward-Takahashi identities, which is explored in section \ref{sec::4ptWT}.

\subsubsection{Massless fields}
\label{subsec::massless3pt}

For a massless spin-$J$ field we have depth $r=0$ and $\nu_3 = -i \left(\frac{x-4}{2}\right)$. From the helicity $J-1$ component of the Mellin-Barnes amplitude \eqref{f1}, it is straightforward to see that a gauge invariant three-point function only exists when the scalars have equal mass. In particular, only if $\nu_1 = \nu_2$ does the polynomial $f^{\left(\nu_1,\nu_2,\nu_3\right)}_{1}\left(s_1,s_2,s_3\right)$ contain the factor \eqref{fansatz} required to generate a boundary term. This is consistent with existing results on cubic couplings of massless spinning fields to scalar matter, where it is well known that consistent cubic couplings, both in flat and in (A)dS space, require the scalars to have equal mass \cite{Berends:1985xx,Bekaert:2010hk}. In the following we shall therefore take $\nu_1 = \nu_2 = \mu$.

The corresponding Ward-Takahashi identities are non-trivial and are generated by the residues of poles \eqref{wtpoles} with
\begin{equation}\label{mlWTpoles}
    \left(J-1\right)-m = n_1+n_2+n_3,
\end{equation}
where the lower the helicity component the greater the number of poles that contribute. This is consistent with the fact that the corresponding cubic couplings \cite{Berends:1985xx,Bekaert:2010hk} are gauge invariant up to terms proportional to the free equations of motion.

For the helicity-$\left(J-1\right)$ component, only the poles with $n_1=n_2=n_3=0$ generate the Ward-Takahashi identity, which reads:\footnote{The momentum space two-point of scalar operators $O_{\mu,0}$ reads (in the normalisation of \cite{Sleight:2019hfp}):
\begin{equation}
    \langle O_{\mu,0}\left({\bf k}\right)O_{\mu,0}\left(-{\bf k}\right)\rangle = \frac{\Gamma\left(-i\mu\right)^2}{4\pi}\left(\frac{k}{2}\right)^{2 i \mu }.
\end{equation}}
\begin{multline}\label{mlhj-13pt}
  {}^{\left(J-1\right)}\langle O_{\mu,0}\left({\bf k}_1\right)O_{\mu,0}\left({\bf k}_2\right)O_{-i\left(\frac{x-4}{2}\right),J}\left({\bf k}_3\right)  \rangle^\prime=g^{\left(J,0\right)}_{12}\frac{\sqrt{\pi}}{2}\frac{\Gamma \left(\frac{x}{2}\right) \text{csch}(\pi  \mu )}{\Gamma\left(-i\mu\right)^2}\,\left(-\frac{i\zeta_3 \cdot {\bf k}_{12}}{2}\right)^{J-1}\\ \times \left[\langle O_{\mu,0}\left({\bf k}_1\right)O_{\mu,0}\left(-{\bf k}_1\right)\rangle -\langle O_{\mu,0}\left(-{\bf k}_2\right)O_{\mu,0}\left({\bf k}_2\right)\rangle\right].
\end{multline}
The Ward-Takahashi identities for the lower helicity components \eqref{helicitym} follow in the same way. Since the number of poles that contribute increase as the helicity decreases, they also become more involved. For instance, for the helicity-$\left(J-2\right)$ component we have
\begin{multline}\label{helj2ward}
 {}^{\left(J-2\right)}\langle O_{\mu,0}\left({\bf k}_1\right)O_{\mu,0}\left({\bf k}_2\right)O_{-i\left(\frac{x-4}{2}\right),J}\left({\bf k}_3\right)  \rangle^\prime=g^{\left(J,0\right)}_{12}\frac{\sqrt{\pi}}{4}\frac{\Gamma \left(\frac{x-4}{2}\right) \text{csch}(\pi  \mu )}{\left(x-4\right)\Gamma\left(-i\mu\right)^2}  \left(-\frac{i\zeta_3 \cdot {\bf k}_{12}}{2}\right)^{J-2}\\ \times  \Bigg[ \left(2 \mu \, k_3^2+i \left(x-4\right) (k_1-k_2) (k_1+k_2)\right)\langle O_{\mu,0}\left({\bf k}_1\right)O_{\mu,0}\left(-{\bf k}_1\right)\rangle\\+\left(2 \mu \, k_3^2-i \left(x-4\right) \left(k_1-k_2\right)\left(k_1+k_2\right)\right)\langle O_{\mu,0}\left(-{\bf k}_2\right)O_{\mu,0}\left({\bf k}_2\right)\rangle\Bigg]\,.
\end{multline}

The polynomials $f^{\left(\nu_1,\nu_2,\nu_3\right)}_{J-m}\left(s_1,s_2,s_3\right)$ which encode the above Ward-Takahashi identities at the level of the Mellin-Barnes amplitude \eqref{helicitym} also become increasingly complicated as the helicity decreases. For example, for helicity-$\left(J-2\right)$ component above we have:

{\footnotesize\begin{align}\label{hel0s2poly}
    f^{\left(\mu,\mu,-i\left(\frac{x-4}{4}\right)\right)}_2(s_1,s_2,s_3)&=-\frac{\frac{x-8}{2}-2(s_1+s_2+s_3)}{4(x-4)}\Big[8 s_3 (x-6) \left(i \mu +s^2_1-2 s_1 s_2+s_1+s^2_2+s_2\right)-16 s^2_3 (i \mu +s_1+s_2)\nonumber\\
    &+(x-4) \left(4 \mu ^2+(2 s^2_1-4 s_1 s_2+s_1+2 s^2_2+s_2) (4 (s_1+s_2+2)-x)-i \mu  (x-8)\right)
    \Big]\,.
\end{align}}
\noindent In section \ref{subsec::3ptimp} we will see how this can be simplified using the freedom to add improvement terms.

As highlighted in the previous section, note that both \eqref{mlhj-13pt} and \eqref{helj2ward} depend on $d$ and $J$ only through the combination $x=d+2J$. Therefore, once they are known, say, in general $d$, for some spin-$J$, then they are known for all spins $J$. Likewise, if they are known for all spins-$J$ in some dimension $d$, then they are known for all $d$. This is also illustrated by the results \eqref{002PMWard} and \eqref{pmdep2ward} for partially-massless fields of depth-1 and -2, which are considered in the following section.

\subsubsection{Partially-massless fields}
\label{subsec::pmmassless3pt}

The solutions to the gauge invariance condition \eqref{cubiccon} for cubic couplings involving partially-massless fields were constructed and classified in the works \cite{Joung:2012rv,Joung:2012hz}. For the cubic coupling of a (partially)-massless field of spin-$J$ to scalar fields, it was found that consistent couplings exist only when the following relation holds between the depth $r$ and the scaling dimensions $\Delta_1$ and $\Delta_2$ of the scalar fields:
\begin{align}\label{depthcond}
     r+\Delta_1-\Delta_2&=2\mathbb{Z}\,,& |\Delta_1-\Delta_2|\leq r.
\end{align}

 In particular:
 
 \begin{itemize}
     \item Partially massless fields of odd depth $r$ can only couple to scalars with scaling dimensions $\Delta_1$ and $\Delta_2$ that differ by odd integers no greater than $r$. 
     
     \item For partially massless fields of even depth $r$, the above condition tells us they can only couple to scalars with scaling dimensions that differ by even integers no greater than $r$. This includes scalars of equal mass.

 \end{itemize}
 
 This has interesting implications for the coupling of partially-massless fields to massive scalars, where $\Delta_{1,2} = \frac{d}{2}+i\nu_{1,2}$ with $\nu_{1,2} \in \mathbb{R}$:
 
  \begin{itemize}
     
          \item Partially massless fields of odd depth cannot couple to massive scalars since their scaling dimensions cannot differ by (odd) integers as required by \eqref{depthcond}. Consistent couplings of scalars to odd depth partially massless fields can therefore only exist when both scalars belong to the complementary series. 
     
     \item Partially massless fields of even depth can only couple to massive scalars if they have equal mass, since the scaling dimensions of Principal series representations can only differ by imaginary values.
 \end{itemize}

 The Ward-Takahashi identities associated to \eqref{depthcond} are non-trivial since this condition coincides with that \eqref{relbdryterm} required to generate a finite, non-zero, boundary term. This is consistent with the analysis \cite{Joung:2012rv,Joung:2012hz} of the corresponding partially-massless cubic couplings, which are gauge invariant up to terms proportional to the free equations of motion and hence induce a deformation in the gauge transformation. 
 
  In the following we give some examples for even and odd depths separately, focusing for simplicity on partially massless fields with depths $r=1$ and $r=2$. For $r=1$ we take $\nu_1=\mu$ and $\nu_2=\mu+i$ where $i$ is the imaginary unit, so that $\Delta_1-\Delta_2=1$, while for $r=2$ we will take the scalars to have equal mass, $\nu_1=\nu_2=\mu$. These choices are the simplest ones consistent with the constraint \eqref{depthcond}.

\paragraph{Partially-massless of depth $r=1$.}

Taking $\nu_1=\mu$ and $\nu_2=\mu+i$, the helicity-$\left(J-2\right)$ component \eqref{helicitym} of the Mellin-Barnes amplitude for a spin-$J$ partially-massless field of depth 1 is given by 
\begin{multline}\nonumber
   f^{\left(\mu,\mu+i,-i\left(\frac{x-6}{2}\right)\right)}_2(s_1,s_2,s_3)=-\frac{(4 (s_1+s_2+s_3+2)-x)}{16 (x-4)}\\ \times \Big[-4 s^2_1 \left(4 s_2 (x-4)-4 (s_3+5) x+8 (4 s_3+9)+x^2\right)\\\nonumber
   -4 s_1 \left(4 s^2_2 (x-4)+s_2 (8 s_3 (x-6)-2 (x-14) x-88)+(x-4 s_3)^2+56 s_3-17 x+60\right)\nonumber\\\nonumber
-8 i \mu  s_1 (4 s_3+x-2)+\left(4 s^2_2-1\right) (x-4) (4 (s_2+s_3+3)-x)+16 s^3_3 (x-4)\\
   +2 i \mu  \left(4 s_2 (4 s_3+x-2)-(x-4 s_3)^2-56 s_3+18 x-64\right)\Big].
\end{multline}
The factor outside of the square brackets ensures that this is a boundary term \eqref{fansatz}. The corresponding Ward-Takahashi identity generated by the residues of poles \eqref{relbdryterm} reads
\begin{multline}\label{002PMWard}
 \hspace*{-0.5cm}   {}^{\left(J-2\right)}\langle O_{\mu,0}\left({\bf k}_1\right)O_{\mu+i,0}\left({\bf k}_2\right)O_{-i\left(\frac{x-6}{2}\right),J}\left({\bf k}_3\right)  \rangle^\prime\\=
    -g^{\left(J,1\right)}_{12}\,\frac{i \sqrt{\pi}\text{csch}(\pi  \mu ) \Gamma \left(\frac{x}{2}-3\right)}{\Gamma\left(-i\mu\right)^2}\left(-\frac{i\zeta_3 \cdot {\bf k}_{12}}{2}\right)^{J-2} \left[\langle O_{\mu,0}\left({\bf k}_1\right)O_{\mu,0}\left(-{\bf k}_1\right)\rangle\right. \\\left.-\frac{i\left(-\mu (x-4)k^2_1+(\mu +i) (x-4)k^2_2+2 \mu  (1-i \mu ) k^2_3\right)}{4 (x-4)\mu^2}\langle O_{\mu+i,0}\left(-{\bf k}_2\right)O_{\mu+i,0}\left({\bf k}_2\right)\rangle \right].
\end{multline}\\

\paragraph{Partially-massless of depth $r=2$.} Taking $\nu_1=\nu_2=\mu$, the helicity-$\left(J-3\right)$ component \eqref{helicitym} of the Mellin-Barnes amplitude for a spin-$J$ partially-massless field of depth 2 is given by

{\footnotesize\begin{multline}\label{r23ptpm}
  f^{\left(\mu,\mu,-i\left(\frac{x-8}{4}\right)\right)}_3(s_1,s_2,s_3)= -\frac{i (s_1-s_2) (4 (s_1+s_2+s_3+3)-x) (4 (s_1+s_2+s_3+4)-x)}{16 (x-4)} \\ \left[-48 s^2_1 (i \mu +s_3+s_2+1)+8 s_1 (x-10) \left(3 i \mu +s^2_3+s_3 (3-2 s_2)+s_2 (s_2+3)+2\right)\right.\\\left.-2 s^2_3 (4 s_2 (x-4)+(x-30) x+128)+8 s^3_3 (x-4)+s_3 \left((4 s_2-3) x^2-8 s_2 (s_2+9) x+32 (s_2-1) (s_2+11)+76 x\right)\right.\\\left.+(s_2+1) \left(-(2 s_2+1) x^2+4 (s_2+6) (2 s_2+1) x-32 (s_2 (s_2+7)+4)\right)+12 \mu ^2 (x-4)-3 i \mu  (x-12) (x-8)\right].
\end{multline}}

\noindent This is again a boundary term \eqref{fansatz}, owing to the factor $(\frac{x-12}{4}-(s_1+s_2+s_3))$. The corresponding Ward-Takahashi identity generated by the residues of poles \eqref{relbdryterm} reads
\begin{multline}\label{pmdep2ward}
   {}^{\left(J-3\right)}\langle O_{\mu,0}\left({\bf k}_1\right)O_{\mu,0}\left({\bf k}_2\right)O_{-i\left(\frac{x-8}{2}\right),J}\left({\bf k}_3\right)  \rangle^\prime= g^{\left(J,2\right)}_{12}\,\frac{\sqrt{\pi}}{2}\frac{\text{csch}(\pi  \mu ) \Gamma \left(\frac{x}{2}-4\right)}{ (x-4)\Gamma\left(-i\mu\right)^2}\left(-\frac{i\zeta_3 \cdot {\bf k}_{12}}{2}\right)^{J-3}\\\times \Bigg[\left((i\mu -1) k^2_1 (x-4)+(\mu -i) \left(2 (\mu +i) k^2_3-i k^2_2 (x-4)\right)\right)\,\langle O_{\mu,0}\left({\bf k}_1\right)O_{\mu,0}\left(-{\bf k}_1\right)\rangle\\
    +\left((i \mu+1 ) k^2_1 (x-4)+(\mu +i ) \left(-2 (\mu -i) k^2_3-ik^2_2 (x-4)\right)\right)\langle O_{\mu,0}\left(-{\bf k}_2\right)O_{\mu,0}\left({\bf k}_2\right)\rangle\Bigg].
\end{multline}

\noindent In the next section \ref{subsec::3ptimp} we will use the freedom to add improvement terms to reduce \eqref{r23ptpm} -- which is a degree 6 polynomial -- to a degree 5 one.

\subsection{On improvement terms}
\label{subsec::3ptimp}

As discussed in section \ref{subsec::MBrep} one can consider adding improvement terms to the canonical cubic vertex \eqref{00Jcanonicalcoupling} which vanish on-shell. These give vanishing boundary term contributions to the corresponding three-point function. In particular, in the helicity-$m$ component \eqref{helicitym} of the three-point function \eqref{00J3pt}, an improvement generates a contribution of the form
\begin{multline}
    {}^{\left(m\right)}{\cal A}^{\text{impr.}}_{\nu_1,0;\nu_2,0;\nu_3,J}\left(s_1,{\bf k}_1;s_2,{\bf k}_2;s_3,{\bf k}_3,\zeta_3\right) = g^{\left(J,r\right)}_{12}\, i\pi \delta\left(\tfrac{x-4\left(J-m\right)}{4}-s_1-s_2-s_3\right) \\ \times \left(-\frac{i}{2}\right)^m \left(\zeta_3 \cdot {\bf k}_{12}\right)^m\,\left(\tfrac{x-4\left(J-m\right)}{4}-s_1-s_2-s_3\right)p^{\text{impr.}}_{J-m}\left(s_1,s_2,s_3\right),
\end{multline}
which we know from section \ref{subsec::MBrep} gives a boundary term contribution to the three-point function. The function $p^{\text{impr.}}_{J-m}\left(s_1,s_2,s_3\right)$ is a polynomial in $s_1$, $s_2$ and $s_3$ which is constrained to ensure that the boundary term is vanishing and thus leaves the three-point Ward-Takahashi identity unaffected.

It is useful to expand the above polynomial in the following basis
\begin{equation}\label{3ptimpbasis}
    p^{\text{impr.}}_{J-m}\left(s_1,s_2,s_3\right) = \sum\limits_{n_i} c_{n_1,n_2,n_3}\left(s_1-\tfrac{i\nu_1}{2}\right)_{n_1}\left(s_2-\tfrac{i\nu_2}{2}\right)_{n_2}\left(s_3-\tfrac{x-4-2r}{4}\right)_{n_3}.
\end{equation}
From section \ref{subsec::WSop} we see that such basis elements are in one-to-one correspondence with the differential operators \eqref{actionop}: The the degree of the polynomial in the $s_j$ corresponds to the order of the differential operator that generates it. In other words:
\begin{equation}\label{polytoop}
    \left(s_1-\tfrac{i\nu_1}{2}\right)_{n_1}\left(s_2-\tfrac{i\nu_2}{2}\right)_{n_2}\left(s_3-\tfrac{i\nu_3}{2}\right)_{n_3} \quad \leftrightarrow \quad \widetilde{\mathcal{O}}_{k_1,\nu_1}^{\left(n_1\right)}\circ\widetilde{\mathcal{O}}_{k_2,\nu_2}^{\left(n_2\right)}\circ\widetilde{\mathcal{O}}_{k_3,\nu_3}^{\left(n_3\right)}.
\end{equation}
The higher the degree of the improvement as a polynomial in $s_j$ the higher the derivative of the (on-shell vanishing) cubic vertex it represents.\footnote{See section \ref{ContactTerms} for more details on this statement.} The coefficients $c_{n_1,n_2,n_3}$ in \eqref{3ptimpbasis} are constrained to give a vanishing boundary term contribution. In particular, the requirement is that the polynomial vanishes on the values \eqref{relbdryterm} of $s_1$, $s_2$, $s_3$. This does not fix the coefficients $c_{n_1,n_2,n_3}$ completely and the leftover freedom can be used to simplify the functions $f^{\left(\nu_1,\nu_2,\nu_3\right)}_{J-m}\left(s_1,s_2,s_3\right)$ in  \eqref{helicitym}.

For example, for massless fields, $\nu_1=\nu_2=\mu$, the improvements \eqref{3ptimpbasis} of degree 4 that we can add to the helicity-$\left(J-2\right)$ component \eqref{hel0s2poly} are:
\begin{subequations}
\begin{align}
c_{000} &\to 0,\\
c_{100} &\to (\mu +i) (i c_{200}+(\mu +2 i) (c_{300}+c_{400} (3-i \mu ))),\\
c_{010} &\to (\mu +i) (i c_{020}+(\mu +2 i) (c_{030}+c_{040} (3-i \mu ))),\\
   c_{110}&\to (\mu +i) (i c_{210}+c_{310} (\mu +2 i)+i c_{120}+c_{220} \mu +i c_{220}+c_{130} \mu +2 i c_{130}).
\end{align}
\end{subequations}
Using this freedom we can choose simpler representatives for $f^{\left(\mu,\mu,-i\left(\frac{x-4}{4}\right)\right)}_2(s_1,s_2,s_3)$. For example,
\begin{multline}
    f^{\left(\mu,\mu,-i\left(\frac{x-4}{4}\right)\right)}_2(s_1,s_2,s_3)=(x-4 (s_1+s_2+s_3+2))\Big[s_1^2(1+s_1-s_2)+s_2^2(1-s_1+s_2)\\+\frac{(x-6) (x-4-4 s_3) \left(\mu ^2+4 s_1 s_2\right)}{8 (x-4)}+\tfrac{\mu ^2}{2}\Big]\,,
\end{multline}
which is simplified compared to \eqref{hel0s2poly}.

Similarly, for the partially massless field of depth 2 with $\nu_1=\mu$ and $\nu_2=\mu+i$, by considering improvements \eqref{3ptimpbasis} of degree 5 one can find the following simpler degree-5 representative for the helicity-$\left(J-3\right)$ component \eqref{r23ptpm}:
\begin{multline}\label{simpr23ptpm}
 f^{\left(\mu,\mu,-i\left(\frac{x-8}{4}\right)\right)}_3(s_1,s_2,s_3)=-i(x-4 (s_3+1))\left(\tfrac{x-12}{4}-(s_1+s_2+s_3)\right)\\ \times \left[  \left(i+\mu \right) \left(2 i+\mu \right) \left(\tfrac{\mu}{2} +i s_1\right) \left(\tfrac{\mu}{2} +i \left(s_1+1\right)\right)\right.\\-\left(i+\mu \right) \left(\tfrac{\mu}{2} +i s_1\right) \left(\tfrac{\mu}{2} +i \left(s_1+1\right)\right) \left(\tfrac{\mu}{2} +i \left(s_1+2\right)\right)-3 \left(i+\mu \right) \left(\tfrac{\mu}{2} +i s_2\right) \left(\tfrac{\mu}{2} +i s_1\right)\left(\tfrac{\mu}{2} +i \left(s_1+1\right)\right)\\\left.+\frac{i(x-10) \left(\mu ^2+1\right)(x-4 (s_3+2)) (\mu +2 i s_1)}{16 (x-4)}-\left(s_1 \leftrightarrow s_2\right)\right]\\
  -\tfrac{i}{2}\left(\tfrac{x-12}{4}-(s_1+s_2+s_3)\right)(\mu +2 i s_2)  (\mu +2 i s_1)  (\mu +2 i (s_1+1))(\mu +2i( s_1+2)).
\end{multline}

\subsection{Special case: Conformally coupled scalars}
\label{subsec::cc3pt}

In the previous sections we studied three-point functions of two scalar operators and a spinning operator, in particular the Ward-Takahashi identities that must be satisfied when the spinning operator is (partially-)conserved and how they arise in the Mellin-Barnes formalism. As noted in \cite{Sleight:2019hfp} (section 4.6), for certain scaling dimensions the Mellin-Barnes representation is not needed to capture the full analytic structure of the correlator and the Mellin-Barnes integrals can be straightforwardly evaluated to give simple closed form expressions. This includes correlators involving conformally coupled scalars which, via the weightshifting operators of section \ref{subsec::WSop}, can then be used to obtain explicit closed form expressions for correlators of certain spinning operators and also scalar operators with scaling dimensions that differ from those of conformally coupled scalars by integers (which are $\Delta=\tfrac{d}{2}+i\nu, \, \nu=\pm \tfrac{i}{2}$). In particular, recall that for partially conserved operators we have
\begin{equation}
   \Delta_3 = \frac{d}{2}+i\nu_3, \quad i\nu_3 = \frac{d-4}{2}+J-r,
\end{equation}
whose three-point functions, via the weight-shifting identity \eqref{spinJ3fromsc}, can be generated from those with the partially conserved operator replaced by a scalar operator with scaling dimension
\begin{equation}\label{pmsred}
    {\bar \Delta}_3 = \frac{d}{2}+i{\bar \nu}_3, \quad i{\bar \nu}_3 = \frac{d-4}{2}-r.
\end{equation}
For $d$ odd this differs by an integer from the scaling dimension of conformally coupled scalars, so the three-point function involving the scalar operator with \eqref{pmsred} can, in turn, be obtained from that of a conformally coupled scalar via application of the differential operators \eqref{Mlowcor} and \eqref{Praisecor}.

The action of the differential operators \eqref{spinJ3fromsc}, \eqref{Mlowcor} and \eqref{Praisecor} are straightforward to implement in Mathematica. Below we give some examples of how this can be used to obtain such expressions for correlators of a (partially-)massless field with conformally coupled scalars, focusing on the case $d=3$. We will also obtain the corresponding Ward-Takahashi identities, which serves as a consistency check for the more general results obtained at the level of the Mellin-Barnes representation in the previous section.

\paragraph{Massless spinning field and two conformally coupled scalars.}

 Consistent three-point functions of a massless spinning field and two scalar fields necessarily require that the scalars have the same mass (reviewed in the sections above). For two conformally coupled scalar operators of the same scaling dimension given by $\nu_1 = \nu_2 = -\tfrac{i}{2}$, following the discussion above, their three-point function with a spin-$J$ conserved current in $d=3$ can be obtained by from the following scalar three-point function with ${\bar \nu}_3=\frac{i}{2}$:
\begin{equation}\label{3ptcc}
    \langle O_{-\frac{i}{2},0}\left({\bf k}_1\right)O_{-\frac{i}{2},0}\left({\bf k}_2\right)O_{\frac{i}{2},0}\left({\bf k}_3\right)  \rangle^\prime = 2\pi i\, \Gamma\left(\frac{d-3}{2}\right)\frac{1}{ k_3}\frac{1}{\left(k_1+k_2+k_3\right)^{\frac{d-3}{2}}}.
\end{equation}
This expression can be obtained from the Mellin-Barnes representation \eqref{scalar3pt} simply by evaluating the integrals in $s_1$, $s_2$ and $s_3$, see \cite{Sleight:2019mgd,Sleight:2019hfp}. The three-point correlation of the massless spin-$J$ field can then be obtained for $d=3$ by acting on the above with the differential operator \eqref{spinJ3fromsc} and setting $d=3$. Below we give some examples for spins $J=1,2,3$.\footnote{Note that the helicity-$J$ component of the 3pt function for a massless spin-$J$ field and two conformally coupled scalars in $d=3$ was shown to be given by a Gauss hypergeometric function in (3.46) of \cite{Sleight:2019hfp}, while the lower helicity components were left implicit through the action of a differential operator. In the view of extracting the corresponding Ward-Takahashi identity, in the following examples we give the lower helicity components explicitly. For massless spin-1 and spin-2, the explicit 3pt functions were given in \cite{Bzowski:2013sza,Baumann:2020dch} which agree with the expressions we obtain in \eqref{d3ml1} and \eqref{d3ml2}.}\\

\newpage
\emph{Massless spin-1 field} ($\nu_3=-\frac{i}{2}$):\footnote{We remind the reader that in order to have a non-vanishing 3pt function for odd spins $J$ the two scalars should carry a colour index, which we leave implicit throughout.}

Applying the differential operator \eqref{spinJ3fromsc} to \eqref{3ptcc} with $J=1$ and $d=3$ we obtain:
\begin{multline}\label{d3ml1}
    \langle O_{-\frac{i}{2},0}\left({\bf k}_1\right)O_{-\frac{i}{2},0}\left({\bf k}_2\right)O_{-\frac{i}{2},1}\left({\bf k}_3;\boldsymbol{\epsilon}_3\right)  \rangle^\prime\\=g^{\left(1,0\right)}_{12}\left[\frac{i}{8 (k_1+k_2+k_3)}\,\boldsymbol{\epsilon}_3\cdot {\bf k}_2+\frac{i(k_1-k_2+k_3)}{16 k_3 (k_1+k_2+k_3)}\,\boldsymbol{\epsilon}_3\cdot{\bf k}_3\right]\,.
\end{multline}
The helicity-0 component is extracted by acting with the projector \eqref{hp}, giving:
\begin{equation}
     \widehat{\mathcal{E}}^{({\bf k}_3)}_{1,0}\left[\langle O_{-\frac{i}{2},0}\left({\bf k}_1\right)O_{-\frac{i}{2},0}\left({\bf k}_2\right)O_{-\frac{i}{2},1}\left({\bf k}_3\right)  \rangle^\prime\right]\Big|_{\boldsymbol{\epsilon}_3 \to \zeta_3} = g^{\left(1,0\right)}_{12}\frac{i}{16} (k_1-k_2),
\end{equation}
which matches the Ward-Takahashi identity \eqref{mlhj-13pt} upon setting $d=3$, $J=1$ and $\mu=- \tfrac{i}{2}$.\\

\emph{Massless spin-2 field} $(\nu_3=-\frac{3i}{2})$:
Applying the differential operator \eqref{spinJ3fromsc} to \eqref{3ptcc} with $J=2$ and $d=3$ we obtain:
\begin{multline}\label{d3ml2}
    \langle O_{-\frac{i}{2},0}\left({\bf k}_1\right)O_{-\frac{i}{2},0}\left({\bf k}_2\right)O_{-\frac{3i}{2},2}\left({\bf k}_3;\boldsymbol{\epsilon}_3\right)  \rangle^\prime\\=g^{\left(2,0\right)}_{12}\left[\frac{(\boldsymbol{\epsilon}_3\cdot {\bf k}_3)^2 (k_1-k_2+k_3)^2}{64 k_3 (k_1+k_2+k_3)^2}+\frac{(\boldsymbol{\epsilon}_3\cdot {\bf k}_2)^2 (-k_1+k_2+k_3)^2}{64 k_3 (k_1+k_2+k_3)^2}\right.\\+\left.\frac{(\boldsymbol{\epsilon}_3\cdot {\bf k}_3) (\boldsymbol{\epsilon}_3\cdot {\bf k}_2) \left[(k_1-k_2-k_3) (k_1-k_2+k_3)-2 k_3 (k_1+k_2+k_3)\right]}{32 k_3 (k_1+k_2+k_3)^2}\right]\,.
\end{multline}
The helicity-1 and helicity-0 components are 
\begin{subequations}
\begin{align}
    \widehat{\mathcal{E}}^{({\bf k}_3)}_{2,1}\left[\langle O_{-\frac{i}{2},0}\left({\bf k}_1\right)O_{-\frac{i}{2},0}\left({\bf k}_2\right)O_{-\frac{3i}{2},2}\left({\bf k}_3;\boldsymbol{\epsilon}_3\right)  \rangle^\prime\right]\Big|_{\boldsymbol{\epsilon}_3 \to \zeta_3} &=  -g^{\left(2,0\right)}_{12}\frac{i}{16} \left(\frac{-i\zeta_3 \cdot {\bf k}_{12}}{2}\right) \left(k_1-k_2\right),\\
     \widehat{\mathcal{E}}^{({\bf k}_3)}_{2,0}\left[\langle O_{-\frac{i}{2},0}\left({\bf k}_1\right)O_{-\frac{i}{2},0}\left({\bf k}_2\right)O_{-\frac{3i}{2},2}\left({\bf k}_3;\boldsymbol{\epsilon}_3\right)  \rangle^\prime\right]\Big|_{\boldsymbol{\epsilon}_3 \to \zeta_3} &=  g^{\left(2,0\right)}_{12}\frac{1}{192} \left(k_1+k_2\right) \left(3 \left(k_1-k_2\right)^2-k_3^2\right).
\end{align}
\end{subequations}
These match the Ward-Takahashi identities \eqref{mlhj-13pt} and \eqref{helj2ward} setting $d=3$, $J=2$ and $\mu = -\tfrac{i}{2}$.\\

\emph{Massless spin-3 field} $(\nu_3=-\frac{5i}{2})$:
Applying the differential operator \eqref{spinJ3fromsc} to \eqref{3ptcc} with $J=3$ and $d=3$ we obtain:
\begin{multline}
    \langle O_{-\frac{i}{2},0}\left({\bf k}_1\right)O_{-\frac{i}{2},0}\left({\bf k}_2\right)O_{-\frac{5i}{2},3}\left({\bf k}_3;\boldsymbol{\epsilon}_3\right)  \rangle^\prime\\=g^{\left(3,0\right)}_{12}\left[\frac{i (\boldsymbol{\epsilon}_3\cdot{\bf k}_1)^3 (k_1-k_2+k_3)^3}{128 k_3 (k_1+k_2+k_3)^3}+\frac{i (\boldsymbol{\epsilon}_3\cdot{\bf k}_2)^3 (k_1-k_2-k_3)^3}{128 k_3 (k_1+k_2+k_3)^3}\right.\\\nonumber
    +\frac{3 i (\boldsymbol{\epsilon}_3\cdot  {\bf k}_1) (\boldsymbol{\epsilon}_3\cdot{\bf k}_2)^2 \left(k_1^3-k_1^2 (3 k_2+k_3)+3 k_1 (k_2+k_3)^2-(k_2-5 k_3) (k_2+k_3)^2\right)}{128 k_3 (k_1+k_2+k_3)^3}\\\left.-\frac{3 i (\boldsymbol{\epsilon}_3\cdot{\bf k}_1)^2 (\boldsymbol{\epsilon}_3\cdot{\bf k}_2) \left(k_3 \left(3 k_1^2+6 k_1 k_2-k_2^2\right)+3 k_3^2 (3 k_1+k_2)-(k_1-k_2)^3+5 k_3^3\right)}{128 k_3 (k_1+k_2+k_3)^3}\right]\nonumber\,.
\end{multline}
To the best of our knowledge this explicit expression for spin-$3$ is new.\footnote{Expressions for any spin-$J$ can be obtained similarly, acting with the differential operator \eqref{spinJ3fromsc} on \eqref{3ptcc}, but due to the increasing complexity of the result we do not give them explicitly here.} The helicity-2 and helicity-1 components are:
\begin{subequations}
\begin{align}
&  \widehat{\mathcal{E}}^{({\bf k}_3)}_{3,2}\left[\langle O_{-\frac{i}{2},0}\left({\bf k}_1\right)O_{-\frac{i}{2},0}\left({\bf k}_2\right)O_{-\frac{5i}{2},3}\left({\bf k}_3;\boldsymbol{\epsilon}_3\right)  \rangle^\prime\right]\Big|_{\boldsymbol{\epsilon}_3 \to \zeta_3}=  g^{\left(3,0\right)}_{12}\frac{3i}{64} \left(\frac{-i \zeta_3 \cdot {\bf k}_{12} }{2}\right)^2 \left(k_1-k_2\right),\\ \nonumber
 & \widehat{\mathcal{E}}^{({\bf k}_3)}_{3,1}\left[\langle O_{-\frac{i}{2},0}\left({\bf k}_1\right)O_{-\frac{i}{2},0}\left({\bf k}_2\right)O_{-\frac{5i}{2},3}\left({\bf k}_3;\boldsymbol{\epsilon}_3\right)  \rangle^\prime\right]\Big|_{\boldsymbol{\epsilon}_3 \to \zeta_3} \\ & \hspace*{2cm} = - g^{\left(3,0\right)}_{12} \frac{9}{10} \frac{1}{192} \left(\frac{-i \zeta_3 \cdot {\bf k}_{12} }{2}\right)  \left(k_1+k_2\right) \left(5 (k_1-k_2)^2-k^2_3\right).
\end{align}
\end{subequations}
These match the Ward-Takahashi identities \eqref{mlhj-13pt} and \eqref{helj2ward} setting $d=3$, $J=3$ and $\mu = -\tfrac{i}{2}$. What we did not give previously is the helicity-0 component, which reads:
\begin{multline}
   \widehat{\mathcal{E}}^{({\bf k}_3)}_{3,0}\left[\langle O_{-\frac{i}{2},0}\left({\bf k}_1\right)O_{-\frac{i}{2},0}\left({\bf k}_2\right)O_{-\frac{5i}{2},3}\left({\bf k}_3;\boldsymbol{\epsilon}_3\right)  \rangle^\prime\right]\Big|_{\boldsymbol{\epsilon}_3 \to \zeta_3}\\ =  -g^{\left(3,0\right)}_{12}\frac{i \left(k_1-k_2\right) \left(-2 k^2 _3\left(7 k^2_1+4 k_1k_2+7 k^2_2\right)+15 \left(k^2_1-k^2_2\right)^2+3 k^4_3\right)}{1280}.
\end{multline}
\vspace*{0.1cm}
\paragraph{Partially-massless spinning field and two conformally coupled scalars.} Recall that consistent three-point functions involving a partially conserved operator and two scalar operators only exist when the scaling dimensions of the scalar operators satisfy \eqref{depthcond}. 

For example, for partially conserved operators of depth-1, according to \eqref{depthcond} the scaling dimensions of the scalar operators in the three-point function must differ by $\pm 1$. The simplest case is if they both correspond to conformally coupled scalars, one with $\nu_1=-\frac{i}{2}$ and the other with $\nu_2=\frac{i}{2}$. In $d=3$ the three-point function of a partially conserved operator of depth-1 is generated via \eqref{spinJ3fromsc} from the three-point function of the latter scalars and a third scalar operator with ${\bar \nu}_3 = \frac{3i}{2}$, given explicitly by:
\begin{multline}\label{3ptpmseed}
    \langle O_{-\frac{i}{2},0}\left({\bf k}_1\right)O_{\frac{i}{2},0}\left({\bf k}_2\right)O_{\frac{3i}{2},0}\left({\bf k}_3\right)  \rangle^\prime \\ = \frac{1}{2} \Gamma \left(\frac{d-5}{2}\right)\frac{1}{k_2 k^3_3}  (2 k_1+2 k_2+k_3 (d-3)) \frac{1}{\left(k_1+k_2+k_3\right)^{\frac{d-3}{2}}},
\end{multline}
which can either be obtained by directly evaluating the Mellin-Barnes integrals in \eqref{scalar3pt} or by acting on the three-point function \eqref{3ptcc} of conformally coupled scalars with the differential operator \eqref{Mlowcor}. 

Below we give some examples, which to the best of our knowledge were not previously given explicitly in the literature.\\

\emph{Partially-massless spin-2 depth 1} ($\nu_3 = -\tfrac{i}{2}$)
Applying the differential operator \eqref{spinJ3fromsc} to \eqref{3ptpmseed} with $J=2$ and $d=3$ we obtain:
\begin{align}
   & \langle O_{-\frac{i}{2},0}\left({\bf k}_1\right)O_{\frac{i}{2},0}\left({\bf k}_2\right)O_{-\frac{i}{2},2}\left({\bf k}_3;\boldsymbol{\epsilon}_3\right)  \rangle^\prime \\& \nonumber \hspace*{3cm}=g^{\left(2,1\right)}_{12}\left[\frac{(\boldsymbol{\epsilon}_3\cdot {\bf k}_1)^2 (k_1-k_2+k_3) \left((k_1-k_2) (k_1+k_2+k_3)-2 k_2 k_3\right)}{16 k_2 k_3^3 (k_1+k_2+k_3)^2}\right.\\\nonumber
    &\hspace*{3cm}+\frac{(\boldsymbol{\epsilon}_3\cdot {\bf k}_1) (\boldsymbol{\epsilon}_3\cdot {\bf k}_2) (k_1-k_2+k_3) (k_1-k_2-k_3) (k_1+k_2+2 k_3)}{8 k_2 k_3^3 (k_1+k_2+k_3)^2}\\\nonumber
    &\hspace*{3cm}\left.+\frac{(\boldsymbol{\epsilon}_3\cdot {\bf k}_2)^2 (k_1-k_2-k_3) \left((k_1-k_2) (k_1+k_2+k_3)+2 k_2 k_3)\right)}{16 k_2 k_3^3 (k_1+k_2+k_3)^2}\right]. 
\end{align}
The helicity-$0$ component is 
\begin{align}
    \widehat{\mathcal{E}}^{({\bf k}_3)}_{2,0}\left[\langle O_{-\frac{i}{2},0}\left({\bf k}_1\right)O_{\frac{i}{2},0}\left({\bf k}_2\right)O_{-\frac{i}{2},2}\left({\bf k}_3;\boldsymbol{\epsilon}_3\right)  \rangle^\prime\right]\Big|_{\boldsymbol{\epsilon}_3 \to \zeta_3}=-g^{\left(2,1\right)}_{12}\frac{k^2_3-3 (k_1-k_2)^2}{48 k_2},
\end{align}
which matches \eqref{002PMWard} with $d=3$, $J=2$ and $\mu = -\tfrac{i}{2}$.\\

\emph{Partially-massless spin-3 depth 1} ($\nu_3 = -\tfrac{3i}{2}$)
Applying the differential operator \eqref{spinJ3fromsc} to \eqref{3ptpmseed} with $J=3$ and $d=3$ we obtain:
\begin{align}
    & \langle O_{-\frac{i}{2},0}\left({\bf k}_1\right)O_{\frac{i}{2},0}\left({\bf k}_2\right)O_{-\frac{3i}{2},2}\left({\bf k}_3;\boldsymbol{\epsilon}_3\right)  \rangle^\prime \\
    & \hspace*{2cm} = g^{\left(3,1\right)}_{12}\left[ \frac{3 i \left(\boldsymbol{\epsilon_3} \cdot {\bf k}_1\right)\left(\boldsymbol{\epsilon_3} \cdot {\bf k}_2\right)^2 (-k_1+k_2+k_3)^2 \left(k^2_1+5 k_1 k_3-k^2_2-k_2 k_3+4 k^2_3\right)}{64 k_2 k^3_3 (k_1+k_2+k_3)^3}\nonumber \right. \\
    &\hspace*{2cm}+\frac{3 i \left(\boldsymbol{\epsilon_3} \cdot {\bf k}_1\right)^2 \left(\boldsymbol{\epsilon_3} \cdot {\bf k}_2\right) (k_1-k_2+k_3)^2 \left(k^2_1+k_1 k_3-(k_2+k_3) (k_2+4k_3)\right)}{64 k_2 k^3_3 (k_1+k_2+k_3)^3}\nonumber \\&\hspace*{2cm}+\frac{i \left(\boldsymbol{\epsilon_3} \cdot {\bf k}_1\right)^3 (k_1-k_2+k_3)^2 \left(k^2_1+k_1 k_3-k_2 (k_2+5 k_3)\right)}{64 k_2 k^3_3 (k_1+k_2+k_3)^3}\nonumber\\&\hspace*{2cm}\left.+\frac{i \left(\boldsymbol{\epsilon_3} \cdot {\bf k}_2\right)^3  (-k_1+k_2+k_3)^2 \left(k^2_1+5 k_1 k_3-k_2 (k_2+k_3)\right)}{64 k_2 k^3_3 (k_1+k_2+k_3)^3}\right]. \nonumber
\end{align}
The helicity-1 component matches with \eqref{002PMWard} upon setting $d=3$, $J=3$ and $\mu = -\tfrac{i}{2}$:
\begin{multline}
    \widehat{\mathcal{E}}^{({\bf k}_3)}_{3,1}\left[\langle O_{-\frac{i}{2},0}\left({\bf k}_1\right)O_{\frac{i}{2},0}\left({\bf k}_2\right)O_{-\frac{3i}{2},2}\left({\bf k}_3;\boldsymbol{\epsilon}_3\right)  \rangle^\prime\right]\Big|_{\boldsymbol{\epsilon}_3 \to \zeta_3} \\ = g^{\left(3,1\right)}_{12}\left(-\frac{i \zeta_3 \cdot {\bf k}_1}{2}\right)\frac{1}{80k_2}\left(5 (k_1-k_2)^2-k^2_3\right).
\end{multline}
The helicity-0 component was not given previously and reads:
\begin{multline}
   \widehat{\mathcal{E}}^{({\bf k}_3)}_{3,0}\left[\langle O_{-\frac{i}{2},0}\left({\bf k}_1\right)O_{\frac{i}{2},0}\left({\bf k}_2\right)O_{-\frac{3i}{2},2}\left({\bf k}_3;\boldsymbol{\epsilon}_3\right)  \rangle^\prime\right]\Big|_{\boldsymbol{\epsilon}_3 \to \zeta_3} \\ = g^{\left(3,1\right)}_{12} \frac{i}{320 k_2} (k_1-k_2) (k_1+k_2) \left(5 (k_1-k_2)^2-3 k^2_3\right).
\end{multline}

\section{Four-point functions}
\label{sec::4ptfunc}

In this section we review and extend some relevant aspects of the Mellin-Barnes representations of four-point functions introduced in \cite{Sleight:2019mgd,Sleight:2019hfp}. The main new result is a systematic study of four-point contact diagrams generated by quartic vertices with and without derivatives in the Mellin formalism, which can be found in section \ref{ContactTerms}.

\subsection{Exchanges}
\label{sec::4ptexch}

We adopt the Mellin-Barnes representation for four-point exchanges introduced in \cite{Sleight:2019hfp}, to which we refer the reader for details and technicalities. The only novelty we present here is an improvement on the notation and presentation. The Mellin-Barnes representation for four-point exchanges is defined as:
\begin{multline}
    \langle O_{\nu_1,J_1}\left({\bf k}_1\right)O_{\nu_2,J_2}\left({\bf k}_2\right)O_{\nu_3,J_3}\left({\bf k}_3\right) O_{\nu_4,J_4}\left({\bf k}_4\right) \rangle^\prime \\= \int^{i\infty}_{-i\infty} \left[ds\right]_4 \left[du d{\bar u}\right]\, \langle O_{\nu_1,J_1}\left({\bf k}_1\right)O_{\nu_2,J_2}\left({\bf k}_2\right)O_{\nu_3,J_3}\left({\bf k}_3\right) O_{\nu_4,J_4}\left({\bf k}_4\right) \rangle^\prime_{s_1,s_2,s_3,s_4;u,{\bar u}},
\end{multline}
which takes the form
\begin{multline}
  \hspace*{-1.25cm}  \langle O_{\nu_1,J_1}\left({\bf k}_1\right)O_{\nu_2,J_2}\left({\bf k}_2\right)O_{\nu_3,J_3}\left({\bf k}_3\right) O_{\nu_4,J_4}\left({\bf k}_4\right) \rangle^\prime_{s_1,\ldots,s_4;u,{\bar u}}
    = \left({\cal A}^{\left({\sf s}\right)}_{\nu_i,J_i}\left(s_i,{\bf k}_i,\boldsymbol{\epsilon}_i;u,{\bar u},{\bf k}_{\sf s}\right)\:+\: {\sf t}\text{-} \: + \: {\sf u}\text{-channel}\right)\\ \times  \rho_{\nu,\nu}\left(u,{\bar u}\right)\rho_{\nu_1,\nu_2,\nu_3,\nu_4}\left(s_1,s_2,s_3,s_4\right)\prod
   ^4_{j=1}\left(\frac{k_j}{2}\right)^{-2s_j+i\nu_j}.
\end{multline}
The poles in the $s_j$ are encoded in 
\begin{equation}\label{rho4}
    \rho_{\nu_1,\nu_2,\nu_3,\nu_4}\left(s_1,s_2,s_3,s_4\right)=\prod^4_{j=1}\frac{1}{2\sqrt{\pi}}\Gamma\left(s_j+\tfrac{i\nu_j}{2}\right)\Gamma\left(s_j-\tfrac{i\nu_j}{2}\right),
\end{equation}
which is the extension of \eqref{rho3} to 4pts. These poles are associated to the external legs and accordingly we refer to the $s_j$ as \emph{external} Mellin variables. The poles in $u$ and ${\bar u}$ are encoded in \begin{equation}
    \rho_{\nu,\nu}\left(u,{\bar u}\right)=\frac{1}{4\pi}\Gamma\left(u+\tfrac{i\nu}{2}\right)\Gamma\left(u-\tfrac{i\nu}{2}\right)\Gamma\left({\bar u}+\tfrac{i\nu}{2}\right)\Gamma\left({\bar u}-\tfrac{i\nu}{2}\right),
\end{equation}
and are associated to the internal leg of the exchange with momentum ${\bf k}_{\sf s}={\bf k}_1+{\bf k}_2$, so we refer to $u$ and ${\bar u}$ as \emph{internal} Mellin variables.

The Mellin-Barnes amplitude for the ${\sf s}$-channel exchange is given by
\begin{multline}  {\cal A}^{\left({\sf s}\right)}_{\nu_i,J_i}\left(s_i,{\bf k}_i,\boldsymbol{\epsilon}_i;u,{\bar u},{\bf k}_I\right) ={\cal A}^{\left({\sf s}\right)}_{\odot|\nu_i,J_i}\left(s_i,{\bf k}_i,\boldsymbol{\epsilon}_i;u,{\bar u},{\bf k}_{\sf s}\right)\\-{\cal A}^{\left({\sf s}\right)}_{<|\nu_i,J_i}\left(s_i,{\bf k}_i,\boldsymbol{\epsilon}_i;u,{\bar u},{\bf k}_{\sf s}\right)-{\cal A}^{\left({\sf s}\right)}_{>|\nu_i,J_i}\left(s_i,{\bf k}_i,\boldsymbol{\epsilon}_i;u,{\bar u},{\bf k}_{\sf s}\right)\,, \label{MBexch}
\end{multline}
where each of the three terms can be identified with a specific term in the corresponding bulk-bulk propagators.\footnote{In particular, the terms with subscript $<$ and $>$ correspond to those generated by terms with a specific ordering of the radial components of the two bulk points. See \cite{Sleight:2019hfp} for the details.} Two of these can be expressed as a convolution integral of the constituent three-point Mellin-Barnes amplitudes \eqref{MBamplitude}\footnote{The three-point Mellin-Barnes amplitudes in \eqref{leqgeq} are contracted together which is implemented by the Thomas-D operator $D_{\boldsymbol{\epsilon}}$ given in \eqref{thomasD}. Note that, as detailed in \cite{Sleight:2019hfp}, these 3pt Mellin-Barnes amplitudes are those for EAdS$_{d+1}$ e.g. \eqref{scalar3pt} and \eqref{00J3pt}, and the cosine factors in \eqref{leqgeq} and \eqref{exchfact} account for the analytic continuation to dS$_{d+1}$. The factor ${\cal N}_4$ accounts for the change in 2pt function normalisation from AdS to dS, see (2.93) of \cite{Sleight:2019hfp}.} 
\begin{subequations}\label{leqgeq}
\begin{multline}
  \hspace*{-1cm}  {\cal A}^{\left({\sf s}\right)}_{>|\nu_i,J_i}\left(s_i,{\bf k}_i,\boldsymbol{\epsilon}_i;u,{\bar u},{\bf k}_{\sf s}\right)= \frac{{\cal N}_4}2\int_{-i\infty}^{+i\infty}\frac{dw}{2\pi i}\frac{1}{w+ \epsilon}\,
    \cos\left(\tfrac{\pi}2(4(s_1+s_2+w)-\tfrac{x-\bar{x}}{2}+i\left(\nu_1+\nu_2+\nu_3+\nu_4\right))\right)\\
  \times  {\cal A}^{\left(x-4w\right)}_{\nu_1,J_1;\nu_2,J_2;\nu,J}\left(s_{1,2},{\bf k}_{1,2},\boldsymbol{\epsilon}_{1,2};u,{\bf k}_{\sf s},D_{\boldsymbol{\epsilon}}\right) {\cal A}^{({\bar x}+4w)}_{\nu_3,J_3;\nu_4,J_4;-\nu,J}\left({\bar u},-{\bf k}_{\sf s},\boldsymbol{\epsilon};s_{3,4},{\bf k}_{3,4},\boldsymbol{\epsilon}_{3,4}\right),
\end{multline}
\begin{multline}
    \hspace*{-1cm}  {\cal A}^{\left({\sf s}\right)}_{<|\nu_i,J_i}\left(s_i,{\bf k}_i,\boldsymbol{\epsilon}_i;u,{\bar u},{\bf k}_{\sf s}\right)=  \frac{{\cal N}_4}2\int_{-i\infty}^{+i\infty}\frac{dw}{2\pi i}\frac{1}{w+\epsilon}\,
    \cos\left(\tfrac{\pi}2(4(s_3+s_4+w)+\tfrac{x-\bar{x}}{2}+i\left(\nu_1+\nu_2+\nu_3+\nu_4\right))\right)\\
   \times  {\cal A}^{\left(x+4w\right)}_{\nu_1,J_1;\nu_2,J_2;\nu,J}\left(s_{1,2},{\bf k}_{1,2},\boldsymbol{\epsilon}_{1,2};u,{\bf k}_{\sf s},D_{\boldsymbol{\epsilon}}\right) {\cal A}^{({\bar x}-4w)}_{\nu_3,J_3;\nu_4,J_4;-\nu,J}\left({\bar u},-{\bf k}_{\sf s},\boldsymbol{\epsilon};s_{3,4},{\bf k}_{3,4},\boldsymbol{\epsilon}_{3,4}\right),
\end{multline}
\end{subequations}
where 
\begin{align}
  &  x = d+N, && {\bar x} = d+{\bar N},
\end{align}
are the parameters \eqref{scalecond} associated to each three-point function. The $\epsilon$-prescription ensures that the integration contour passes to the right of the pole $w\sim0$. Note that the cosine factors arise from combining the contributions from each branch of the in-in contour, which differ by phases (see section 4 of \cite{Sleight:2019hfp}).\footnote{In particular, these cosine factors are absent from the Mellin-Barnes representation for the exchange in Euclidean AdS.} The remaining contribution is completely factorised:
\begin{multline}\label{exchfact}
   {\cal A}^{\left({\sf s}\right)}_{\odot|\nu_i,J_i}\left(s_i,{\bf k}_i,\boldsymbol{\epsilon}_i;u,{\bar u},{\bf k}_{\sf s}\right)= \frac{{\cal N}_4}2\cos\left(\tfrac{\pi}{2}\left(\tfrac{x-{\bar x}}{2}+i\left(\nu_1+\nu_2-\nu_3-\nu_4\right)\right)\right)\\ \times {\cal A}^{\left(x\right)}_{\nu_1,J_1;\nu_2,J_2;\nu,J}\left(s_{1,2},{\bf k}_{1,2},\boldsymbol{\epsilon}_{1,2};u,{\bf k}_{\sf s},D_{\boldsymbol{\epsilon}}\right) {\cal A}^{({\bar x})}_{\nu_3,J_3;\nu_4,J_4;-\nu,J}\left({\bar u},-{\bf k}_{\sf s},\boldsymbol{\epsilon};s_{3,4},{\bf k}_{3,4},\boldsymbol{\epsilon}_{3,4}\right).
\end{multline}
The expressions for the corresponding ${\sf t}$- and ${\sf u}$-channel exchanges follow from the ${\sf s}$-channel expressions above via the appropriate interchange of ${\sf s}_i$, ${\bf k}_i$, $\boldsymbol{\epsilon}_i$, $J_i$ and $\nu_i$. 

Notice that contributions \eqref{leqgeq} factorise on the simple pole at $w=0$. This is the on-shell factorisation of the exchange into its constituent three-point Mellin-Barnes amplitudes, which appears in a way that is reminiscent of the on-shell factorisation of exchanges in flat space -- the simple pole in the variable $w$ plays an analogous role to the simple pole in the appropriate Mandelstam variable. 

In section \ref{3ptimpbasis} we saw that improvement terms in three-point functions give (vanishing) boundary term contributions. This is no longer the case for exchange diagrams because the internal leg is off-shell. In this case such terms generate bulk contact terms since the improvement terms -- which are proportional to the free equations of motion -- cancel with the bulk-bulk propagator. Let's consider the most general improvement term we can add to the three-point Mellin-Barnes amplitude \eqref{MBamplitude}, 
\begin{multline}
    {\cal A}^{(x)}_{\nu_1,J_1;\nu_2,J_2;\nu,J}\left(s_j,{\bf k}_j,\boldsymbol{\epsilon}_j\right) \to \delta\left(\tfrac{x}{4}-s_1-s_2-s_3\right)\left[1+\left(\tfrac{x}{4}-s_1-s_2-s_3\right)p^{\text{impr.}}\left(s_1,s_2,s_3\right)\right]\\ \times  \mathfrak{C}_{\nu_1,J_1;\nu_2,J_2;\nu_3,J_3}\left(s_1,s_2,s_3|\boldsymbol{\epsilon}_k \cdot {\bf k}_j,\boldsymbol{\epsilon}_k \cdot \boldsymbol{\epsilon}_j\right),
\end{multline}
where $p^{\text{impr.}}\left(s_1,s_2,s_3\right)$ is a polynomial in the Mellin variables $s_1$, $s_2$ and $s_3$, and the factor $\left(\tfrac{x}{4}-s_1-s_2-s_3\right)$ that multiplies it generates the boundary term in the way we saw in section \ref{sec::3pt}. In the Mellin-Barnes exchange amplitude \eqref{MBexch} however, such improvement terms generate terms proportional to $w$ in the contributions \eqref{leqgeq}:
\begin{subequations}\label{4ptimprove}
\begin{multline}\label{4ptimprovep}
    {\cal A}^{\left(x \pm 4w\right)}_{\nu_1,J_1;\nu_2,J_2;\nu,J}\left(s_{1,2},{\bf k}_{1,2},\boldsymbol{\epsilon}_{1,2};u,{\bf k}_{\sf s},\boldsymbol{\epsilon}\right) \to \delta\left(\tfrac{x\pm 4w}{4}-s_1-s_2-u\right)\left[1\mp w \, p^{\text{impr.}}\left(s_1,s_2,u\right)\right]\\ \times  \mathfrak{C}_{\nu_1,J_1;\nu_2,J_2;\nu,J}\left(s_1,s_2,u\right),
\end{multline}
\begin{multline}\label{4ptimprovepbar}
   {\cal A}^{({\bar x} \pm 4w)}_{\nu_3,J_3;\nu_4,J_4;-\nu,J}\left({\bar u},-{\bf k}_{\sf s},\boldsymbol{\epsilon};s_{3,4},{\bf k}_{3,4},\boldsymbol{\epsilon}_{3,4}\right) \to \delta\left(\tfrac{{\bar x}\pm 4w}{4}-s_3-s_4-{\bar u}\right)\left[1\mp w \, {\bar p}^{\text{impr.}}\left(s_3,s_4,{\bar u}\right)\right]\\ \times  \mathfrak{C}_{\nu_3,J_3;\nu_4,J_4;-\nu,J}\left(s_3,s_4,{\bar u}\right).
\end{multline}
\end{subequations}
These terms then cancel the simple pole in \eqref{leqgeq} at $w=0$. The residue of this pole in the contributions \eqref{leqgeq} is therefore universal i.e. blind to improvements and, together with the purely factorised contribution \eqref{exchfact}, gives the on-shell exchange. The improvement terms instead correspond to bulk contact terms that can be uplifted to local quartic vertices in a Lagrangian. This is the Mellin-Barnes counterpart of the fact that one gets a bulk contact term when you act on a bulk-bulk propagator with the operator corresponding to the free equations of motion. 

\subsection{On bulk quartic contact terms}\label{ContactTerms}

At the end of the previous section we argued that improvement terms in cubic vertices contribute bulk quartic contact terms in their corresponding 4pt exchange diagrams four-point exchange diagrams \eqref{4ptimprove}. In this section we explore this relation in more detail, focusing for ease of illustration on the four-point functions involving only scalar fields -- where we can take ${\mathfrak{C}_{\nu_i,0}\left(s_i\right)=1}$. The discussion for spinning fields follows in the same way using the corresponding expressions \eqref{MBamplitude} for the spinning 3pt Mellin-Barnes amplitudes. The total contribution to the exchange generated by, say, the improvement \eqref{4ptimprovep} reads 
\begin{multline}\label{4ptimprov}
   \hspace*{-0.7cm} \langle O_{\nu_1,0}\left({\bf k}_1\right)O_{\nu_2,0}\left({\bf k}_2\right)O_{\nu_3,0}\left({\bf k}_3\right) O_{\nu_4,0}\left({\bf k}_4\right) \rangle^{\text{impr.}} = \int^{i\infty}_{-i\infty} \left[ds\right]_4 \left[{}^{({\sf s})}{\cal A}^{\text{impr.}}_{\nu_i,0}\left(s_i,{\bf k}_i\right)+{\sf t}\text{-}+{\sf u}\text{-channel}\right] \\ \times \rho_{\nu_1,\nu_2,\nu_3,\nu_4}\left(s_1,s_2,s_3,s_4\right) \prod
   ^4_{j=1}\left(\frac{k_j}{2}\right)^{-2s_j+i\nu_j},
\end{multline}
where 
\begin{multline}\label{totalimp}
      {}^{({\sf s})}{\cal A}^{\text{impr.}}_{\nu_1,0;\nu_2,0;\nu_3,0;\nu_4,0}\left(s_1,{\bf k}_1,s_2,{\bf k}_2,s_3,{\bf k}_3,s_4,{\bf k}_4\right)=  {\cal N}_4\int^{+i\infty}_{-i\infty}\left[du d{\bar u}\right] \rho_{\nu,\nu}\left(u,{\bar u}\right)\left(\frac{k_{\sf s}}{2}\right)^{-2\left(u+{\bar u}\right)} \\ \times  \sin \left(\pi \left({\bar u}-u\right)\right)\sin \left(\tfrac{\pi}{2}  \left(i \nu_1+i \nu_2+i \nu_3+i \nu_4+\tfrac{x+{\bar x}}{2}-2\left(u+{\bar u}\right)\right)\right)\\ \times \int_{-i\infty}^{+i\infty}\frac{dw}{2\pi i}\, 2\pi i\, \delta\left(\tfrac{x-4w}{4}-s_1-s_2-u\right) 2\pi i\, \delta\left(\tfrac{{\bar x}+4w}{4}-s_3-s_4-{\bar u}\right) \,p^{\text{impr.}}\left(s_1,s_2,u\right),
\end{multline}
which is obtained simply by combining the contributions \eqref{leqgeq} with the replacement \eqref{4ptimprove}. 

To make the connection with bulk quartic contact terms more explicit the integrals in \eqref{totalimp} need to be evaluated. This simply amounts to evaluating the integral in $w$ since the integral in $u$ and ${\bar u}$ are eliminated by the presence of the two Dirac delta functions. To evaluate the $w$ integral, the key is that in the basis \eqref{3ptimpbasis} the improvement $p^{\text{impr.}}\left(s_1,s_2,u\right)$ takes the following form
\begin{equation}\label{impansatz}
    p^{\text{impr.}}\left(s_1,s_2,u\right) = \sum_{n}c_{n}\left(s_1,s_2\right)\left(u-\tfrac{i\nu}{2}\right)_n,
\end{equation}
where the coefficients $c_{n}\left(s_1,s_2\right)$ are polynomials in $s_1$ and $s_2$, which in the basis \eqref{3ptimpbasis} takes the form
\begin{equation}
    c_{n}\left(s_1,s_2\right) = \sum\limits_{n_1,n_2} c_{n_1,n_2,n}\left(s_1-\tfrac{i\nu_1}{2}\right)_{n_1}\left(s_2-\tfrac{i\nu_2}{2}\right)_{n_2}.
\end{equation}
The integral in $w$ can then be evaluated using the following identity:\footnote{This is proven in appendix \ref{App:U_int}, together with its generalisation to include the possibility of adding improvements ${\bar p}^{\text{impr.}}\left(s_3,s_4,{\bar u}\right)$ in \eqref{4ptimprovepbar} as well.}
\begin{align}\label{Uint}
\int^{i\infty}_{-i\infty} & \frac{dw}{2\pi i}\,\sin \left(\pi \left({\bar u}-u\right)\right)\sin \left(\tfrac{\pi}{2}  \left(i \nu_1+i \nu_2+i \nu_3+i \nu_4+\tfrac{x+{\bar x}}{2}-2\left(u+{\bar u}\right)\right)\right)\\ \nonumber
& \hspace*{2cm}\times \left(u-\tfrac{i\nu}{2}\right)_n\,\rho_{\nu,\nu}\left(u,{\bar u}\right)\left(\frac{k_{\sf s}}{2}\right)^{-2\left(u+{\bar u}\right)}\Bigg|_{{}^{{\bar u}=\tfrac{x+4w}{4}-s_3-s_4}_{u=\tfrac{x-4w}{4}-s_1-s_2}} \\\nonumber&=  \sin \left(\tfrac{\pi}{2}  \left(i \nu_1+i \nu_2+i \nu_3+i \nu_4+\tfrac{x+{\bar x}}{2}\right)\right)\\
&\times \left[\sum_{j=0}^{n-1}\frac{(-1)^{j+1}  (j-(n-1))_j \Gamma (i\nu+n-j )}{j! \Gamma (1+j-i \nu )}\left(\frac{k^2_{\sf s}}{4}\right)^{j}i\pi \delta\left(\tfrac{x+\bar{x}}4+j-s_1-s_2-s_3-s_4\right)\right]\,.\nonumber
\end{align}
The result \eqref{Uint}, inserted in \eqref{totalimp}, gives the Mellin-Barnes representation of a four-point contact diagram. These are polynomials in the four external Mellin variables $s_j$ which multiply a Dirac delta function in their sum $s_1+s_2+s_3+s_4$. The latter encode the bulk contact singularities for $E_T = k_1+k_2+k_3+k_4 \to 0$. To see this let us first consider improvements with $n_1=n_2=0$ and take the external scalars to be conformally coupled, keeping the exchanged scalar generic. In this case it is straightforward to lift in integrals \eqref{4ptimprov} in the external Mellin variables $s_j$. In particular, each term in the sum over $j$ in \eqref{Uint} is equal to the Mellin-Barnes representation of the $\phi^4$ contact diagram with conformally coupled scalars $\phi$ (which have $\nu_{1,2,3,4}=\tfrac{i}{2}$) and boundary dimension $d^\prime=\tfrac{x+{\bar x}}{2}+2j$, which is given by
\begin{equation}
  {\cal A}^{\phi^4}_{\tfrac{i}{2},0;\tfrac{i}{2},0;\tfrac{i}{2},0;\tfrac{i}{2},0}\left(s_j,{\bf k}_j\right) = {\cal N}_4\sin \left(\tfrac{\pi}{2}  \left(\tfrac{d^\prime-4}{2}\right)\right) i\pi \delta\left(\tfrac{d^\prime}{2}-s_1-s_2-s_3-s_4\right).
\end{equation}
The Mellin-Barnes integrals in $s_{1,2,3,4}$ were evaluated in \cite{Sleight:2019mgd} to give the expression \cite{Arkani-Hamed:2015bza}:
\begin{multline}\label{ccphi4}
     \langle O_{\tfrac{i}{2},0}\left({\bf k}_1\right)O_{\tfrac{i}{2},0}\left({\bf k}_2\right)O_{\tfrac{i}{2},0}\left({\bf k}_3\right) O_{\tfrac{i}{2},0}\left({\bf k}_4\right) \rangle^{\phi^4}=2{\cal N}_4\sin \left(\tfrac{\pi}{2}  \left(\tfrac{d^\prime-4}{2}\right)\right)\frac{1}{k_1k_2k_3k_4}\frac{\Gamma\left(d^\prime-2\right)}{\left(E_T\right)^{d^\prime-2}}.
\end{multline}
Combining this with \eqref{Uint} gives us the following bulk contact term generated by the improvement \eqref{impansatz} with $n_1=n_2=0$:
\begin{multline}\label{ccimpcontact}
    \langle O_{\tfrac{i}{2},0}\left({\bf k}_1\right)O_{\tfrac{i}{2},0}\left({\bf k}_2\right)O_{\tfrac{i}{2},0}\left({\bf k}_3\right) O_{\tfrac{i}{2},0}\left({\bf k}_4\right) \rangle^{\text{impr.},\,n_1=n_2=0}=2{\cal N}_4\sin \left(\tfrac{\pi}{2}  \left(\tfrac{x+{\bar x}}{4}+j-2\right)\right) \\ \times \frac{1}{k_1k_2k_3k_4}\sum_{j=0}^{n-1}\frac{(-1)^{j+1}  (j-(n-1))_j \Gamma (i\nu+n-j )}{j! \Gamma (1+j-i \nu )}\frac{\Gamma\left(\tfrac{x+{\bar x}-4}{2}+2j\right)}{\left(E_T\right)^{\tfrac{x+{\bar x}}{2}+2j-2}}\left(\frac{k^2_{\sf s}}{4}\right)^j.
\end{multline}
This is a quartic contact diagram for a local quartic vertex of conformally coupled scalars $\phi$ with $\left(n-1\right)$ derivatives, where each term in the sum (i.e. for fixed $j$) involves no more than $j$ derivatives -- which one can read off from the degree of the singularity in $E_T$ \cite{Arkani-Hamed:2018kmz}.\footnote{Recall that when all the fields in the exchange are scalars, which we are considering here, we have $x={\bar x}=d$. If we have a single external leg of spin-$J$ recall that we have $x=d+2J$ or ${\bar x}=d+2J$. From \eqref{ccimpcontact} we can conclude that spinning external legs only increase the degree of the singularity in $E_T$.} Notice that for $n=0$ the contact term \eqref{ccimpcontact} is vanishing, meaning that they can only be generated by improvements \eqref{impansatz} with a non-trivial $u$-dependence where, the higher the degree of the polynomial in $u$, the greater the number of derivatives that appear in the contact interaction it generates. 

For improvements that also depend on $s_1, s_2$, i.e. with non-zero $n_1,n_2$ in \eqref{impansatz}, the basis \eqref{3ptimpbasis} is extremely useful. As explained in section \ref{subsec::3ptimp}, each basis element can be recast as a differential operator \eqref{polytoop} in the momentum, meaning that contact terms generated by improvements with non-zero $n_1, n_2$ can be obtained by acting with $\widetilde{\mathcal{O}}_{k_1,\nu_1}^{\left(n_1\right)}$, $\widetilde{\mathcal{O}}_{k_2,\nu_2}^{\left(n_2\right)}$ on that \eqref{ccimpcontact} generated by improvements with $n_1=n_2=0$. This increases the degree of the singularity in $E_T$, where the $\widetilde{\mathcal{O}}_{k_j,\nu_j}^{\left(n_j\right)}$ adds derivatives to the field described by the external Mellin variable $s_j$ in the quartic vertex.

From the above we can make the following observations about improvements $p^{\text{impr.}}\left(s_1,s_2,u\right)$, which we shall make use of later on:
\begin{enumerate}
\item Improvements can only generate non-trivial bulk quartic contact terms if they have a non-trivial dependence on $u$. The higher the degree the improvement is as a polynomial in $s_1$, $s_2$ and $u$, the higher the derivative of the quartic vertex that it corresponds to.

\item Improvements that are linear in $u$ and have no $s_1, s_2$ dependence generate the bulk quartic contact term with Mellin-Barnes representation
\begin{equation}
 i \pi\, \delta\left(\tfrac{x+\bar{x}}4-s_1-s_2-s_3-s_4\right),
\end{equation}
which is that of a quartic contact diagram given by the vertex $\phi_1\phi_2\phi_3\phi_4$ of scalar fields $\phi_j$ with no derivatives. If there is also a dependence on $s_1, s_2$ the bulk contact term has the form
\begin{equation}
        c_{1}\left(s_1,s_2\right)i \pi\,\delta\left(\tfrac{x+\bar{x}}4-s_1-s_2-s_3-s_4\right),
    \end{equation}
    where $c_{1}\left(s_1,s_2\right)$ is a polynomial in $s_1$ and $s_2$. Through the correspondence \eqref{polytoop} we can understand this is generated by a quartic vertex $\phi_1\phi_2\phi_3\phi_4$ with derivatives acting on $\phi_1$ and $\phi_2$.

    \item Improvements that have a $u$-dependence given by $u^2$ do not generate bulk contact terms -- their integral \eqref{Uint} is vanishing. This can be understood by noting that the integral for the $n=2$ basis element reads
    \begin{equation}\label{n2basis}
        \left(u-\tfrac{i\nu}{2}\right)_2=u^2+\left(1-i\nu\right) u -\tfrac{i\nu}{2}\left(1-\tfrac{i\nu}{2}\right),
    \end{equation}
    and plugging $n=2$ into the integral \eqref{Uint} gives $\left(1-i\nu\right)$ times the result for $n=1$. The contact term generated by the constant term in \eqref{n2basis} is vanishing, as established in point 1, hence the contact term generated by the $u^2$ term must be vanishing.
\item More generally, an improvement that is degree $n$ in $u$ generates a derivative contact term that is degree-$\left(n-2\right)$ in $k^2_{{\sf s}}$. The reason this is not degree $\left(n-1\right)$, as the formula \eqref{Uint} naively seems to indicate, is that for $n>1$ the term in the sum with $j=\left(n-1\right)$ is vanishing by virtue of the Pochhammer factor $\left(j-\left(n-1\right)\right)_j$. 

\item The lowest derivative quartic vertex that generates a contact term proportional to $k^2_{{\sf s}}$ is given by the improvement
\begin{equation}
     p^{\text{impr.}}\left(s_1,s_2,u\right) = u\left(u-\tfrac{i\nu}{2}\right)\left(u+\tfrac{i\nu}{2}\right),
\end{equation}
which one can confirm by plugging it into \eqref{Uint}.

\end{enumerate}

As a final comment we emphasise that for external (partially-)massless fields the improvements \eqref{impansatz} are further constrained by the requirement that they do not affect the three-point Ward-Takahashi identity -- see section \ref{subsec::3ptimp}. As we shall see in section \ref{sec::4ptWT}, this implies that some of the above possibilities cannot be realised in this case as it imposes a lower bound on the degree of the polynomial \eqref{impansatz} and hence, via the analysis above, also on the degree of the singularity in $E_T$.

\section{Consistency of (partially)-massless matter couplings}
\label{sec::4ptWT}

\begin{figure}[t]
    \centering
    \captionsetup{width=0.95\textwidth}
    \includegraphics[width=1\textwidth]{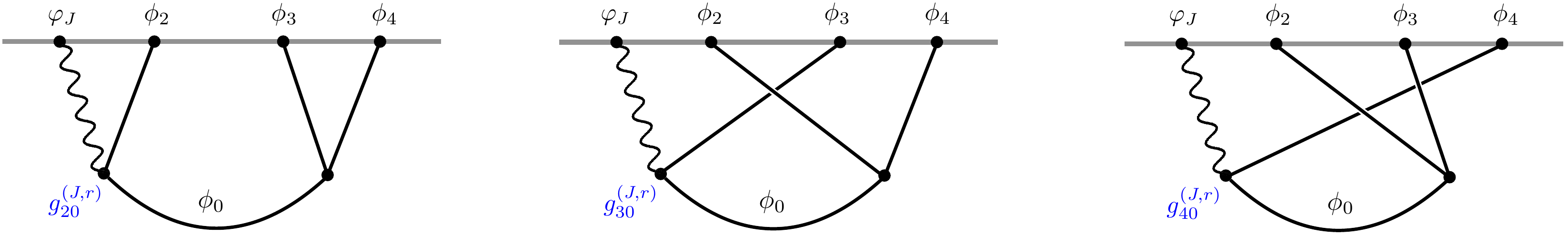}
    \caption{${\sf s}$-, ${\sf t}$- and ${\sf u}$-channel exchange of a scalar $\phi_0$ between scalars $\phi_{2,3,4}$ and a single partially-massless spin-$J$ field of depth-$r$ in de Sitter space.}
    \label{fig::exch}
\end{figure}

In section \ref{sec::3pt} we saw that at the three-point level the Ward-Takahashi identities constrain the masses of scalars that can interact with a (partially)-massless field. The coupling constant, however, is not constrained by such a three-point analysis and for this one must go to four-points. In particular, considering the four-point function of three scalars $\phi_i$ with a single (partially-)conserved operator, for the tree-level exchange of a scalar field $\phi_0$ of mass $m^2_0 = -(\tfrac{d}{2}+i\nu)(\tfrac{d}{2}-i\nu)$ in dS$_{d+1}$, in the following we will explore how the Ward-Takahashi identity can be used to constrain its coupling $g^{\left(J,r\right)}_{i0}$ with a spin-$J$ partially massless field of depth $r$ and one other scalar $\phi_i$. See figure \ref{fig::exch}. Previous works on momentum space four-point functions of (partially)-massless fields in (A)dS include \cite{Raju:2010by,Raju:2011mp,Raju:2012zs,Albayrak:2018tam,Albayrak:2019asr,Albayrak:2019yve,Baumann:2020dch,Sleight:2020obc,Meltzer:2020qbr,Albayrak:2020fyp,Armstrong:2020woi,Melville:2021lst,Meltzer:2021bmb}.

The full four-point function is the sum of the ${\sf s}$-, ${\sf t}$- and ${\sf u}$-channel contributions:
\begin{equation}\label{full4pt}
    {\cal A}_{\nu_1,J;\nu_2,0;\nu_3,0;\nu_4,0} = {\cal A}^{\left({\sf s}\right)}_{\nu_1,J;\nu_2,0;\nu_3,0;\nu_4,0}+ {\sf t}\text{-channel}+ {\sf u}\text{-channel}. 
\end{equation}
To study the consequences of Ward-Takahashi identities, as for the three-point functions in section \ref{sec:3ptWard}, it is useful to consider the decomposition into helicities $m=0, 1, \ldots, J$, 
\begin{equation}
   {\cal A}^{\left({\sf s}\right)}_{\nu_1,J;\nu_2,0;\nu_3,0;\nu_4,0} =\sum^J_{m=0}\Upsilon_{J-m}(\boldsymbol{\epsilon}_1,{\bf k}_1)\, k^{m-J}_1\, {}^{\left(m\right)}{\cal A}^{\left({\sf s}\right)}_{\nu_1,J;\nu_2,0;\nu_3,0;\nu_4,0},
\end{equation}
where the spin-$J$ operator ${\cal O}_{\nu_1,J}$ with scaling dimension $\Delta_1 = \frac{d}{2}+i\nu_1$ has momentum ${\bf k}_1$ and auxiliary vector $\boldsymbol{\epsilon}_1$. The helicity-$m$ component of the ${\sf s}$-channel exchange \eqref{MBexch} is
\begin{multline}  {}^{\left(m\right)}{\cal A}^{\left({\sf s}\right)}_{\nu_1,J;\nu_2,0;\nu_3,0;\nu_4,0} ={}^{\left(m\right)}{\cal A}^{\left({\sf s}\right)}_{\odot|\nu_1,J;\nu_2,0;\nu_3,0;\nu_4,0} -{}^{\left(m\right)}{\cal A}^{\left({\sf s}\right)}_{<|\nu_1,J;\nu_2,0;\nu_3,0;\nu_4,0}-{}^{\left(m\right)}{\cal A}^{\left({\sf s}\right)}_{>|\nu_1,J;\nu_2,0;\nu_3,0;\nu_4,0}\,.  \nonumber
\end{multline}
This is in fact inherited from the helicity-$m$ component of the constituent 3pt function with spin-$J$ via \eqref{leqgeq} and \eqref{exchfact}:
\begin{subequations}
\begin{multline}\label{0>}
\hspace*{-0.5cm}  {}^{\left(m\right)}{\cal A}^{\left({\sf s}\right)}_{>|\nu_1,J;\nu_2,0;\nu_3,0;\nu_4,0}
  = \frac{{\cal N}_4}2\int_{-i\infty}^{+i\infty}\frac{dw}{2\pi i}\frac{1}{w+ \epsilon}\,
    \cos\left(\tfrac{\pi}2(4(s_1+s_2+w)-\tfrac{x-\bar{x}}{2}+i\left(\nu_1+\nu_2+\nu_3+\nu_4\right))\right)\\
  \times   {}^{\left(m\right)}{\cal A}^{\left(x-4w\right)}_{\nu_1,J;\nu_2,0;\nu,0}\left(s_{1},{\bf k}_{1},\boldsymbol{\epsilon}_{1};s_{2},{\bf k}_{2};u,{\bf k}_{\sf s}\right) {\cal A}^{({\bar x}+4w)}_{\nu_3,0;\nu_4,0;-\nu,0}\left({\bar u},-{\bf k}_{\sf s};s_{3,4},{\bf k}_{3,4}\right),
\end{multline}
and
\begin{multline}\label{0<}
   \hspace*{-0.5cm} {}^{\left(m\right)}{\cal A}^{\left({\sf s}\right)}_{<|\nu_1,J;\nu_2,0;\nu_3,0;\nu_4,0}= \frac{{\cal N}_4}2\int_{-i\infty}^{+i\infty}\frac{dw}{2\pi i}\frac{1}{w+\epsilon}\,
    \cos\left(\tfrac{\pi}2(4(s_3+s_4+w)+\tfrac{x-\bar{x}}{2}+i\left(\nu_1+\nu_2+\nu_3+\nu_4\right))\right)\\
   \times   {}^{\left(m\right)}{\cal A}^{\left(x+4w\right)}_{\nu_1,J;\nu_2,0;\nu,0}\left(s_{1},{\bf k}_{1},\boldsymbol{\epsilon}_{1};s_{2},{\bf k}_{2};u,{\bf k}_{\sf s}\right) {\cal A}^{({\bar x}-4w)}_{\nu_3,0;\nu_4,0;-\nu,0}\left({\bar u},-{\bf k}_{\sf s};s_{3,4},{\bf k}_{3,4}\right),
\end{multline}
\end{subequations}
and 
\begin{multline}\label{odotm}
   \hspace*{-0.5cm} {}^{\left(m\right)}{\cal A}^{\left({\sf s}\right)}_{\odot|\nu_1,J;\nu_2,0;\nu_3,0;\nu_4,0}=\frac{{\cal N}_4}2\cos\left(\tfrac{\pi}{2}\left(\tfrac{x-{\bar x}}{2}+i\left(\nu_1+\nu_2-\nu_3-\nu_4\right)\right)\right)\\ \times {}^{\left(m\right)}{\cal A}^{\left(x\right)}_{\nu_1,J;\nu_2,0;\nu,0}\left(s_{1},{\bf k}_{1},\boldsymbol{\epsilon}_{1};s_{2},{\bf k}_{2};u,{\bf k}_{\sf s}\right){\cal A}^{({\bar x})}_{\nu_3,0;\nu_4,0;-\nu,0}\left({\bar u},-{\bf k}_{\sf s};s_{3,4},{\bf k}_{3,4}\right).
\end{multline}
The helicity decomposition of the ${\sf t}$- and ${\sf u}$-channel exchanges follow similarly: The ${\sf t}$-channel expressions follow from the ${\sf s}$-channel ones above through the interchanges ${\sf s}_2 \leftrightarrow {\sf s}_3$, ${\bf k}_2 \leftrightarrow {\bf k}_3$ and $\nu_2 \leftrightarrow \nu_3$. The ${\sf u}$-channel expressions follow from ${\sf s}_2 \leftrightarrow {\sf s}_4$, ${\bf k}_2 \leftrightarrow {\bf k}_4$ and $\nu_2 \leftrightarrow \nu_4$.

Let us now suppose that the spin-$J$ operator ${\cal O}_{\nu_1,J}$ is partially conserved \eqref{conscondt}, which occurs for: 
\begin{equation}
    \nu_1 = - i \left(\frac{x}{2}-2-r\right), \qquad r = 0, 1, 2, \ldots, J-1.
\end{equation}
In order for the four-point function \eqref{full4pt} to be consistent, as for the 3pt functions involving a partially conserved operator considered in section \ref{sec:3ptWard}, the components with helicity $m=0, \ldots, J-1-r$ must not contain bulk quartic contact terms -- meaning that there are no singularities in the four-point total energy variable $E_T = k_1+k_2+k_3+k_4$ as $E_T \to 0$.

Note that such bulk quartic contact terms cannot be generated by the contributions ${}^{\left(m\right)}{\cal A}^{\left({\sf s}, {\sf t}, {\sf u}\right)}_\odot$, which from \eqref{odotm} we see are completely factorised into the product of the three-point function of scalar operators and the helicity-$m$ component of the three-point function involving the spin-$J$ operator ${\cal O}_{\nu_1,J}$. Owing to this property, for helicities $m=0, \ldots, J-1-r$ their contribution to the four-point Ward-Takahashi identity is in fact \emph{inherited} from the Ward-Takahashi identity for the helicity-$m$ component of the three-point function considered in section \ref{sec:3ptWard}.

The remaining contributions ${}^{\left(m\right)}{\cal A}^{\left({\sf s}, {\sf t}, {\sf u}\right)}_{<}$ and ${}^{\left(m\right)}{\cal A}^{\left({\sf s}, {\sf t}, {\sf u}\right)}_{>}$ only give bulk quartic contact terms and are thus the only source of such terms in the exchange. To see this, note that 
\begin{multline}\label{03ptaw}
    {}^{\left(m\right)}{\cal A}^{\left(x\pm4w\right)}_{\nu_1,J;\nu_2,0;\nu,0}\left(s_{1},{\bf k}_{1},\boldsymbol{\epsilon}_{1};s_{2},{\bf k}_{2};s_3,{\bf k}_{\sf s}\right)= g^{\left(J,r\right)}_{20}\, i \pi\, \delta\left(\tfrac{x\pm 4w-4\left(J-m\right)}{4}-s_1-s_2-s_3\right)\\ \times \left(-\frac{i}{2}\right)^m \left(\zeta_1 \cdot {\bf k}_{2{\sf s}}\right)^m f^{\left(\nu_1,\nu_2,\nu_3\right)}_{J-m}\left(s_1,s_2,s_3\right),
\end{multline}
where ${\bf k}_{2{\sf s}} = {\bf k}_{2}-{\bf k}_{\sf s}$ and gauge invariance requires that for $m=0, \ldots, J-1-r$ the function $f^{\left(\nu_1,\nu_2,\nu_3\right)}_{J-m}$ has the form (see section \ref{sec:3ptWard}):
\begin{equation}\label{oI}
    f^{\left(\nu_1,\nu_2,\nu_3\right)}_{J-m}\left(s_1,s_2,s_3\right)= \left(\frac{x-4\left(J-m\right)}{4}-s_1-s_2-s_3\right)p^{\text{W-T}}_{J-m}\left(s_1,s_2,s_3\right).
\end{equation}
 The factor of $\left(\frac{x-4\left(J-m\right)}{4}-s_1-s_2-s_3\right)$ in \eqref{oI} implies that the three-point factors \eqref{03ptaw} are proportional to $w$:
\begin{multline}
    {}^{\left(m\right)}{\cal A}^{\left(x\pm4w\right)}_{\nu_1,J;\nu_2,0;\nu,0}\left(s_{1},{\bf k}_{1},\boldsymbol{\epsilon}_{1};s_{2},{\bf k}_{2};s_3,{\bf k}_3\right)=\mp w\,g^{\left(J,r\right)}_{20}\, i \pi\, \delta\left(\tfrac{x\pm 4w-4\left(J-m\right)}{4}-s_1-s_2-s_3\right) \\ \times \left(-\frac{i}{2}\right)^m \left(\zeta_1 \cdot {\bf k}_{2{\sf s}}\right)^m\,p^{\text{W-T}}_{J-m}\left(s_1,s_2,s_3\right).
\end{multline}
As we saw at the end of section \ref{sec::4ptexch}, when inserted into both ${}^{\left(m\right)}{\cal A}^{\left({\sf s}, {\sf t}, {\sf u}\right)}_{<}$ and ${}^{\left(m\right)}{\cal A}^{\left({\sf s}, {\sf t}, {\sf u}\right)}_{>}$, this factor of $w$ cancels the simple poles at $w=0$ in \eqref{0>} and \eqref{0<} and with it generates bulk quartic contact terms with Mellin-Barnes representation given by (\emph{c.f.} \eqref{4ptimprov}):
\begin{multline}\label{WTcontact}
     {}^{\left(m\right)}{\cal A}^{\left({\sf s}\right)}_{\nu_1,J;\nu_2,0;\nu_3,0;\nu_4,0}\left(\boldsymbol{\epsilon}_1,s_1,{\bf k}_1;s_2,{\bf k}_2,s_3,{\bf k}_3,s_4,{\bf k}_4\right)\Bigg|_{\text{contact}}\\= g^{\left(J,r\right)}_{20}\left(-\frac{i}{2}\right)^m \left(\zeta_1 \cdot {\bf k}_{2{\sf s}}\right)^m{\cal N}_4 \int^{+i\infty}_{-i\infty}\left[du d{\bar u}\right] \rho_{\nu,\nu}\left(u,{\bar u}\right)\left(\frac{k_{\sf s}}{2}\right)^{-2\left(u+{\bar u}\right)} \\ \times  \sin \left(\pi \left({\bar u}-u\right)\right)\sin \left(\tfrac{\pi}{2}  \left(i \nu_1+i \nu_2+i \nu_3+i \nu_4+\tfrac{x+{\bar x}}{2}-2\left(u+{\bar u}\right)\right)\right)\\ \times \int_{-i\infty}^{+i\infty}\frac{dw}{2\pi i}\, 2\pi i\, \delta\left(\tfrac{x-4w-4\left(J-m\right)}{4}-s_1-s_2-u\right) 2\pi i\, \delta\left(\tfrac{{\bar x}+4w}{4}-s_3-s_4-{\bar u}\right) \,p^{\text{W-T}}_{J-m}\left(s_1,s_2,u\right),
\end{multline}
which can be evaluated using \eqref{Uint}. In total we therefore have
\begin{equation}
     {}^{\left(m\right)}{\cal A}_{\nu_1,J;\nu_2,0;\nu_3,0;\nu_4,0} = {}^{\left(m\right)}{\cal A}_{\nu_1,J;\nu_2,0;\nu_3,0;\nu_4,0}\Bigg|_{\text{W-T}}+{}^{\left(m\right)}{\cal A}_{\nu_1,J;\nu_2,0;\nu_3,0;\nu_4,0}\Bigg|_{\text{contact}},
\end{equation}
where, as explained above, the four-point Ward-Takahashi identity is given by the factorised contributions \eqref{odotm}:
\begin{align}
    {}^{\left(m\right)}{\cal A}_{\nu_1,J;\nu_2,0;\nu_3,0;\nu_4,0}\Bigg|_{\text{W-T}} = {}^{\left(m\right)}{\cal A}^{\left({\sf s}\right)}_{\odot|\nu_1,J;\nu_2,0;\nu_3,0;\nu_4,0}\,+\,{\sf t}\text{-channel}\,+\,{\sf u}\text{-channel}.
\end{align}
This would appear to be violated by the contributions \eqref{WTcontact}, where
\begin{equation}
    {}^{\left(m\right)}{\cal A}_{\nu_1,J;\nu_2,0;\nu_3,0;\nu_4,0}\Bigg|_{\text{contact}}={}^{\left(m\right)}{\cal A}^{\left({\sf s}\right)}_{\nu_1,J;\nu_2,0;\nu_3,0;\nu_4,0}\Bigg|_{\text{contact}}\,+\,{\sf t}\text{-channel}\,+\,{\sf u}\text{-channel}.
\end{equation}
At this point there are two possibilities that might restore the four-point Ward-Takahashi identity:
\begin{enumerate}
    \item By adding bulk quartic contact terms that would cancel the offending ones generated by ${}^{\left(m\right)}{\cal A}^{\left({\sf s}, {\sf t}, {\sf u}\right)}_{<}$ and ${}^{\left(m\right)}{\cal A}^{\left({\sf s}, {\sf t}, {\sf u}\right)}_{>}$. This would correspond to adding quartic contact interactions involving the three external scalars and the (partially-)massless spin-$J$ field to the Lagrangian.
    \item Fixing the cubic coupling $g^{\left(J,r\right)}_{ij}$ of the (partially-)massless field to scalars $\phi_i$ and $\phi_j$ so that the offending bulk contact singularities cancel among themselves.
\end{enumerate}
Using the language of the Mellin-Barnes representation it does not take much to see that the first possibility would not work, at least assuming locality of quartic contact interactions. In particular,
quartic contact terms that have a tensorial structure given by a power of $\left(\zeta_1 \cdot {\bf k}_{2{\sf s}}\right)$ as in \eqref{WTcontact} can be reinterpreted as improvements $p^{\text{impr.}}\left(s_1,s_2,s_3\right)$ in the cubic vertices that mediate the scalar exchange in the ${\sf s}$-channel (and similar for the ${\sf t}$- and ${\sf u}$-channels).\footnote{Contact terms that have a mixed tensorial structure involving (in addition to $\left(\zeta_1 \cdot {\bf k}_{2{\sf s}}\right)$) also $\left(\zeta_1 \cdot {\bf k}_{3}\right)$ and/or $\left(\zeta_1 \cdot {\bf k}_{4}\right)$, cannot be written as improvements in a scalar exchanges --- only as improvements in exchanges of spinning fields.} These however cannot cancel all the bulk quartic contact terms generated by $p^{\text{W-T}}\left(s_1,s_2,s_3\right)$ in \eqref{oI} which encodes the three-point Ward-Takahashi identity -- otherwise the identity could be violated by improvements. See the analysis in section \ref{subsec::3ptimp}. The four-point Ward-Takahashi identity can therefore only be restored by constraining the cubic coupling $g^{\left(J,r\right)}_{ij}$ of the (partially-)massless field with two scalars. This will be studied more rigorously in the following sections. 

\subsection{Coupling massless spinning fields to scalar matter}
\label{subsecc::genml4pt}

At the four-point level, for a massless spin-$J$ field the Ward-Takahashi identity requires the cancellation of bulk quartic contact terms \eqref{WTcontact} in the helicity $m=0, \ldots, J-1$ components of the exchange \eqref{full4pt}. This constrains the coupling $g^{\left(J,0\right)}_{i0}$ of generic scalar fields $\phi_0$ and $\phi_i$ of equal mass to a massless field of spin-$J$ in (A)dS$_{d+1}$. For the helicity-$\left(J-1\right)$ component we have (from \eqref{f1} with $\nu_1=\nu_2$):
\begin{equation}\label{impWT3pt}
    p^{\text{W-T}}_{J-1}\left(s_1,s_2,u\right) = -2\left(s_2-u\right),
\end{equation}
which is linear in $u$. Using \eqref{WTcontact}, the corresponding bulk quartic contact term is (where $\nu_1=- i \left(\frac{x-4}{2}\right)$):\footnote{If instead we were considering the same exchange process but in AdS$_{d+1}$ we would obtain the same result for the for the helicity-$\left(J-1\right)$-component but without the factor: 
\begin{equation}\label{sinfactor}
    \sin \left(\tfrac{\pi}{2} \left( i\nu_1+i\nu_2+i\nu_3+i\nu_4+\tfrac{x+{\bar x}}{2}\right)\right).
\end{equation}
In \cite{Sleight:2019mgd} it was shown that the bulk contact terms of in-in four-point functions in dS$_{d+1}$ differ from those in AdS$_{d+1}$ precisely by the factor \eqref{sinfactor}. The conclusions that we draw therefore hold for AdS$_{d+1}$.} 
\begin{multline}\label{mlcontact}
{}^{\left(J-1\right)}{\cal A}^{\left({\sf s}\right)}_{\nu_1,J;\nu_2,0;\nu_3,0;\nu_4,0}\left(\boldsymbol{\epsilon}_1,s_1,{\bf k}_1;s_2,{\bf k}_2,s_3,{\bf k}_3,s_4,{\bf k}_4\right)+ {\sf t}\text{-channel}+ {\sf u}\text{-channel}\:\Bigg|_{\text{contact}}\\
= {\cal N}_4 \sin \left(\tfrac{\pi}{2} \left( i\nu_1+i\nu_2+i\nu_3+i\nu_4+\tfrac{x+{\bar x}}{2}\right)\right)\,i \pi \delta\left(\tfrac{x-4+{\bar x}}{4}-s_1-s_2-s_3-s_4\right) \\ \times \left(g_{20}^{(J,0)}(\zeta_1\cdot\bold{k}_2)^{J-1}+g_{30}^{(J,0)}(\zeta_1\cdot\bold{k}_3)^{J-1}+g_{40}^{(J,0)}(\zeta_1\cdot\bold{k}_4)^{J-1}\right),
\end{multline}
As argued at the end of the last section, this bulk contact term cannot be cancelled by adding improvement terms to the cubic vertices that mediate the exchange. In more detail, the Mellin-Barnes representation of the above bulk contact term is given by the Dirac delta function:
\begin{equation}\label{ddcontactml}
   i \pi\, \delta\left(\tfrac{x-4+{\bar x}}{4}-s_1-s_2-s_3-s_4\right).
\end{equation}
Such a contact term could only be compensated by an improvement \eqref{impansatz} that is linear in $u$, see section \ref{ContactTerms}. As discussed in section \ref{subsec::3ptimp}, the improvement itself cannot be chosen arbitrarily and is constrained to vanish for the values \eqref{wtpoles}\footnote{When using equation \eqref{wtpoles} we are taking $s^{\text{there}}_3$ to be $s^{\text{here}}_1$ and $s^{\text{there}}_1$ to be $u^{\text{here}}$.} of $s_{1}$, $s_{2}$ and $u$ so as to give a vanishing boundary term in the corresponding three-point function. This can be achieved in various ways. In particular, in this case, for the helicity-$\left(J-1\right)$ component we have $n_1=n_2=n_3=0$ in \eqref{wtpoles}. Therefore, to give a vanishing boundary term, such an improvement is restricted to take one of the following forms: 
\begin{subequations}\label{impmlwein}
\begin{align}
    p^{\text{impr.}}\left(s_1,s_2,u\right) &= u\left(s_1-\tfrac{x-4}{4}\right)q^{\text{impr.}}_1\left(s_1,s_2\right),\\
    p^{\text{impr.}}\left(s_1,s_2,u\right) &= u \left(s_2+ \tfrac{i\nu_2}{2}\right)\left(s_2- \tfrac{i\nu_2}{2}\right)q^{\text{impr.}}_2\left(s_1,s_2\right),
\end{align}
\end{subequations}
where $q^{\text{impr.}}_1$ and $q^{\text{impr.}}_2$ are polynomials in $s_1$ and $s_2$. Each of the above forms are linear in $u$ so that they generate the Dirac delta function \eqref{ddcontactml}. Neither of them however can cancel the offending contact term \eqref{mlcontact}, since the latter is given by a constant multiplying the Dirac delta function \eqref{ddcontactml}, while the above improvements ensure that the contact terms they generate dress the Dirac delta function \eqref{ddcontactml} with a polynomial in $s_1$ and $s_2$ which is at least degree 1. In other words, from the analysis of section \ref{ContactTerms}, the improvements \eqref{impmlwein} generate contact terms with a degree of singularity in $E_T$ that is \emph{higher} than that of \eqref{mlcontact}. 
The offending contact term \eqref{mlcontact} therefore cannot be cancelled by a finite number of local quartic contact terms and therefore must vanish by itself. In particular, assuming that the factor \eqref{sinfactor} is non-vanishing -- which we can do for generic $d$ or generic scaling dimensions $\nu_i$ -- the bulk quartic contact terms \eqref{mlcontact} can only vanish if
\begin{equation}\label{wtml}
    \left(g_{20}^{(J,0)}(\zeta_1\cdot\bold{k}_2)^{J-1}+g_{30}^{(J,0)}(\zeta_1\cdot\bold{k}_3)^{J-1}+g_{40}^{(J,0)}(\zeta_1\cdot\bold{k}_4)^{J-1}\right) = 0.
\end{equation}
Setting spin $J=1$ this recovers \emph{conservation of charge}:
\begin{shaded}
\begin{equation}\label{cc}
    g^{\left(1,0\right)}_{20}+g^{\left(1,0\right)}_{30}+g^{\left(1,0\right)}_{40}=0,
\end{equation}
\end{shaded}
\noindent and, for spin $J=2$, the \emph{Equivalence Principle}:
\begin{shaded}
\begin{equation}\label{ep}
    g^{\left(2,0\right)}_{20}=g^{\left(2,0\right)}_{30}=g^{\left(2,0\right)}_{40}.
\end{equation}
\end{shaded}
 \noindent For spin-$J>2$ we find that the gauge invariance condition \eqref{wtml} can only be satisfied if there is \emph{no consistent coupling of massless higher-spin fields to scalar matter}:
 \begin{shaded}
\begin{equation}
    g^{\left(J>2,0\right)}_{20}=g^{\left(J>2,0\right)}_{30}=g^{\left(J>2,0\right)}_{40}=0.\label{hsconstr}
\end{equation} 
 \end{shaded}
\noindent We emphasise that this result assumes locality of interactions, as in Weinberg's flat-space analysis \cite{Weinberg:1964ew}.\footnote{By now it is well known that this conclusion of Weinberg's result for higher spins $J>2$ in flat space do not hold if one allows quartic contact interactions that are as non-local as the exchange amplitude \cite{Taronna:2011kt}. The same has also been shown to be true in AdS$_{d+1}$ \cite{Sleight:2017pcz}. Allowing such non-localities in field theory would however render them ill defined in the absence of a guiding principle that would replace space-time locality, see e.g. \cite{Barnich:1993vg}.} This is complementary to the result \cite{Sleight:2017pcz}, which showed that Ward identities of an underlying global higher-spin symmetry require quartic interactions that are as non-local as exchanges if consistent interactions of higher-spin gauge fields are to exist in AdS$_{d+1}$. It is clear that, if we allow ourselves to add quartic contact interactions that are as non-local as the exchange, the obstruction \eqref{wtml} -- which itself is generated by the exchange -- can in principle be cancelled. 

For de Sitter space, strictly speaking our analysis does not cover the values of $d$ and scaling dimensions $\nu_j$ that give a vanishing sine factor \eqref{sinfactor}, in which case the Ward-Takahashi identity is satisfied without any constraint on the couplings $g^{\left(J,0\right)}_{i0}$. This vanishing of the sine factor \eqref{sinfactor} is actually a consequence of unitarity in dS \cite{Goodhew:2020hob}. For completeness it should be clarified if this could allow for non-trivial couplings of massless higher-spin fields to scalars of certain mass in de Sitter space, though it would be unexpected. We expect this possibility to be ruled out by a similar analysis at the level of the Wave Function, where the exchange can be obtained from the AdS result by a simple Wick rotation \cite{Maldacena:2002vr} -- so the dS couplings would be constrained just as they are in the AdS case, which is covered by our analysis above simply by dividing by the factor \eqref{sinfactor}. The same statements apply to the partially-massless case \eqref{PMcontactquartic} considered in the next section.

\subsection{Coupling partially-massless spinning fields to scalar matter}
\label{subsec::pmcouplings}

In a similar fashion the couplings $g^{\left(J,r\right)}_{ij}$ of a spin-$J$ partially massless field of depth-$r$ can be constrained by requiring that the bulk contact terms all cancel in the helicity $m=0,\ldots,J-1-r$ components of the exchange \eqref{full4pt}. 

In the following we focus on partially massless fields of depth $r=2$, since it is the lowest depth at which there exist matter couplings to generic equal mass scalars,\footnote{The exception is the coupling of a depth-1 partially massless field to conformally coupled scalars in dS$_4$, which can have scaling dimensions that differ by 1 -- as consistent with the constraint \eqref{depthcond}. See discussion above equation \eqref{3ptpmseed}.} where in the following we take $\nu=\nu_2=\nu_3=\nu_4=\mu$. Consistency of depth-2 partially massless couplings requires that there are no bulk contact terms starting from the helicity-$\left(J-3\right)$ downwards. The three-point functions of depth-2 partially massless fields  were studied in section \ref{subsec::pmmassless3pt}. In particular, from \eqref{simpr23ptpm} we have
\begin{multline}
 p^{\text{W-T}}_{J-3}\left(s_1,s_2,u\right)=-i(x-4 (s_1+1))\left[  \tfrac{1}{4} (i+\mu ) (2 i+\mu ) (\mu +2 i u) (\mu +2 i (u+1))\right.\\-\tfrac{1}{8}  \left(i+\mu \right) (\mu +2 i u) (\mu +2 i (u+1)) (\mu +2 i (u+2))-\tfrac{3}{8} (i+\mu ) (\mu +2 i s_2) (\mu +2 i u)(\mu +2 i (u+1))\\\left.+\frac{i(x-10) \left(\mu ^2+1\right)(x-4 (s_1+2)) (\mu +2 i u)}{16 (x-4)}-\left(u \leftrightarrow s_2\right)\right]\\
  -\tfrac{i}{2}(\mu +2 i s_2)  (\mu +2 i u)  (\mu +2 i (u+1))(\mu +2i(u+2)),
\end{multline}
which is a degree-4 polynomial. Using the analysis of section \ref{ContactTerms}, the bulk contact term contribution to the helicity-$\left(J-3\right)$ component of the exchange then has the following form (where $\nu_1 =- i \left(\tfrac{x-8}{2}\right)$ and $\nu=\nu_2=\nu_3=\nu_4=\mu$):\footnote{The reason $j$ runs from 0 to 1 and not 0 to 2 is that the polynomial $p^{\text{W-T}}\left(s_1,s_2,u\right)$, although degree 4, is only degree 3 in $u$.}
\begin{multline}\label{PMcontactquartic}
{}^{\left(J-3\right)}{\cal A}^{\left({\sf s}\right)}_{\nu_1,J;\nu_2,0;\nu_3,0;\nu_4,0}\left(\boldsymbol{\epsilon}_1,s_1,{\bf k}_1;s_2,{\bf k}_2,s_3,{\bf k}_3,s_4,{\bf k}_4\right)+ {\sf t}\text{-channel}+ {\sf u}\text{-channel}\:\Bigg|_{\text{contact}}\\
= {\cal N}_4 \sin \left(\tfrac{\pi}{2} \left( i\nu_1+i\nu_2+i\nu_3+i\nu_4+\tfrac{x+{\bar x}}{2}\right)\right)\,\sum^1\limits_{j=0}i \pi \delta\left(\tfrac{x-12+{\bar x}}{4}+j-s_1-s_2-s_3-s_4\right) \\ \hspace*{-0.75cm} \times \left(g^{\left(J,2\right)}_{20|j}(s_1,s_2)\left(k^2_{\sf s}\right)^{j}(\zeta_1\cdot\bold{k}_2)^{J-3}+g^{\left(J,2\right)}_{30|j}(s_1,s_3)\left(k^2_{\sf t}\right)^{j}(\zeta_1\cdot\bold{k}_3)^{J-3}+g^{\left(J,2\right)}_{40|j}(s_1,s_4)\left(k^2_{\sf u}\right)^{j}(\zeta_1\cdot\bold{k}_4)^{J-3}\right),
\end{multline}
where $g^{\left(J,2\right)}_{i0|j}(s_1,s_i)$, $i=2,3,4$, are polynomials of at most degree $3-j$ in $s_1$ and $s_i$. For instance, for the $j=0$ contribution, by evaluating \eqref{WTcontact} using \eqref{Uint} we have:

{\footnotesize\begin{multline}\label{obspm}
  \frac{g^{\left(J,2\right)}_{20|0}(s_1,s_2)}{g^{\left(J,2\right)}_{20}} =  -\frac{i \left[8 (1-i \mu ) s_1 \left(12 s^2_2 (x-4)-x^2+3 \mu ^2 (x-4)-i \mu  (x-10) (x-6)+16 x-60\right)\right]}{8 (x-4)}\\
    -\frac{i \left[(x-4) \left(16 s_2 \mu^2+x \left(x-18\right)+24 s^2_2 (i \mu  (x-8)-x+12)+64 s^3_2+6 i \mu ^3 (x-8)+\mu ^2 ((x-24) x+152)+80\right)\right]}{8 (x-4)}\\
    -\frac{i \left[16 \left(\mu ^2+1\right) s^2_1 (x-10)\right]}{8 (x-4)},
\end{multline}}

\noindent which is degree 3 in $s_1$ and $s_2$. 

We can then ask if a contact term of the above form can be cancelled by adding improvement terms to the cubic vertex. It is sufficient to focus on those improvements which could cancel the contact term with $j=0$ which, as we saw in the previous section, can only be linear in $u$. In the case of a partially massless-field, the improvement  $p^{\text{impr.}}\left(s_1,s_2,u\right)$ must vanish on more values of $s_1$, $s_2$ and $u$ compared to the massless case considered previously. For $r=2$ and helicity-$m=J-3$, these are given by \eqref{wtpoles} with
\begin{subequations}
\begin{align}
    n_1=1, \quad n_2=0, \quad n_3=0,\\
    n_1=0, \quad n_2=1, \quad n_3=0,\\
    n_1=0, \quad n_2=0, \quad n_3=1.
\end{align}
\end{subequations}
This constrains improvements that are linear in $u$ to take one of the following forms:
\begin{subequations}
\begin{align}
    p^{\text{impr.}}\left(s_1,s_2,u\right)&=u\left(s_1-\tfrac{x-8}{4}\right)\left(s_1-\tfrac{x-8}{4}+1\right)q^{\text{impr.}}_1\left(s_1,s_2\right)\label{impq1}\\\label{impq2}
    p^{\text{impr.}}\left(s_1,s_2,u\right)&=u\left(s_1-\tfrac{x-8}{4}\right)\left(s_2+ \tfrac{i\nu_2}{2}\right)\left(s_2- \tfrac{i\nu_2}{2}\right)q^{\text{impr.}}_2\left(s_1,s_2\right),\\
     p^{\text{impr.}}\left(s_1,s_2,u\right) &=u\left(s_2+ \tfrac{i\nu_2}{2}+1\right)\left(s_2- \tfrac{i\nu_2}{2}+1\right)\left(s_2+ \tfrac{i\nu_2}{2}\right)\left(s_2- \tfrac{i\nu_2}{2}\right)q^{\text{impr.}}_3\left(s_1,s_2\right),\label{impq3}
\end{align}
\end{subequations}
where $q^{\text{impr.}}_1$, $q^{\text{impr.}}_2$ and $q^{\text{impr.}}_3$ are polynomials in $s_1$ and $s_2$. Note that improvements of the form \eqref{impq3} are  polynomials of at least degree 4 in $s_2$ and are therefore not useful to cancel the contact term \eqref{obspm}, which is degree 3 in $s_1$ and $s_2$. The other two possible forms of improvement \eqref{impq1} and \eqref{impq2} are of at least degree-2 and degree-3 respectively, but they both have zeros at $s_1=\tfrac{x-4}{4}$. It is straightforward to check that the contact term \eqref{obspm} does not have a zero at $s_1=\tfrac{x-4}{4}$ and therefore no combination of \eqref{impq1} and \eqref{impq2} can be chosen to cancel it. The contact term must therefore vanish by itself, giving the constraint:
\begin{equation}\label{pmobsconstr}
    \left(g^{\left(J,2\right)}_{20|0}(s_1,s_2)(\zeta_1\cdot\bold{k}_2)^{J-3}+g^{\left(J,2\right)}_{30|0}(s_1,s_3)(\zeta_1\cdot\bold{k}_3)^{J-3}+g^{\left(J,2\right)}_{40|0}(s_1,s_4)(\zeta_1\cdot\bold{k}_4)^{J-3}\right)=0.
\end{equation}
The difference between this constraint and that \eqref{wtml} for the massless case is that in the above each tensorial structure is multiplied by a function $g^{\left(J,2\right)}_{i0|0}(s_1,s_i)$ of the Mellin variables rather than a constant. For $J>4$ this requires:
\begin{equation}
    g^{\left(J>4,2\right)}_{20|0}(s_1,s_2)=g^{\left(J>4,2\right)}_{30|0}(s_1,s_3)=g^{\left(J>4,2\right)}_{40|0}(s_1,s_4)=0,
\end{equation}
which in turn implies
\begin{equation}
    g^{\left(J>4,2\right)}_{20}=g^{\left(J>4,2\right)}_{30}=g^{\left(J>4,2\right)}_{40}=0.
\end{equation}
This is the partially-massless depth-2 analogue of the higher-spin constraint \eqref{hsconstr} in the massless case. For spin $J=4$ the constraint \eqref{pmobsconstr} requires:
\begin{equation}\label{J4constpm}
    g^{\left(J=4,2\right)}_{20|0}(s_1,s_2)=g^{\left(J=4,2\right)}_{30|0}(s_1,s_3)=g^{\left(J=4,2\right)}_{40|0}(s_1,s_4),
\end{equation}
while for spin $J=3$ we have
\begin{equation}\label{J3constpm}
    g^{\left(J=3,2\right)}_{20|0}(s_1,s_2)+g^{\left(J=3,2\right)}_{30|0}(s_1,s_3)+g^{\left(J=3,2\right)}_{40|0}(s_1,s_4)=0.
\end{equation}
We have checked explicitly that neither of \eqref{J4constpm} and \eqref{J3constpm} hold for the expression \eqref{obspm} for $g^{\left(J,2\right)}_{i0|0}(s_1,s_i)$, taking into account that the external Mellin variables are related via the constraint ${s_1+s_2+s_3+s_4=\tfrac{x-12+{\bar x}}{4}}$ and the possibility to add improvement terms. We therefore have
\begin{equation}
    g^{\left(J,2\right)}_{20|0}(s_1,s_2)=g^{\left(J,2\right)}_{30|0}(s_1,s_3)=g^{\left(J,2\right)}_{40|0}(s_1,s_4)=0,
\end{equation}
for all spins $J$,\footnote{Note that partially massless fields of depth-2 only have spin $J>2$.} which in turn implies that there is \emph{no consistent cubic coupling of a depth-2 partially-massless field to scalar matter}:
\begin{shaded}
\begin{equation}
    g^{\left(J,2\right)}_{20}=g^{\left(J,2\right)}_{30}=g^{\left(J,2\right)}_{40}=0.
\end{equation}
\end{shaded}
\noindent Like for the massless case in the previous section this assumes locality of quartic interactions. As a further confirmation of this result, in the following we will recover it (and those of the previous section) using an alternative approach for conformally coupled scalars and $d=3$.

\subsection{Special case: Conformally coupled scalars}

In section \ref{subsec::cc3pt} we saw that three-point functions of conformally coupled scalars have simple explicit expressions that do not involve Mellin-Barnes integrals. For the four-point exchanges, this implies that the representation given by \eqref{0>}, \eqref{0<} and \eqref{odotm} can be reduced to simpler form when the scalars are conformally coupled upon evaluating the integrals in $s_{1,2,3,4}$, $u$ and ${\bar u}$. If we also replace the spin-$J$ field with a conformally coupled scalar, these read (see section 4.6 of \cite{Sleight:2019hfp}): 
\begin{subequations}\label{ccc4ptexch}
\begin{align}
    \mathcal{A}^{({\sf  s})}_{>|\tfrac{i}{2},0;\tfrac{i}{2},0;\tfrac{i}{2},0;\tfrac{i}{2},0}&=\frac{{\cal N}_4}{{\bar k}_{\sf s}\,k_1 k_2 k_3 k_4}\int^{+i\infty}_{-i\infty} \frac{dw}{2\pi i}\,\frac1{w+\epsilon}\,\cos \left(2\pi\omega+\tfrac{\pi (\bar{x}-x)}{4} \right)\Gamma \left(-2 w+\tfrac{x-3}{2}\right) \Gamma \left(2 w+\tfrac{\bar{x}-3}{2}\right)\\\nonumber&\hspace{100pt}\times (-k_{\sf s}+k_1+k_2)^{2 w+\frac{3-x}{2}} (k_3+k_4+\bar{k}_{\sf s})^{-2 w+\frac{3-\bar{x}}{2}}\\
    \mathcal{A}^{({\sf  s})}_{<|\tfrac{i}{2},0;\tfrac{i}{2},0;\tfrac{i}{2},0;\tfrac{i}{2},0}&=-\frac{{\cal N}_4}{{\bar k}_{\sf s}\,k_1 k_2 k_3 k_4}\int^{+i\infty}_{-i\infty} \frac{dw}{2\pi i}\,\frac1{w+\epsilon}\,\cos \left(2\pi w+\tfrac{\pi (x-\bar{x})}{4}\right)\,\Gamma \left(-2 w+\tfrac{\bar{x}-3}{2}\right) \Gamma \left(2 w+\tfrac{x-3}{2}\right)\nonumber\\&\hspace{100pt}\times (k_{\sf s}+k_1+k_2)^{-2 w+\frac{3-\bar{x}}{2}} (-k_3-k_4+\bar{k}_{\sf s})^{2 w+\frac{3-\bar{x}}{2}}\\
    \mathcal{A}^{({\sf  s})}_{\odot|\tfrac{i}{2},0;\tfrac{i}{2},0;\tfrac{i}{2},0;\tfrac{i}{2},0}&=\frac{{\cal N}_4}{{\bar k}_{\sf s} k_1 k_2 k_3 k_4}\,\cos\left(\tfrac{\pi}4(x-\bar{x})\right)\Gamma \left(\frac{d-3}{2}\right)^2 (k_{\sf s}+k_1+k_2)^{\frac{3-x}{2}} (k_3+k_4+\bar{k}_{\sf s})^{\frac{3-\bar{x}}{2}}\,.
\end{align}
\end{subequations}
The final $w$-integral in \eqref{ccc4ptexch} can be evaluated to give an explicit expression for the exchange in terms of the Gauss hypergeometric function (see \cite{Sleight:2019mgd} equation (4.54)):
\begin{subequations}
\begin{multline}\label{exchseed1}
   \mathcal{A}^{({\sf  s})}_{>|\tfrac{i}{2},0;\tfrac{i}{2},0;\tfrac{i}{2},0;\tfrac{i}{2},0}= {\cal N}_4\frac{\sin \left(\tfrac{\pi(x+\bar{x})}{4} \right) \Gamma \left(\tfrac{x+\bar{x}}2-3\right)}{8 {\bar k}_{\sf s}(x-3)} \left(k_3+k_4+{\bar k}_{\sf s}\right)^{3-\tfrac{x+\bar{x}}2} \,\\ \times  _2F_1\left(\frac{x-3}{2},\frac{x+\bar{x}}2-3;\frac{x-1}{2};\frac{k_{\sf s}-k_1-k_2}{k_3+k_4+{\bar k}_{\sf s}}\right),
\end{multline}
and
\begin{multline}\label{exchseed2}
   \mathcal{A}^{({\sf  s})}_{<|\tfrac{i}{2},0;\tfrac{i}{2},0;\tfrac{i}{2},0;\tfrac{i}{2},0} = {\cal N}_4\frac{\sin \left(\tfrac{\pi(x+\bar{x}}{4} \right)\Gamma \left(\tfrac{x+\bar{x}}2-3\right) }{8 {\bar k}_{\sf s} (x-3)}\\ \times (-k_3-k_4+{\bar k}_{\sf s})^{3-\tfrac{x+\bar{x}}2} \, _2F_1\left(\frac{x-3}{2},\frac{x+\bar{x}}2-3;\frac{x-1}{2};\frac{k_{\sf s}+k_1+k_2}{{\bar k}_{\sf s}-k_3-k_4}\right).
\end{multline}
\end{subequations}
Above we kept ${\bar k}_{\sf s} \ne k_{{\sf s}}$ and $x, {\bar x}$, so that one can then act with the weight-shifting operators of section \ref{subsec::WSop} on the constituent 3pt structures to change the scaling dimension and the spin of the external fields -- in the same spirit as section \ref{subsec::cc3pt}.  Afterwards one should set $k_{{\sf s}} = {\bar k}_{\sf s}$. In this way, one can generate the exchange with a single external (partially-)massless field and conformally coupled scalars starting from the above scalar seed.

In the following we will take this approach to derive some examples of explicit expressions for the exchange \eqref{full4pt} and with a single external (partially-)massless spinning field and conformally coupled scalars for $d=3$. We also study the constraints coming from gauge-invariance, confirming the more general analysis given in sections \ref{subsecc::genml4pt} and \ref{subsec::pmcouplings}. In particular, we will show that there are contact singularities in $E_T$ in the lower helicity components of exchange diagrams that cannot be generated by local quartic contact interactions, which can only be higher order singularities in $E_T$.

\paragraph{Coupling massless spinning fields to conformally coupled scalars.}

The exchange for an external massless spinning field is generated in the same spirit as the corresponding three-point function considered in section \ref{subsec::cc3pt}: We act with the weight-shifting differential operator \eqref{spinJ3fromsc} on the explicit expression for the exchange involving only conformally coupled scalars (given by \eqref{exchseed1} and \eqref{exchseed2}), then set $d=3$ and ${\bar k}_{\sf s}=k_{\sf s}$. This is straightforwardly implemented in Mathematica. \\

\noindent \emph{External Massless spin-1 field} ($\nu_1=-\frac{i}{2}$): Following the procedure described above we obtain:
\begin{multline}
     {\cal A}^{\left({\sf s}\right)}_{-\tfrac{i}{2},1;\tfrac{i}{2},0;\tfrac{i}{2},0;\tfrac{i}{2},0}\left(\boldsymbol{\epsilon}_1, {\bf k}_1, {\bf k}_2,{\bf k}_3, {\bf k}_4\right)={\cal N}_4\,g_{20}^{(1,0)}\left[\frac{i \boldsymbol{\epsilon}_1\cdot\bold{k}_1 }{k_1 k_3 k_4 (k_{\sf s}+k_{12}) (k_{12}-k_{\sf s})}\log \left(\frac{k_{\sf s}+k_{34}}{E_T}\right)\right.\\\left.-\frac{i \boldsymbol{\epsilon}_1\cdot\bold{k}_2 }{k_2 k_3 k_4 (k_{12}-k_{\sf s}) (k_{\sf s}+k_{12})}\log \left(\frac{k_{\sf s}+k_{34}}{E_T}\right)\right]\,.
\end{multline}
The helicity-1 part of this exchange was given in \cite{Baumann:2020dch}, which matches with the second line of our expression above. As before, the helicity-0 component is extracted by acting with the projector \eqref{hp}, giving:
\begin{multline}\label{hel0-1000}
    \widehat{\mathcal{E}}_{1,0}^{(\bold{k}_1)}\left[{\cal A}^{\left({\sf s}\right)}_{-\tfrac{i}{2},1;\tfrac{i}{2},0;\tfrac{i}{2},0;\tfrac{i}{2},0}\left(\boldsymbol{\epsilon}_1, {\bf k}_1, {\bf k}_2,{\bf k}_3, {\bf k}_4\right)\right]\Big|_{\boldsymbol{\epsilon}_1 \to \zeta_1}\\=-{\cal N}_4\,g_{20}^{(1,0)}\frac{i}{2 k_1 k_2 k_3 k_4}\,\Bigg[\log \left(k_{\sf s}+k_{34}\right)-\log E_T\Bigg]\,.
\end{multline}
Note that this is zeroth order in $k^2_{\sf s}$, as consistent with \eqref{mlcontact} which is proportional to the Dirac delta function \eqref{ddcontactml}.
The term on the left in the square bracket is proportional to the three-point function of conformally coupled scalars in $d=3$ (see e.g. (3.35) in \cite{Sleight:2019mgd}) and so gives the four-point Ward-Takahashi identity. The term on the second line has a singularity in $E_T = k_1+k_2+k_3+k_4$ and so is a bulk quartic contact term that violates the Ward-Takahashi identity. This singularity is however logarithmic, i.e. proportional to $\log E_T$, which cannot be generated by a local quartic vertex involving a single massless spin-1 field and three conformally coupled scalars. This can be understood from the following simple argument:\footnote{We have also checked this explicitly by extracting the explicit contact terms generated by the lowest derivative admissible improvements \eqref{impmlwein} which have $q^{\text{impr.}}_1=q^{\text{impr.}}_1=\text{const.}$ and evaluating the integrals in $s_j$, confirming that there is no $\log E_T$ singularity.} The singularity of the contact diagram \eqref{ccphi4} generated by the $\phi^4$ interaction where $\phi$ is a conformally coupled scalar is a simple pole in $E_T$ for $d=3$. Derivatives only increase the order of the singularity in $E_T$. Therefore, for contact diagrams of conformally coupled scalars the lowest order singularity is a simple pole and so, in particular, they cannot contain $\log E_T$ terms. Now, for $d=3$, conformally coupled scalars are in the same higher-spin multiplet as massless spinning fields \cite{Fradkin:1986ka} and their contact diagrams are therefore related by higher-spin symmetry. Contact diagrams involving a massless spinning field and three conformally coupled scalars therefore cannot contain singularities in $E_T$ that are lower order than those of four conformally coupled scalars.\footnote{The higher-spin symmetry transformation can be realised as derivative operators, an example of which is the operator \eqref{spinJ3fromsc}.} The $\log E_T$ singularity in \eqref{hel0-1000} must therefore cancel upon summing the ${\sf s}$-, ${\sf t}$- and ${\sf u}$-channel exchanges, giving:
\begin{equation}
    \left(g^{\left(1,0\right)}_{20}+g^{\left(1,0\right)}_{30}+g^{\left(1,0\right)}_{40}\right)\log E_T=0,
\end{equation}
which recovers charge conservation \eqref{cc}.

\noindent \emph{External Massless spin-2 field} ($\nu_1=-\frac{3i}{2}$): In this case we obtain
\begin{align}\label{graviton4pt}
  &  {\cal A}^{\left({\sf s}\right)}_{-\tfrac{3i}{2},2;\tfrac{i}{2},0;\tfrac{i}{2},0;\tfrac{i}{2},0}\left(\boldsymbol{\epsilon}_1, {\bf k}_1, {\bf k}_2,{\bf k}_3, {\bf k}_4\right)\\&\hspace*{2.5cm}={\cal N}_4\,g_{20}^{(2,0)}\frac{1}{k_2 k_3 k_4}\left[(\boldsymbol{\epsilon}_1\cdot\bold{k}_2)^2\left(\frac{k^2_{\mathsf{s}}-k_{12}^2-2k_{12}k_1}{2\left(k_{\sf s}+k_{12}\right)^2\left(k_{\sf s}-k_{12}\right)^2}\,\log \left(\frac{k_{34}+k_{\mathsf{s}}}{E_T}\right)\right.\right. \nonumber \\&\hspace*{10cm} \left.\left.-\frac{k_1}{2\left(k_{12}+k_{\mathsf{s}}\right)\left(k_{12}-k_{\mathsf{s}}\right) E_T}\right)\right.\nonumber\\
    &\hspace*{-0.25cm}+(\boldsymbol{\epsilon}_1\cdot\bold{k}_2)(\boldsymbol{\epsilon}_1\cdot\bold{k}_1)\left(\frac{k^2_{\mathsf{s}}-k^2_1+k^2_2}{k_{\mathsf{s}}^2\left(k_{\sf s}+k_{12}\right)^2\left(k_{\sf s}-k_{12}\right)^2}\,\log \left(\frac{k_{34}+k_{\sf s}}{E_T}\right)-\frac{k_2}{k_{\mathsf{s}}^2 \left(k_{\sf s}-k_{12}\right)\left(k_{\sf s}+k_{12}\right)}\frac{1}{E_T}\right)\nonumber\\\nonumber
    &\hspace*{-0.25cm}+(\boldsymbol{\epsilon}_1\cdot\bold{k}_1)^2\left(\frac{k^2_1-k^2_2-k^2_{\mathsf{s}}}{2k_{\mathsf{s}}^2\left(k_{\sf s}+k_{12}\right)^2\left(k_{\sf s}-k_{12}\right)^2}\log \left(\frac{k_{34}+k_{\sf s}}{E_T}\right)-\frac{k_2}{2k_{\mathsf{s}}^2\left(k_{12}+k_{\sf s}\right)\left(k_{12}-k_{\sf s}\right)}\frac{1}{E_T}\right.\\
    &\hspace{10cm}\left.\left.+\frac{1}{12 k_2k_{\mathsf{s}}^2}\frac{1}{E_T}+\frac{1}{24k_1 k_2k_{\mathsf{s}}^2}\right)\right]\,.\nonumber
\end{align}
The helicity-2 component was given in \cite{Baumann:2020dch} which matches with the first line of the expression above. For the helicity-1 component we have:
\begin{multline}
      \widehat{\mathcal{E}}_{2,1}^{(\bold{k}_1)}\left[ {\cal A}^{\left({\sf s}\right)}_{-\tfrac{3i}{2},2;\tfrac{i}{2},0;\tfrac{i}{2},0;\tfrac{i}{2},0}\left(\boldsymbol{\epsilon}_1, {\bf k}_1, {\bf k}_2,{\bf k}_3, {\bf k}_4\right)\right]\Big|_{\boldsymbol{\epsilon}_1 \to \zeta_1}
      \\ ={\cal N}_4\,g_{20}^{(2,0)}\frac{\zeta_1\cdot\bold{k}_2}{4k_2 k_3 k_4}\Bigg[\log (k_3+k_4+k_{\mathsf{s}})-\log E_T+\frac{k_1}{E_T}\Bigg].
\end{multline}
 Like for the massless spin-1 case above, the left term in the square brackets is proportional to the three-point function of conformally coupled scalars and so gives the corresponding 4pt Ward-Takahashi identity. In addition to this we have singularities in $E_T$ which violate the Ward-Takahashi identity. One of this is a simple pole in $E_T$ which in principle can be compensated by adding a local quartic contact interaction. The other is $\log E_T$ which, like for the massless spin-1 case, cannot be compensated and must therefore vanish upon summing the ${\sf s}$-, ${\sf t}$- and ${\sf u}$-channel exchanges, giving:
\begin{equation}
    \left(g^{\left(2,0\right)}_{20}\left(\zeta_1 \cdot {\bf k}_2\right)+g^{\left(2,0\right)}_{30}\left(\zeta_1 \cdot {\bf k}_3\right)+\left(\zeta_1 \cdot {\bf k}_4\right)g^{\left(2,0\right)}_{40}\right)\log E_T=0,
\end{equation}
which recovers the equivalence principle \eqref{ep}. The helicity-0 component is given by:
\begin{multline}
      \widehat{\mathcal{E}}_{2,0}^{(\bold{k}_1)}\left[ {\cal A}^{\left({\sf s}\right)}_{-\tfrac{3i}{2},2;\tfrac{i}{2},0;\tfrac{i}{2},0;\tfrac{i}{2},0}\left(\boldsymbol{\epsilon}_1, {\bf k}_1, {\bf k}_2,{\bf k}_3, {\bf k}_4\right)\right]\Big|_{\boldsymbol{\epsilon}_1 \to \zeta_1}
      \\= {\cal N}_4\,\frac{g_{20}^{(2,0)}}{k_2 k_3 k_4}\left[-\frac{1}{24}\left(k_{1}^2+3 (k_{2}-k_{\mathsf{s}}) (k_{2}+k_{\mathsf{s}})\right)\log \left(k_{3}+k_{4}+k_{\mathsf{s}}\right)\nonumber\right.\\
    \left.+\frac{k_{1}}{72 E_T}\left(-3 k_{1}^2+2 k_{1} (-2 k_{2}+k_{3}+k_{4})-9 k_{2}^2+9 k_{\mathsf{s}}^2\right)+\frac{\left(k_{1}^2+3 (k_{2}-k_{\mathsf{s}}) (k_{2}+k_{\mathsf{s}})\right)}{24}\log E_T \right]\,,
\end{multline}
where we recognise the first term on the r.h.s. gives the four-point Ward-Takahashi identity, since it is proportional to the three-point function of conformally coupled scalars in $d=3$. The other terms contain singularities in $E_T$ and, using that 
\begin{align}
    k_{\mathsf{s}}^2+k_{\mathsf{t}}^2+k_{\mathsf{u}}^2-k_1^2-k_2^2-k_3^2-k_4^2=0\,,
\end{align}
where
\begin{align}
    k_\mathsf{s}&=|{\bf k}_1+{\bf k}_2|\,,& k_\mathsf{t}&=|{\bf k}_1+{\bf k}_3|\,,& k_\mathsf{u}&=|{\bf k}_1+{\bf k}_4|\,,
\end{align}
it is straightforward to show that these cancel upon summing the ${\sf s}$-, ${\sf t}$- and ${\sf u}$-channel exchanges only if the equivalence principle \eqref{ep} holds.\\

\noindent \emph{External Massless spin-J field} ($\nu_1=-\left(J-\tfrac{1}{2}\right)i$): For higher spins $J$ the action of the operator \eqref{spinJ3fromsc} which generates the exchange from the scalar seed \eqref{ccc4ptexch} gets increasingly involved, though it is straightforward to generate explicit results for a given spin-$J$ by implementing its action in Mathematica. Such explicit expressions, as one can already understand from the corresponding three-point functions in section \ref{subsec::cc3pt}, get more and more complicated as the spin-$J$ increases. In the following we therefore just focus on the constraints coming from gauge-invariance where, following the analysis in section \ref{subsecc::genml4pt}, it is sufficient to focus on the helicity-$\left(J-1\right)$ contribution \eqref{mlcontact}. For conformally coupled scalars, the Mellin-Barnes integrals that appear in the helicity-$\left(J-1\right)$ contribution \eqref{mlcontact} are equivalent to that of the quartic contact diagram generated by the non-derivative interaction of three conformally coupled scalars and a scalar with $\nu_1 = -\left(J-\tfrac{1}{2}\right)i$ and boundary dimension:\footnote{This is read off from the Dirac delta function \eqref{ddcontactml} and in the second equality we made the replacements: $x=d+2J$ and ${\bar x}=d$.}
\begin{equation}
  d^\prime=\frac{x+{\bar x}-4}{2}=d+J-2.
\end{equation} 
Such a contact diagram can be generated by acting $J$ times with the differential operator \eqref{Praise} on the four-point contact diagram \eqref{ccphi4} of conformally coupled scalars where $x$ in $d^\prime$ shifted by $x \to x-2J$, giving boundary dimension $d^\prime=d-2$. From \eqref{ccphi4}, the latter is given explicitly by
\begin{align}\label{4ptexchseedcc}
   {\cal N}_4\, \frac{2}{k_1k_2k_3k_4} \sin \left(\frac{d \pi}{2}\right) \Gamma \left(d-4\right)  E_T^{4-d}.
\end{align}
For $d=3$, by carefully expanding, one obtains
\begin{align}
 {\cal N}_4\, \frac{2}{k_1k_2k_3k_4}\left[  E_T\frac{ 1-(\gamma -1)(d-3)}{(d-3)}-E_T \log E_T+O\left(d-3\right)\right],
\end{align}
which in particular contains the non-analytic term $E_T \log E_T$. Upon acting $J$ times with the differential operator \eqref{Praise}, this term is responsible for $\log E_T$ singularities in the exchange for external massless spin-$J$, for all $J$. To illustrate, for external massless spin-1, the helicity-0 component is obtained upon acting with \eqref{Praise} on \eqref{4ptexchseedcc}, which gives
\begin{multline}
   {}^{\left(0\right)}{\cal A}^{\left({\sf s}\right)}_{-\frac{i}{2},1;\frac{i}{2},0;\frac{i}{2},0;\frac{i}{2},0}\left(\boldsymbol{\epsilon}_1,{\bf k}_1;{\bf k}_2,{\bf k}_3,{\bf k}_4\right)\Bigg|_{\text{contact}} \\ = {\cal N}_4\, g_{20}^{(1,0)} \frac{1}{k_2 k_3 k_4}\left[\frac{1-\gamma\left(d-3\right)}{\left(d-3\right)}+\log E_T+O\left(d-3\right)\right].
\end{multline}
For massless spin-$2$, acting twice with \eqref{Praise} on \eqref{4ptexchseedcc}, for the helicity-1 component we have
\begin{multline}
  \hspace*{-0.5cm} {}^{\left(1\right)}{\cal A}^{\left({\sf s}\right)}_{-\frac{3i}{2},2;\frac{i}{2},0;\frac{i}{2},0;\frac{i}{2},0}\left(\boldsymbol{\epsilon}_1,{\bf k}_1;{\bf k}_2,{\bf k}_3,{\bf k}_4\right)\Bigg|_{\text{contact}} = {\cal N}_4\, g_{20}^{(2,0)}\left(\zeta_1 \cdot {\bf k}_2\right)\frac{1}{k_2 k_3 k_4} \left[-\frac{d-4}{2(d-3)}+ (\gamma -1)E_T\right.\\\left.+\frac{1}{2}\log E_T-\frac{ k_1}{2E_T}+O\left(d-3\right)\right].
\end{multline}
For massless spin-$3$, acting three times with \eqref{Praise} on \eqref{4ptexchseedcc}, for the helicity-2 component we have
\begin{multline}
   {}^{\left(2\right)}{\cal A}^{\left({\sf s}\right)}_{-\frac{5i}{2},2;\frac{i}{2},0;\frac{i}{2},0;\frac{i}{2},0}\left(\boldsymbol{\epsilon}_1,{\bf k}_1;{\bf k}_2,{\bf k}_3,{\bf k}_4\right)\Bigg|_{\text{contact}} =  {\cal N}_4\,g_{20}^{(3,0)}\left(\zeta_1 \cdot {\bf k}_2\right)^2\frac{1}{k_2k_3k_4}\left[\frac{3}{4 (d-3)}\right.\\+\left((4-3 \gamma ) k^2_1-3 (2 \gamma -1) k_1 (k_2+k_3+k_4)-3 \gamma  (k_2+k_3+k_4)^2\right)\frac{1}{4 E^2_T}\\\left.-\frac{3}{4}\log E_T+O\left(d-3\right)\right],
\end{multline}
and so on for higher spin $J$ which is obtained by acting $J$ times with the operator \eqref{Praise}, where each application generates a $\log E_T$ singularity as demonstrated above. As we saw in the above examples these must cancel upon summing the ${\sf s}$-, ${\sf t}$- and ${\sf u}$-channels, giving rise to the following constraint for general spin-$J$:
\begin{align}
    \left(g^{\left(J,0\right)}_{20}\left(\zeta_1 \cdot {\bf k}_2\right)^{J-1}+g^{\left(J,0\right)}_{30}\left(\zeta_1 \cdot {\bf k}_3\right)^{J-1}+\left(\zeta_1 \cdot {\bf k}_4\right)^{J-1}g^{\left(J,0\right)}_{40}\right)\log E_T=0\,,
\end{align}
which once again recovers charge conservation \eqref{cc} and the equivalence principle \eqref{ep} for $J=1$ and $J=2$, while for $J>2$ that there can be no consistent coupling of massless higher-spin fields to scalar matter (in local theories).

\paragraph{Coupling depth-2 P-M spinning fields to conformally coupled scalars.} From section \ref{subsec::cc3pt} we know that the $d=3$ exchange with a single external depth-2 partially massless field and conformally coupled scalars can be generated, via the weight-shifting identity \eqref{spinJ3fromsc}, from the exchange with the partially massless field replaced by a scalar of scaling dimension ${\bar \nu_1}=\frac{5i}{2}$. 

In the following we will focus on the constraints coming from gauge invariance, which only requires to study the bulk quartic contact terms \eqref{PMcontactquartic} in the helicity-$\left(J-3\right)$ component of the exchange and to identify any terms which cannot be generated by local quartic vertices. We first note that the contact terms in \eqref{PMcontactquartic} with $j=1$, that are linear in $k^2_{\sf s}$, $k^2_{\sf t}$ and $k^2_{\sf u}$, contain a $\log E_T$ singularity. This can be understood by making a similar argument to that which we gave for the massless case above. In particular, by decomposing the polynomial $g^{\left(J,2\right)}_{20|1}\left(s_1,s_2\right)$ in the basis $\left(s_1-\tfrac{i\nu_1}{2}\right)_{n_1}\left(s_2-\tfrac{i\nu_2}{2}\right)_{n_2}$, one notes that for the non-zero $n_1=n_2=0$ component the Mellin-Barnes integrals are equivalent to that of the quartic contact diagram generated by the zero-derivative interaction of three-conformally coupled scalars and a scalar with $\nu_1=-i\left(J-\tfrac{5}{2}\right)$, and with boundary dimension\footnote{This is read off from the Dirac delta function with $j=1$ in \eqref{PMcontactquartic} and in the second equality we replaced $x=d+2J$ and ${\bar x}=d$.}
\begin{equation}
  d^\prime=\frac{x+{\bar x}-12+4}{2}=d+J-4.
\end{equation} 
The latter contact diagram can be generated by acting $\left(J-2\right)$ times with the differential operator \eqref{Mlow} on the zero-derivative four-point contact diagram \eqref{ccphi4} of conformally coupled scalars, where $x$ in the boundary dimension $d^\prime$ is shifted by $x \to x-2\left(J-2\right)$, giving $d^\prime=d-2$. The latter is precisely \eqref{4ptexchseedcc} which, by the same argument as the massless case above, is responsible for $\log E_T$ singularities in the exchange with a single external partially massless spin-$J$, for all $J$, upon acting on \eqref{4ptexchseedcc} with the differential operator \eqref{Praise}. We can therefore conclude that the helicity-$\left(J-3\right)$ component of the ${\sf s}$-channel exchange contains the following singularity in $E_T$ (where $\nu_1=-i\left(J-\tfrac{5}{2}\right)$):
\begin{equation}
 \hspace*{-0.3cm}   {}^{\left(J-3\right)}{\cal A}^{\left({\sf s}\right)}_{\nu_1,J;\frac{i}{2},0;\frac{i}{2},0;\frac{i}{2},0}\left(\boldsymbol{\epsilon}_1,{\bf k}_1;{\bf k}_2,{\bf k}_3,{\bf k}_4\right)\Bigg|_{\text{contact}} \supset {\cal N}_4\,g^{\left(J,2\right)}_{20}\left(\zeta_1 \cdot {\bf k}_2\right)^{J-3} \frac{k^2_{\sf s}}{k_2 k_3 k_4}\,\log E_T.
\end{equation}

Like for the massless case considered above, a simple argument shows that such a singularity cannot be generated by a local quartic vertex involving a single depth-2 partially-massless field and three conformally coupled scalars. The latter can be generated, via the application of the differential operator \eqref{spinJ3fromsc}, from contact diagrams involving a scalar field with ${\bar \nu}_1=\tfrac{5i}{2}$ and three conformally coupled scalars. The contact diagram with the external partially-massless field cannot have a lower order singularity in $E_T$ than the scalar seed from which it is generated. For $d=3$ the contact diagram generated by the zero-derivative quartic vertex involving a scalar field with ${\bar \nu}_1=\tfrac{5i}{2}$ and three conformally coupled scalars is given by\footnote{This expression was obtained by acting twice with the differential operator \eqref{Mlow} on the contact diagram \eqref{ccphi4} generated by the non-derivative quartic vertex $\phi^4$ with $\phi$ a conformally coupled scalar and boundary dimension $d^\prime=d-4$. One then expands around $d=3$.} 
\begin{equation}
    \frac1{k_2k_3k_4}\left(\frac{1}{k^2_1 E^{2}_T}+ \frac{1}{k^3_1 E_T} +{\cal O}\left(d-3\right)\right),
\end{equation}
which does not have a $\log E_T$ singularity. Diagrams generated by derivative vertices can only increase the singularity in $E_T$. We can therefore conclude that the $\log E_T$ singularity in the helicity-$\left(J-3\right)$ component of the exchange cannot be compensated by adding a local quartic vertex to the Lagrangian. It must therefore vanish by itself upon summing the ${\sf s}$-, ${\sf t}$- and ${\sf u}$-channels, which gives the constraint:
\begin{equation}
  \left( g^{\left(J,2\right)}_{20}k^2_{\sf s}\left(\zeta_1 \cdot {\bf k}_2\right)^{J-3}+g^{\left(J,2\right)}_{30}k^2_{\sf t}\left(\zeta_1 \cdot {\bf k}_3\right)^{J-3}+g^{\left(J,2\right)}_{40}k^2_{\sf u}\left(\zeta_1 \cdot {\bf k}_4\right)^{J-3}\right)\log E_T=0
\end{equation}
This can only be satisfied if, for all $J$,
\begin{equation}
  g^{\left(J,2\right)}_{20}=g^{\left(J,2\right)}_{30}=g^{\left(J,2\right)}_{40}=0,
\end{equation}
which recovers the result of section \ref{subsec::pmcouplings}, i.e. that there can be no consistent coupling of a depth-2 partially-massless field of any spin $J$ to scalar matter (in local theories).

\section*{Acknowledgments}

The research of C.S. was partially supported by l'Universit\'e libre de Bruxelles and the European Union's Horizon 2020 research and innovation programme under the Marie Sk\l odowska-Curie grant agreement No 793661. C.S. is supported by the STFC grant ST/T000708/1. The research of M.T. was partially supported by the program  “Rita  Levi  Montalcini”  of the MIUR (Minister for Instruction, University and Research) and the INFN initiative STEFI. C.S. would like to thank the organisers of the workshop ``Cosmology Meets CFT Correlators 2020” at the National Taiwan University for their hospitality and the participants for stimulating discussions.

\appendix

\section{Helicity projection operators}
\label{appendix::helciityproj}

In this appendix we introduce a convenient formalism to project a given spinning conformal structure into its helicity components. 

The main tools are the following Thomas-D operators \cite{Thomas352}:
\begin{subequations}\label{ThomasDs}
\begin{align}\label{thomasD}
    \left(D_{\boldsymbol{\epsilon}}\right)^i&=\left(\tfrac{d}{2}-1+ \boldsymbol{\epsilon}\cdot\pl_{\boldsymbol{\epsilon}}\right)\pl_{\boldsymbol{\epsilon}^i}-\tfrac12\boldsymbol{\epsilon}^i\pl_{\boldsymbol{\epsilon}^2}\,,\\
    \left(\mathcal{D}^{(\bold{p})}_{\boldsymbol{\epsilon}}\right)^i&=\left(\tfrac{d-3}{2}+ \boldsymbol{\epsilon}\cdot\pl_{\boldsymbol{\epsilon}}\right)\left[\pl_{\boldsymbol{\epsilon}^i}-\hat{p}^i \,\hat{{\bf p}}\cdot\pl_{\boldsymbol{\epsilon}}\right]-\tfrac12\boldsymbol{\epsilon}^i\left[\pl_{\boldsymbol{\epsilon}^2}-(\hat{{\bf p}}\cdot\pl_{\boldsymbol{\epsilon}})^2\right]\,,\label{thomasDtrans}
\end{align}
\end{subequations}
where $\hat{{\bf p}}\equiv{\bf p}/|{\bf p}|$. The first differential operator \eqref{thomasD} is the standard Thomas-D operator which acts on equivalence classes of $\boldsymbol{\epsilon}$-polynomials, modulo the relation $\boldsymbol{\epsilon}^2\sim0$. It returns the traceless/harmonic representative of such polynomials/tensors. The second operator \eqref{thomasDtrans} is still a Thomas-D operator but its action is now defined on equivalence classes of polynomials modulo $\boldsymbol{\epsilon}^2\sim0$ and $\boldsymbol{\epsilon}\cdot \bold{p}\sim0$. It returns the transverse (with respect to the vector $\bold{p}$) and traceless representative. Since polynomials in $\boldsymbol{\epsilon}$ are in one to one correspondence with tensors, in the following we shall loosely refer to such polynomials as tensors and to the corresponding traceless and transverse representatives as helicity components.

The helicity-$m$ component of a traceless symmetric rank-$J$ tensor with respect to a vector ${\bf p}$ then can be evaluated by combining the above two differential operators. One first projects $\left(J-m\right)$ symmetric indices in the longitudinal direction using \eqref{thomasD} and then the remaining indices in the transverse direction via \eqref{thomasDtrans}. This is implemented by the following operator:
\begin{align}\label{helicity_proj}
    {}^{\zeta}\mathcal{E}^{\left({\bf p}\right)}_{J,m}=\frac{\mathfrak{c}_{J,m}}{J!\left(\tfrac{d-2}2+m\right)_{J-m}\left(\tfrac{d-3}2\right)_{m}}\,(\zeta\cdot \mathcal{D}_{\boldsymbol{\epsilon}}^{\left({\bf p}\right)})^{m}(\hat{{\bf p}}\cdot D_{\boldsymbol{\epsilon}})^{J-m}\,,
\end{align}
where completeness fixes the coefficient $\mathfrak{c}_{J,m}$ to be:
\begin{align}
    \mathfrak{c}_{J,m}=\frac{2^{d+J+m-3} \Gamma \left(\tfrac{d}{2}+J-1\right) \Gamma \left(\tfrac{d}{2}+m-\frac{1}{2}\right)}{\sqrt{\pi } \Gamma (d+J+m-2)}\,.
\end{align}
This is manifestly transverse and traceless with respect to the vector $\zeta$ and has by definition helicity-$m$.

The action of \eqref{helicity_proj} on the monomial $(\boldsymbol{\epsilon}\cdot \zeta_2)^J$ is given by
\begin{align}
    {}^{\zeta_1}\mathcal{E}_{J,m}^{({\bf p})}\left[(\boldsymbol{\epsilon}\cdot \zeta_2)^J\right]=\mathfrak{c}_{J,m}\,\Xi_{m}(\zeta_1,\zeta_2)\,{\Upsilon}_{J-m}(\zeta_2)\,,
\end{align}
where we have defined the following normalised Gegenbauer polynomials:
\begin{subequations}
\begin{align}
        {\Upsilon}_{n}(\zeta)&=\frac{n!}{2^{n}\left(\frac{d-2}{2}+J-n\right)_n}\,\zeta^n\,C_{n}^{(\frac{d}{2}+J-n-1)}\left(\hat{\zeta}\cdot\hat{p}\right)\,,\\
        {\Xi}_{n}(\zeta_1,\zeta_2)&=\frac{n!}{2^{n}\left(\frac{d-3}{2}\right)_n}\left[\zeta_1^2-({\zeta_1}\cdot \hat{p})^2\right]^{n/2}\left[\zeta_2^2-({\zeta_2}\cdot \hat{p})^2\right]^{n/2}\label{XiPoly}\\\nonumber
        &\hspace{100pt}\times\,C_{n}^{(\frac{d-3}2)}\left(\tfrac{\zeta_1\cdot \zeta_2-(\zeta_1\cdot\hat{p})(\zeta_2\cdot\hat{p})}{\sqrt{\left[\zeta_1^2-(\zeta_1\cdot \hat{p})^2\right]\left[\zeta_2^2-(\zeta_2\cdot \hat{p})^2\right]}}\right).
\end{align}
\end{subequations}
From this follows the completeness relation:
\begin{align}\label{completeness}
    \frac{J!}{2^J\left(\frac{d-2}{2}\right)_J}(\boldsymbol{\epsilon}_1\boldsymbol{\epsilon}_2)^J C_{J}^{\tfrac{d-2}2}(\hat{\boldsymbol{\epsilon}}_1\cdot\hat{\boldsymbol{\epsilon}}_2)&=\sum_{m=0}^J {\Upsilon}_{J-m}(\boldsymbol{\epsilon}_1)\  {}^{\boldsymbol{\epsilon}_1}\mathcal{E}_{J,m}^{({\bf p})}\left[(\boldsymbol{\epsilon}\cdot\boldsymbol{\epsilon}_2)^J\right]\\&=\sum_{m=0}^J\mathfrak{c}_{J,m} {\Upsilon}_{J-m}(\boldsymbol{\epsilon}_1){\Xi}_{m}(\boldsymbol{\epsilon}_1,\boldsymbol{\epsilon}_2){\Upsilon}_{J-m}(\boldsymbol{\epsilon}_2)\,,\nonumber
\end{align}
which expresses the traceless contraction, encoded by the Gegenbauer polynomial on the left-hand side, in terms of a sum of transverse-traceless contractions encoded by the corresponding helicity harmonics $\Xi_m$. The above decomposition is equivalent to the following representation of the identity on traceless tensors:
\begin{align}
    \phi_J(\boldsymbol{\epsilon}_1)=\sum_{m=0}^{J}\Upsilon_{J-m}(\boldsymbol{\epsilon}_1)\ {}^{\boldsymbol{\epsilon}_1}\mathcal{E}_{J,m}^{({\bf p})}\left[\phi_{J}(\boldsymbol{\epsilon})\right]\,.
\end{align}

Employing on the left-hand side of \eqref{completeness} the equivalence relation $\boldsymbol{\epsilon}_i^2\sim 0$ and on the right hand side the transverse and traceless equivalence relation $\boldsymbol{\epsilon}_i^2-(\boldsymbol{\epsilon}_i\cdot p)^2\sim0$ one can write the suggestive identity:
\begin{align}
    (\boldsymbol{\epsilon}_1\cdot\boldsymbol{\epsilon}_2)^J\sim 
    \sum_{m=0}^J\mathfrak{c}_{J,m} {\Upsilon}_{J-m}(\boldsymbol{\epsilon}_1)(\widetilde{\boldsymbol{\epsilon}_1\cdot\boldsymbol{\epsilon}_2})^m{\Upsilon}_{J-m}(\boldsymbol{\epsilon}_2)\,,
\end{align}
where $\widetilde{\boldsymbol{\epsilon}_1\cdot\boldsymbol{\epsilon}_2}$ denotes the transverse and traceless contraction of $\boldsymbol{\epsilon}_1$ and $\boldsymbol{\epsilon}_2$. The above form suggests to define the following transverse projection operations:
\begin{align}\label{hp}
   \widehat{\mathcal{E}}^{({\bf p})}_{J,m}=\frac{1}{(m+1)_{J-m}\left(\tfrac{d-2}2+m\right)_{J-m}}\,(\bold{p}\cdot D_{\boldsymbol{\epsilon}})^{J-m}\,,
\end{align}
which can be used to define the $m$-th transverse-traceless component of an arbitrary tensorial structure. 

In view of the above decomposition it is therefore useful to work in terms of the helicity components so defined:
\begin{align}
    \phi_J(\boldsymbol{\epsilon}_1)=\begin{pmatrix}
    \phi_J^{(J)}(\boldsymbol{\epsilon}_1)\\
    \vdots\\
    \phi^{(0)}_{J}(\boldsymbol{\epsilon}_1)
    \end{pmatrix}\equiv\begin{pmatrix}
    \widehat{\mathcal{E}}_{J,J}^{({\bf p})}\left[\phi\right]\\
    \vdots\\
    \widehat{\mathcal{E}}_{J,0}^{({\bf p})}\left[\phi\right]
    \end{pmatrix}.
\end{align}
In terms of the above components the traceless contraction between traceless tensors reads:
\begin{align}
    \phi_1(\boldsymbol{\epsilon}_1)\circ\phi_2(\boldsymbol{\epsilon}_2)=\sum_{m=0}^{J}\frac{\mathfrak{c}_{J,m}}{p^{2m}}\left(\widehat{\mathcal{E}}_{J,m}^{({\bf p})}\left[\phi_1(\boldsymbol{\epsilon}_1)\right]\right)\hat{\circ}\left(\widehat{\mathcal{E}}_{J,m}^{({\bf p})}\left[\phi_2(\boldsymbol{\epsilon}_2)\right]\right)\,,
\end{align}
where $\hat{\circ}$ is the transverse and traceless contraction defined on the transverse subspace. E.g.:
\begin{align}
    \phi_1\,{\circ}^{(\bold{p})}\,\phi_2=\frac{1}{m!\left(\frac{d-3}2\right)_m}\phi_1\left(\mathcal{D}_{\zeta}^{({\bf p})}\right)\ \phi_2(\zeta)\,.
\end{align}

\section{Mellin-Barnes representation of 3pt functions with a spin-$J$ operator}
\label{APP::00J}

The Mellin-Barnes amplitude for the three-point correlation function of a spin-$J$ operator $O_{\nu_3,J}$ with auxiliary vector $\boldsymbol{\epsilon}_3$ and two scalar operators $O_{\nu_{1,2},0}$ reads (see section 3.2 of \cite{Sleight:2019hfp}):
\begin{multline}\label{00J3ptA}
 \hspace*{-0.5cm}    {\cal A}^{\left(d+2J\right)}_{\nu_1,0;\nu_2,0;\nu_3,J}\left(s_1,{\bf k}_1;s_2,{\bf k}_2;s_3,{\bf k}_3,\boldsymbol{\epsilon}_3\right) = g^{\left(J\right)}_{12}\, i \pi \delta\left(\tfrac{d+2J}{4}-s_1-s_2-s_3\right)\, \\ \times  \mathfrak{C}_{\nu_1,0;\nu_2,0;\nu_3,J}\left(s_1,s_2,s_3|\boldsymbol{\epsilon}_3 \cdot {\bf k}_1,\boldsymbol{\epsilon}_3 \cdot {\bf k}_2,\boldsymbol{\epsilon}_3 \cdot {\bf k}_3\right).
\end{multline}
The tensorial structure $\mathfrak{C}_{\nu_1,0;\nu_2,0;\nu_3,J}\left(s_j|\boldsymbol{\epsilon}_3 \cdot {\bf k}_j\right)$ is a degree $J$ polynomial in the contractions $\left(\boldsymbol{\epsilon}_3 \cdot {\bf k}_j\right)$:
\begin{multline}\label{tensor00JA}
 \mathfrak{C}_{\nu_1,0;\nu_2,0;\nu_3,J}\left(s_j|\boldsymbol{\epsilon}_3 \cdot {\bf k}_j\right)=\sum^J_{\alpha=0}\binom{J}{\alpha} \left(- \boldsymbol{\epsilon}_3 \cdot {\bf k}_3\right)^\alpha \sum^\alpha_{\beta=0}\binom{\alpha}{\beta} {\cal Y}^{\left(J\right)}_{\nu_1,\nu_2,\nu_3|\alpha,\beta}\left(\boldsymbol{\epsilon}_3 \cdot {\bf k}_1,\boldsymbol{\epsilon}_3 \cdot {\bf k}_2\right)\\ \times  H_{\nu_1,\nu_2,\nu_3|\alpha,\beta}\left(s_1,s_2,s_3\right),
\end{multline}
where
\begin{equation}\label{Hab}
    H_{\nu_1,\nu_2,\nu_3|\alpha,\beta}\left(s_1,s_2,s_3\right) = \frac{\left(s_1+\frac{i\nu_1}{2}\right)_{\alpha-\beta}\left(s_2+\frac{i\nu_2}{2}\right)_{\beta}}{\left(s_3+\frac{i\nu_3}{2}-\alpha\right)_\alpha},
\end{equation}
and 
\begin{multline}\label{yab}
    \mathcal{Y}^{\left(J\right)}_{\nu_1,\nu_2,\nu_3|\alpha,\beta}\left(\boldsymbol{\epsilon}_3 \cdot {\bf k}_1,\boldsymbol{\epsilon}_3 \cdot {\bf k}_2\right)=(-i)^J\frac{\left(\tfrac{4-d-2 J-2i (\nu_1-\nu_2+\nu_3)}{4} \right)_\beta \left(\tfrac{d-4 \alpha+4 \beta+2 J-2 i (\nu_1-\nu_2-\nu_3)}{4}\right)_{\alpha-\beta}}{\left(\tfrac{d}{2}+i \nu_3-1\right)_{J-\alpha}\left(\frac{d}{2}+i \nu_3+J-\alpha-1\right)_\alpha}\\\times\sum_{n=0}^{J-\alpha}\left(\tfrac{d-4 \beta+2 J-4 n+2 i \nu_1-2 i \nu_2+2 i \nu_3}{4}\right)_n \left(\tfrac{d+4 \beta-2 J+4 n-2 i \nu_1+2 i \nu_2+2 i \nu_3}{4}\right)_{J-\alpha-n} \\\times \binom{J-\alpha}{n}\,({\boldsymbol{\epsilon}_3}\cdot \bold{k}_1)^{J-\alpha-n}\,(-{\boldsymbol{\epsilon}_3}\cdot \bold{k}_2)^n.
\end{multline}

The above representation has the nice property that the helicity-$\left(J-J^\prime\right)$ component can be obtained from the helicity-$0$ component of the three-point function with an operator of spin-$J^\prime$:
\begin{multline}
     {}^{\left(J-J^\prime\right)}{\cal A}^{\left(x\right)}_{\nu_1,0;\nu_2,0;\nu_3,J}\left(s_1,{\bf k}_1;s_2,{\bf k}_2;s_3,{\bf k}_3,\boldsymbol{\epsilon}_3\right)= \left(-\frac{i}{2}\right)^{J-J^\prime} \left(\zeta_3 \cdot {\bf k}_{12}\right)^{J-J^\prime}\\ \times {}^{\left(0\right)}{\cal A}^{\left(x\right)}_{\nu_1,0;\nu_2,0;\nu_3,J^\prime}\left(s_1,{\bf k}_1;s_2,{\bf k}_2;s_3,{\bf k}_3,\boldsymbol{\epsilon}_3\right).
\end{multline}
This identity in particular implies that the highest helicity component of a spin-$J$ three-point function can be obtained from the three-point function in $x$ boundary dimensions with the spin-$J$ operator replaced with a scalar operator $\left(J^{\prime}=0\right)$ of the same scaling dimension.

Likewise, the helicity-$\left(J-1\right)$ and -$\left(J-2\right)$ components can be obtained from the helicity-$0$ component of the spin-1 and -2 three-point functions. This property allowed us in section \ref{sec:3ptWard} to obtain Ward-Takahashi identities for (partially-)massless fields of arbitrary spin-$J$ from a computation that is of no greater complexity than that for spins-1 and 2.

\section{Bulk quartic contact terms from improvements}\label{App:U_int}

In this appendix we detail the evaluation of the $w$-integral in \eqref{leqgeq} for improvement terms \eqref{4ptimprove}. As we saw in section \ref{ContactTerms}, such terms give rise to bulk quartic contact terms in the four-point exchange. The key is to decompose the $u$, ${\bar u}$-dependence of the improvements \eqref{4ptimprove} in the Pochhammer basis \eqref{3ptimpbasis}:
\begin{subequations}
\begin{align}\label{imp1}
    p^{\text{impr.}}\left(s_1,s_2,u\right) &= \sum_{n}c_{n}\left(s_1,s_2\right)\left(u-\tfrac{i\nu}{2}\right)_n,\\
    {\bar p}^{\text{impr.}}\left(s_3,s_4,{\bar u}\right) &= \sum_{{\bar n}}{\bar c}_{{\bar n}}\left(s_3,s_4\right)\left({\bar u}-\tfrac{i\nu}{2}\right)_{\bar n},
\end{align}
\end{subequations}
where $c_{n}\left(s_1,s_2\right)$ and ${\bar c}_{{\bar n}}\left(s_3,s_4\right)$ are polynomials in $s_1, s_2$ and $s_3, s_4$ respectively. The integral over $w$ can then be evaluated using the following identity (which is proven below):
\begin{align}\label{Uint2}
\int^{+i\infty}_{-i\infty} & \frac{dw}{2\pi i}\,\sin \left(\pi \left({\bar u}-u\right)\right)\sin \left(\tfrac{\pi}{2}  \left(i \nu_1+i \nu_2+i \nu_3+i \nu_4+\tfrac{x+{\bar x}}{2}-2\left(u+{\bar u}\right)\right)\right)\\ \nonumber
& \hspace*{2cm}\times \left(u-\tfrac{i\nu}{2}\right)_n\left({\bar u}-\tfrac{i\nu}{2}\right)_{\bar n}\,\rho_{\nu,\nu}\left(u,{\bar u}\right)\left(\frac{k_{\sf s}}{2}\right)^{-2\left(u+{\bar u}\right)}\Bigg|_{{}^{{\bar u}=\tfrac{x+4w}{4}-s_3-s_4}_{u=\tfrac{x-4w}{4}-s_1-s_2}} \\\nonumber&\hspace*{-1cm}=\sin \left(\tfrac{\pi}{2} \left( i\nu_1+i\nu_2+i\nu_3+i\nu_4+\tfrac{x+{\bar x}}{2}\right)\right)\\
\nonumber&\hspace*{-1cm}\times \left[\sum_{j=0}^{\bar{n}-1}\left(\frac{k^2_{\sf s}}{4}\right)^{(n+j)}\frac{(-1)^{\bar{n}+j+1}  (n-\bar{n}+j+1)_j \Gamma (i\nu-n-j)}{j! \Gamma (1-\bar{n}+j+i \nu )}i \pi\,\delta\left(s_1+s_2+s_3+s_4-\tfrac{x+\bar{x}}4-n-j\right)\right.
\\\nonumber&\left.\hspace*{-1cm}+\sum_{j=0}^{{n}-1}\left(\frac{k^2_{\sf s}}{4}\right)^{({\bar n}+j)}\frac{(-1)^{\bar{n}+j+1}  (\bar{n}-n+j+1)_j \Gamma (i\nu+n-j )}{j! \Gamma (1+\bar{n}+j-i \nu )}i \pi\,\delta\left(s_1+s_2+s_3+s_4-\tfrac{x+\bar{x}}4-\bar{n}-j\right)\right]\,.\nonumber
\end{align}
Dirac delta functions of the above form are the signature of bulk quartic contact terms at the level of the Mellin-Barnes representation, which was explained in section \ref{ContactTerms}. In section \ref{ContactTerms} this identity was given only for improvements \eqref{imp1}, which is obtained from the above by setting ${\bar n}=0$.

There is an important subtlety in the evaluation of the $w$-integral \eqref{Uint2}. In particular, closing the integration contour on either the positive or negative real axis and summing the residues of the poles that are enclosed naively gives a vanishing result! However, one notes that the $w$-integral has poor behaviour at infinity of the form
\begin{align}
    \sim R^{\tfrac{x+\bar{x}}{2}-2-2\Re\left(s_1+s_2+s_3+s_4\right)}\,, \qquad R \to \infty,
\end{align}
which gives an additional pole in the remaining Mellin variables $s_i$. This suggest that, rather than vanishing, the $w$-integral should be defined as a distribution.

Looking closer at the sum over the residues and using the Gauss summation theorem, it turns out that the contours for Mellin-Barnes integrals in the variables $s_i$ cannot be chosen in the same way for all terms in the sum.\footnote{This observation was also made in \cite{Sleight:2019hfp}.} For this reason, it is actually incorrect to directly sum the residues coming from evaluating the $w$-integral -- which is what would give a vanishing result. One should first bring the integration contours in $s_i$ to the same path for all terms, for which one should carefully take into account the residues of the poles that are crossed in the process. This is what gives a non-vanishing result and is how distributions are encoded in Mellin-Barnes integrals! More in detail, for the case at hand, after summing over the residues of poles:
\begin{subequations}
\begin{align}
    w&= \frac{1}{4} (-4 \bar{n}-4 j+4 s_3+4 s_4-\bar{x})+\frac{i \nu }{2}\,,& j&\in\mathbb{N}\,,\\
    w&= \frac{1}{4} (-4 j+4 s_3+4 s_4-\bar{x})-\frac{i \nu }{2}\,,& j&\in\mathbb{N}\,,
\end{align}
\end{subequations}
and using the Gauss summation theorem to sum the two series one is left with two contributions. One proportional to
\begin{align}
    \Gamma \left(\tfrac{4 n-4 s_1-4 s_3-4 s_2-4 s_4+x+\bar{x}}{4}\right) \Gamma \left(-n-\bar{n}+2 s_1+2 s_3+2 s_2+2 s_4-\tfrac{x}{2}-\tfrac{\bar{x}}{2}+1\right)\,,
\end{align}
and the second proportional to
\begin{align}
    \Gamma \left(\tfrac{4 n-4 s_1-4 s_3-4 s_2-4 s_4+x+\bar{x}}{4}\right) \Gamma \left(-n-\bar{n}+2 s_1+2 s_3+2 s_2+2 s_4-\tfrac{x}{2}-\tfrac{\bar{x}}{2}+1\right)\,.
\end{align}
We can then explicitly see that in order to bring the contour to the same path for both terms we must pass the following poles:
\begin{align}
    4\left(s_1+s_2+s_3+s_4\right)&= x+\bar{x}+j+n\,,& 0\leq j\leq \bar{n}-1\,,
\end{align}
for the first term and
\begin{align}
    4\left(s_1+s_2+s_3+s_4\right)&= x+\bar{x}+j+\bar{n}\,,& 0\leq j\leq n-1\,,
\end{align}
for the second. Gathering their residues one arrives to eq.~\eqref{Uint2}.

\bibliographystyle{JHEP}
\bibliography{refs}

\end{document}